\documentclass [11pt, proquest] {uwthesis}[2017/05/05]
 
\usepackage{alltt}  %

\usepackage{hyperref,amssymb,amsmath,graphicx,xcolor}
\usepackage{bm}
\usepackage{mathtools}

\def\si{^1 \hskip -0.03in S _0}

\newcommand{\ket}[1]{\left| #1 \right>} 
\newcommand{\braket}[2]{\left< #1 \vphantom{#2} \right|
 \left. #2 \vphantom{#1} \right>} 
\newcommand{\mbraket}[3]{\left< #1 \vphantom{#2#3} \right|
 #2 \left| #3 \vphantom{#1#2} \right>} 
\newcommand{\tr}{\text{Tr}} 
\let\bar=\smallbar 
\newcommand{\bar}[1]{\overline{#1}} 
\newcommand{\fs}[1]{\slashed{#1}} 
\let\tilde=\widetilde

\usepackage{amssymb,latexsym}
\usepackage{amsmath,amsbsy,bbm}
\usepackage{subfigure}
\usepackage{slashed,array,mathtools,amsfonts,multirow,hyperref}
\usepackage[utf8]{inputenc}
\DeclareUnicodeCharacter{00A0}{~}

\renewcommand{\v}[1]{\ensuremath{\mathbf{#1}}} 

\newcommand{\uv}[1]{\ensuremath{\mathbf{\hat{#1}}}} 
\newcommand{\avg}[1]{\left< #1 \right>} 

\begin{document}
 
%
%

\prelimpages
 
%
%
\Title{Statistical Angles on the Lattice QCD Signal-to-Noise Problem}
\Author{Michael L. Wagman}
\Year{2017}
\Program{Physics}

\Chair{Martin J. Savage}{Prof.}{Physics}
\Signature{Silas R. Beane}
\Signature{David B. Kaplan}

\copyrightpage

\titlepage

%
%

%
%

\setcounter{page}{-1}
\abstract{%
  The theory of quantum chromodynamics (QCD) encodes the strong interactions that bind quarks and gluons into nucleons and that bind nucleons into nuclei.
  Predictive control of QCD would allow nuclear structure and reactions as well as properties of supernovae and neutron stars to be theoretically studied from first principles.
  Lattice QCD (LQCD) can represent generic QCD predictions in terms of well-defined path integrals,
  but the sign and signal-to-noise problems have obstructed LQCD calculations of large nuclei and nuclear matter in practice.
  This thesis presents a statistical study of LQCD correlation functions, with a 
  particular focus on characterizing the structure of the noise associated with quantum fluctuations.
  The signal-to-noise problem in baryon correlation functions is demonstrated to arise from a sign problem associated with Monte Carlo sampling of complex correlation functions.
  Properties of circular statistics are used to understand the emergence of a large time noise region where standard energy measurements are unreliable.
  Power-law tails in the distribution of baryon correlation functions, associated with stable distributions and L{\'e}vy flights,
  are found to play a central role in their time evolution.

  Building on these observations, a
  new statistical analysis technique called phase reweighting is introduced
  that allow energy levels to be extracted from large-time correlation functions with time-independent signal-to-noise ratios.
  Phase reweighting effectively includes dynamical refinement of source magnitudes
  but introduces a bias
  associated with the phase.
  This bias can be removed
  by performing 
  an extrapolation,
  but at the expense of
  re-introducing a signal-to-noise problem.
  Lattice QCD calculations 
  of the $\rho^+$ and nucleon masses and of the $\Xi\Xi(\si)$ binding energy
  show consistency between standard results obtained using smaller-time correlation functions and
  phase-reweighted results using large-time correlation functions inaccessible to standard statistical analysis methods.
  A detailed study of the statistics and phase reweighting of isovector meson correlation functions demonstrates that phase reweighting applies to real but non-positive correlation functions and can be used to predict ground-state energies of correlation functions that are too noisy to be analyzed by other methods. 
  The relative precision of phase reweighting compared to standard methods is expected to be increased on 
  lattices with larger time directions than those considered in this thesis, 
  and these preliminary studies suggest phase reweighting of noisy nuclear correlation functions 
  should be investigated on larger lattices. 
  The results of this thesis suggest 
  that phase reweighting may be applicable more broadly to real but non-positive correlation functions in quantum Monte Carlo simulations of particle, nuclear, and condensed matter physics systems as well as to complex correlation functions describing multi-baryon systems in LQCD.
}

%
%
\tableofcontents
 
%
%
%
 
%

\acknowledgments{
 {\narrower\noindent
   An old proverb says that it takes a village to raise a child.
   In my experience, raising a graduate student all the way through twenty-first grade requires not just one but several villages.
   
   First and foremost, I am grateful to my advisor Prof. Martin Savage
   for teaching by example how to
   always stay curious about the physical world,
   turn fuzzy questions into crisp ideas,
   and push through whatever unexpected obstacles arise to
   finish a worthwhile project in however long it takes.
   Martin's focus, commitment, and sense of justice have been inspirational to me,
   and his training has given me a much deeper understanding of
   both how to calculate anything
   and how to assess what is actually worth calculating in order to learn more about nature.

   My earliest teachers were my parents Kathy and Ed and my sister Nicole.
   Their encouragement and
   willingness to keep answering ``why?''
   built the foundation of my education,
   and the worlds of spaceships and superheros that I first explored with Nicole
   taught me to think of science as exciting and anything as possible.

   A number of other teachers have played essential roles in my education.
   I am forever grateful to my middle school teacher Mrs. Graham,
   who opened my eyes to the wonders of mathematics
   and taught me to think critically and independently about every subject from math to literature.
   I am likewise grateful to the high school teachers who nurtured my passion for learning,
   especially Ms. Hawthorn, Mr. Young, Mr. Davis, Mr. Goldsborough,
   and my AP physics teacher Mr. Rowe for confusing and de-confusing me about what an electric field is
   and sharing an infectious enthusiasm for physics.
   I would also like to wholeheartedly thank my classmates,
   especially my friends from Mrs. Graham's class, my collaborators in TSA, and my band mates,
   for making learning a fun collaborative project
   and pondering life's big philosophical questions with me as if we could solve them all.

   My time at Brown University was full of lessons from great teachers.
   I am particularly grateful to Prof. James Valles for
   his encouragement and mentoring
   and thank him, Ilyong Jung, and Shawna Hollen for introducing me to the world of research.
   I am also grateful to Prof. Jan Hesthaven and Scott Field for
   nurturing my interests in theoretical and computational physics research.
   I learned many lessons about life and about physics
   from my friends and roommates at Brown,
   and I cannot thank them enough for their love and support as well as countless stimulating discussions. 
   In particular I thank Matt Dodelson for convincing me to take any math course that looked interesting
   regardless of the prerequisites
   and thank him, Adam Coogan, and Tom Iadecola for helping me survive the ensuing all-nighters.

   During graduate school, I benefited greatly from the teachings of
   Profs. Silas Beane, David Kaplan, Andreas Karch, Gerald Miller, Ann Nelson, Sanjay Reddy, Martin Savage, Steve Sharpe, and Larry Yaffe.
   Much of my learning was done alongside my physics-life-buddy Dorota Grabowska,
   with whom I look forward to many more dinner table conversations about the foundations of quantum field theory,
   and many dear friends from my graduate-school cohort and fellow residents of Physics House.
   I am deeply grateful to the other members of the NPLQCD collaboration:
   Silas Beane, Emmanuel Chang, Zohreh Davoudi, Will Detmold, Kostas Orginos, Assumpta Parre{\~n}o, Martin Savage, Phiala Shanahan, Brian Tiburzi, and Frank Winter
   for welcoming me into an inspirational family of collaborators.
   I thank Michael Buchoff for the countless hours we spent following half of ``shut up and calculate''
   and for guiding me through the early stages of entering the physics community.
   I also thank Ra{\'u}l Brice{\~n}o, Aleksey Cherman, Max Hansen, Natalie Klco, and Alessandro Roggero
   for their guidance and collaboration and for the positive energy that they bring to the world.
   I am also grateful for my opportunities to meet friends and collaborators outside UW,
   in particular Sri Sen and my TASI cohort.
   The research presented in this thesis has also benefited from conversations with
   Tanmony Bhattacharya, Joe Carlson, Tom DeGrand, Leonardo Giusti, Rajan Gupta, Dean Lee, Alessandro Lovato, Steve Peiper, Daniel Trewartha, and Bob Wiringa,
   and would not have been possible without inspiration from David Kaplan to dive into learning about the signal-to-noise problem.

   Last but far from least, I thank my partner Kelly Buckley for her support during the highs and the lows of research,
   for helping me find big picture explanations,
   and for joining me on our next adventure.
 \par}
}

%

\dedication{\begin{center}To teachers who can explain quantum field theory and to students who find it beautiful. \end{center}}

 
\listoffigures

%
%

\textpages
 

\chapter{Introduction}
 
Fluctuations of quantum fields throughout spacetime lead to a seemingly classical universe ultimately governed by probabilistic laws.
Quantum field theory (QFT) describes statistical distributions for experimental results consistent with the laws of quantum mechanics and special relativity.
Experiments to date are well-described by a quantum field theory (QFT) called the Standard Model (SM) of particle physics in conjunction with general relativity.
There are notable exceptions,
namely neutrino masses, dark matter, dark energy, and  quantum gravity, 
and understanding the nature of these exceptions is a central goal of modern physics.
Another aim is to understand the connection between microscopic fields and macroscopic matter.
These are not independent goals, and relating macroscopic properties of matter to statistical functions of SM fields allows SM predictions to be precisely tested and new physics beyond the Standard Model (BSM) to be potentially discovered.
Connecting microscopic fields and macroscopic observables also allows matter in extreme environments to be theoretically understood and its properties predicted;
for example high-frequency gravitational waves emitted from neutron star collisions
that are expected to be observed at LIGO
reflect in part the microscopic equation of state of quarks and gluons.

The electron field in the SM is associated with quantum states containing definite numbers of point-like charged particles called electrons.
The quark field is superficially similar to three copies of the electron field with distinct charges called colors.
However, strong quantum fluctuations in the vacuum confine quarks into color-neutral composite objects called hadrons:
mesons such as pions, baryons such as protons and neutrons, and their bound states such as atomic nuclei.
Much effort in nuclear and particle physics is directed towards understanding the connection between experimentally observable hadrons and nuclei and the theory of fluctuating quark and gluon fields.
Establishing this connection quantitatively allows
reliable theoretical calculations of nuclear structure and reactions
to be made without further assumptions besides the validity of the SM.
Reliable predictions can in turn be used to 
better understand exotic astrophysical environments such as neutron stars and supernovae,
improve models of nuclear forces and theoretical inputs to models of fusion and fission in stars and reactors,
and inform experimental searches for BSM physics relying on precise measurements of hadrons and nuclei.

In high-energy particle collisions, the SM can be accurately solved using perturbative techniques 
where quantum fluctuations are assumed to make small corrections to classical results.
In low energy density regions such as everyday materials,
and even at higher energy density regions like the center of the sun, 
strong nuclear interactions described in the SM by quantum chromodynamics (QCD) cannot be treated perturbatively.
The only known method for reliably solving QCD non-perturbatively for generic low-energy systems is lattice QCD (LQCD).
In LQCD, quark and gluon fields are stochastically sampled on a discrete set of points forming a spacetime lattice using Monte Carlo techniques.
Physical observables are identified with functions of the quark and gluon fields averaged over quantum fluctuations.
Predictions can be directly compared with experiment
in order to identify and understand strong interaction effects in particle and nuclear physics.
LQCD predictions can also be taken as input to effective field theory (EFT) and quantum many-body methods
that can then predict the properties of other systems with controlled uncertainties.

Broad nuclear theory efforts will benefit from the ability of LQCD to
reliably calculate properties of strongly interacting matter 
that are difficult or impossible to access experimentally.
Stellar fusion and other astrophysical reactions responsible for the synthesis of heavy elements take place at high temperatures difficult to create on earth, 
and in some cases fusion cross-sections must be determined phenomenologically by nuclear theory.
Precise LQCD calculations of electroweak fusion reactions would allow these cross-sections to be determined from first principles.
Calculations of other electroweak reactions including neutrino-nucleus scattering would have applications to experimental neutrino efforts such as DUNE and non-proliferation efforts.
LQCD can also provide predictions of the quark and gluon structure of nuclei that can be tested by and used to interpret results from a proposed electron-ion collider (EIC).
Precise LQCD calculations of even simple systems can be used to determine poorly-known parameters in nuclear many-body models,
and for example precise determinations of three-nucleon forces would improve the accuracy of nuclear structure and reaction calculations of heavy nuclei at FRIB.
Future LQCD studies of fusion reactions involving more complex systems
can accurately determine poorly known nuclear reaction rates
needed as inputs to nuclear many-body models 
that in turn inform
macroscopic models of supernovae, neutron stars, and reactors.

Nuclear and hadronic experiments searching for BSM physics need LQCD calculations
to reliably connect experimental results
to constraints on BSM theories.
Dark matter direct detection experiments and searches and for fundamental symmetry violation, 
including neutrinoless double-beta decay, neutron electric dipoles moments, proton decay, and neutron-antineutron oscillations, 
require accurate QCD predictions in order to reliably constrain BSM theory.
In fundamental symmetry searches, QCD results are not needed to see a signal of BSM physics, but they are needed to turn experimental results into quantitative predictions to be verified by other experiments and to establish reliable bounds on theory from null results.

In these and other applications of LQCD to particle and nuclear physics,
precise calculations can be performed for few-particle systems,
but calculations of nuclear matter and of large nuclei face infamous obstacles
called the sign problem and the signal-to-noise (StN) problem respectively.
The sign problem refers to issues in
numerically calculating integrals of oscillatory functions.
It arises in Monte Carlo calculations
where results are sensitive to delicate cancellations between opposite sign contributions that are only apparent with high statistics. 
The StN problem refers to the issue of exponential precision loss in calculations of protons, neutrons, and nuclei.
The sign and StN problems have so far obstructed LQCD calculations of the equation of state of cold, dense matter inside neutron stars and the structure and reactions of large nuclei.

This thesis presents statistical observations
to better understand
and statistical techniques to tame the StN problem for baryon and multi-baryon correlation functions in LQCD.
Building on observations that the nucleon StN problem can be associated with a random walk in the phase of complex correlation functions,
new statistical estimators are proposed that possess constant, rather than exponentially degrading, precision but have a bias that must be removed by extrapolation.
These estimators are shown to reproduce LQCD results for single- and multi-particle systems using only late-time correlation functions too noisy to be analyzed by previous methods.

Chapter~\ref{chap:statistics} describes observations of the statistical distributions of nucleon correlation functions in LQCD.
By considering a decomposition of correlation functions into magnitude and phase, the StN problem is shown to follow from the sign problem obstructing Monte Carlo sampling of non-positive definite functions.
The probability distributions of the magnitudes and phases are correlation functions are shown to possess interesting structure,
and in particular time evolution of the phase of the correlation function is observed to resemble a L{\'e}vy flight on the unit circle.
Empirical evidence for heavy-tailed distributions resembling stable distributions is found for 
differences of phases at dynamically correlated times.
Observations from this chapter were previously described in:

\begin{enumerate}
\item M.~L.~Wagman and M.~J.~Savage, ``Taming the Signal-to-Noise Problem in Lattice QCD by Phase Reweighting,'' arXiv:1704.07356 [hep-lat].
\end{enumerate}

Chapter~\ref{chap:PR} describes phase reweighting, an improved statistical estimator for complex correlation functions
motivated by the preceding observations.
Phase reweighting removes the exponential degradation of precision arising from the StN problem,
at the cost of introducing a bias.
The bias can be removed by extrapolating to a well-defined limit.
As this limit is approached the bias becomes exponentially smaller but precision becomes exponentially worse.
First results of phase reweighting for meson, baryon, and multi-baryon systems are shown to be encouraging.
Phase reweighting and its first applications previously appear in:

\begin{enumerate}
\item M.~L.~Wagman and M.~J.~Savage, ``On the Statistics of Baryon Correlation Functions in Lattice QCD,'' arXiv:1611.07643 [hep-lat].
\end{enumerate}

Chapter~\ref{chap:mesons} describes the statistics of real but sometimes negative meson correlation functions in LQCD.
The real part of the wrapped-normal log-normal distribution introduced in Chapter~\ref{chap:statistics} provides a good description of isovector meson correlation functions.
Lepage-Savage scaling is shown to be a generic property of complex correlation functions, and to apply to isovector meson correlation functions.
The sample mean of an ensemble of $N$ isovector meson correlation functions show systematic deviations from the true average unless $\avg{cos \theta_i} \geq 1/\sqrt{N}$ in accordance with the expectations of circular statistics~\cite{Fisher:1995}.
These observations suggest that the picture of L{\'e}vy Flights on the unit circle used to motivate phase reweighting applies, and real but non-positive-definite meson correlation functions can be viewed as projections of complex correlation functions with similar statistical behavior to the baryons.
Phase reweighting has been used to the StN problem for the $\rho^+$ in Chatper~\ref{chap:PR}, and its success as well as the failure of the ratio estimator introduced in Chapter~\ref{chap:statistics} and further explored in this chapter.
The ground state energies of other isovector meson channels, previously inaccessible to techniques based on spectroscopy in the golden window, are precisely extracted from the noise region of phase-reweighted correlation functions.

The remainder of this chapter provides background on QCD in Sec.~\ref{sec:qcd}, lattice field theory in Sec.~\ref{sec:lattice}, and meson and baryon correlation functions in Sec.~\ref{sec:corr} that is helpful for the subsequent chapters.

\section{Quarks and Gluons}\label{sec:qcd}

Quantum field theories generically describe correlations between various sorts of matter throughout spacetime that
are consistent with the general postulates of quantum mechanics and special relativity.
A particular QFT is specified by the number and symmetry transformation properties of fields included to represent these various sorts of matter.
The SM includes fields representing
spin one-half fermions called quarks and leptons,
spin one gauge bosons called gluons, photons, $W^\pm$, and $Z^0$ bosons,
and a spin-zero Higgs boson
experimentally discovered recently at the LHC~\cite{Aad:2012tfa,Chatrchyan:2012xdj}.
The SM's properties are highly constrained by $SU(3)_C\times SU(2)_L\times U(1)_Y$ gauge invariance,
where the photon, $W^\pm$, and $Z^0$ are associated with the $SU(2)_L\times U(1)_Y$ electroweak gauge group~\cite{Glashow:1961tr,Salam:1964ry,Weinberg:1967tq,Glashow:1970gm}
and the gluons are associated with the $SU(3)_C$ QCD gauge group~\cite{Fritzsch:1972jv,Fritzsch:1973pi,Politzer:1973fx,Gross:1973id}.

The effects of heavy particles in QFT decouple from low-energy physics~\cite{Appelquist:1974tg}, and through renormalization\footnote{See Ref.~\cite{Collins:105730} for a clear and comprehensive review of renormalization in QFT with further references to the original literature.} 
can be implicitly included in an EFT 
that only explicitly includes fields for particles with mass lower than a freely chosen renormalization scale $\mu$.
The benefit of considering a low-energy EFT is typically that low-energy dynamics are described more simply; the cost is that only dynamics at energy scales $\lesssim \mu$ are accurately described.\footnote{Introductions to EFT for a wide range of particle, nuclear, atomic, and condensed matter systems can be found in~\cite{Buras:1998raa,Kaplan:2005es,Bedaque:2002mn,Beane:2000fx,Kaplan:1996nv,Scherer:2002tk,Bernard:1995dp,Braaten:2004rn}.}
Nuclear energy scales are typically measured in MeV, and nuclei should be well-described by an EFT of the SM that includes explicit fields for the gluons, photons, and light quarks and leptons valid for $\mu \ll 80$ GeV, though this remains to be verified experimentally. 
The explicit gauge group of this low-energy EFT is $SU(3)_C\times U(1)_{EM}$ where the photon is associated with the gauge group $U(1)_{EM}$ of quantum electrodynamics (QED).
Quantum fluctuations in the photon field do not change the qualitative behavior of classical photons in many systems and can often be included as perturbative corrections to classical electromagnetism.\footnote{Important cases where QED must be treated non-perturbatively include bound states held together by non-perturbative Coulomb forces~\cite{Caswell:1985ui}, near-threshold scattering states of charged particles~\cite{Kong:1999sf}, and charged particles in finite volumes~\cite{Beane:2014qha}.}
Leptons only interact with quarks`, gluons, and one another through photon exchange, and so as a further simplification the interactions of quarks and gluons with leptons and photons can be perturbatively expanded about the non-interacting limit.
Up to $O(\alpha)$ corrections, the low-energy physics of light nuclei should be accurately described by QCD.

SM matter can exist in qualitatively different thermodynamic phases.
Normal matter exists in a phase
where QCD is confining\footnote{Precisely defining confinement is subtle, see Ref.~\cite{Greensite:2003bk}.} and
electroweak interactions are screened by the Higgs mechanism~\cite{Higgs:1964ia,Higgs:1964pj,Higgs:1966ev,Englert:1964et}.
Other regions of the SM phase diagram
where matter has dramatically different properties exist,
for instance the deconfined quark-gluon plasma experimentally created at RHIC~\cite{Arsene:2004fa,Back:2004je,Adcox:2004mh} and the LHC~\cite{Aamodt:2008zz}, 
which LQCD calculations predict forms above $T\sim 160$ MeV~\cite{McLerran:1980pk,Kuti:1980gh,Engels:1980ty,Kajantie:1981wh,Kogut:1982rt,Pisarski:1983ms,Celik:1983wz,Aoki:2006we,Borsanyi:2010bp,Bazavov:2011nk,Bhattacharya:2014ara}.
Extremely dense astrophysical environments such as the interior of neutron stars may also contain matter in exotic phases of the SM~\cite{Collins:1974ky,Alford:1998mk,Alford:1997zt,Rapp:1997zu,Schafer:1998ef}.
Calculating the properties of cold, dense, strongly interacting matter from first principles
but is an essential step towards understanding nuclei, nuclear matter, and neutron stars
as highly entangled, emergent states
of quarks, gluons, and other SM fields.
Reliable calculations of the QCD equation of state for cold, dense matter will be required to understand gravitational wave signals from neutron star mergers~\cite{Flanagan:2007ix,Damour:2012yf}, 
a timely goal for nuclear theory now that gravitational waves from binary black hole mergers have been observed at LIGO~\cite{Abbott:2016blz}.
LQCD calculations can describe cold, dense, strongly interacting matter in principle,
but have long been obstructed by the sign problem in practice~\cite{Gibbs:1986ut}.

In it's confined phase, the lowest-energy states in QCD describe color-singlet particles:
mesons that have the conserved charges of a quark-antiquark pair,
baryons that have the conserved charges of three quarks,
bound states such as nuclei,
glueballs,
and other exotic particles.
Gluon number is not conserved in QCD, so in addition to the minimum number of quarks that a hadron must have by symmetry, 
hadrons consistent of an indeterminate, fluctuating number of gluons and quark-antiquark pairs.
An unphysical but illuminating version of QCD with $N_f$ massless quark fields has the global symmetry group $SU(N_f)_L\times SU(N_f)_R\times U(1)_Q^{N_f}$.
The $N_f$ copies of $U(1)_Q$ are associated with conservation of the total quark minus antiquark number for each quark flavor.
The QCD action is also invariant under flavor-singlet axial transformations, but the quantum theory is not and quark number is not separately conserved for positive and negative chirality quarks~\cite{Adler:1969gk,Bell:1969ts}.
Weak interactions mix quark flavors, and $U(1)_Q^{N_f}$ is broken to the subgroup $U(1)_B$ that acts identically on all quark flavors when electroweak interactions are included.
The conserved charge $B$ associated with $U(1)_B$ is called baryon number and is equal to one-third the total quark number.
Non-perturbative electroweak effects can mix baryons with leptons, breaking baryon number $U(1)_B$ and lepton number $U(1)_L$ to the subgroup $U(1)_{B-L}$, but violations of baryon number are negligible at accessible energies~\cite{tHooft:1976up}.

Chiral symmetry $SU(N_f)_L\times SU(N_f)_R$ transformations mix different flavors and chiralities of massless quarks.
Not all of these transformations are associated with conserved charges.
The low-temperature QCD vacuum is only invariant under a subgroup $SU(N_f)_V \subset SU(N_f)_L\times SU(N_f)_R$ of transformations that do not mix chirality~\cite{Nambu:1961fr,Nambu:1961tp,Vafa:1983tf,Vafa:1984xg}.
Symmetries that not preserved by the vacuum are said to be spontaneously broken and are associated with massless particles instead of conservation laws.
In nature, quarks are massive and $SU(N_f)_L\times SU(N_f)_R$ is explicitly broken.
When considering the dynamics of massless Nambu-Goldstone bosons, explicit $SU(N_f)\times SU(N_f)_R$ breaking by quark masses can be understood as a small perturbation leading to approximately massless pseudo-Nambu-Goldstone bosons.
The up and down quarks are the lightest quarks, and $SU(2)_L\times SU(2)_R$ is a good approximate symmetry where $SU(2)_V$ can be associated with approximate conservation of isospin.
The lightest mesons, the pions, can be accurately described as pseudo-Goldstone bosons in chiral perturbation theory ($\chi$PT), the low-energy EFT of spontaneously broken chiral symmetry~\cite{Weinberg:1978kz,Gasser:1983yg,Gasser:1984gg}.
The strange quark is heavier, but is still light compared to hadronic scales and the pions, kaons, and eta can be understood as an octet of pseudo-Nambu-Goldstone bosons in three-flavor $\chi$PT~\cite{Gasser:1984gg}.
Since the pion is the lightest state in QCD, it's Compoton wavelength $m_\pi^{-1}$ sets the longest correlation length in the QCD vacuum.
$m_{\pi}^{-1}$ also sets correlation length for widely separated color-singlet hadrons and the long-distance behavior of nuclear forces.
Low-energy baryon dynamics are described by EFT($\fs{\pi}$), an EFT that accurately describes baryon-baryon scattering with energy and momentum transfer much less than $m_\pi$.
Attempts to combine pions and nucleons into a convergent EFT for low-energy nuclear physics have a long history~\cite{Gasser:1987rb,Jenkins:1990jv,Weinberg:1990rz,Weinberg:1991um,Kaplan:1996xu,Kaplan:1998tg,Kaplan:1998we,Bedaque:1998kg,Fleming:1999ee,Beane:2001bc}, and are still under active investigation~\cite{Epelbaum:2008ga}.

The quark and gluon fields of QCD are tensors with components representing each spin, color, and flavor.
Lorentz spinor fields $q(x)$ representing quarks transform in the fundamental representation of the algebra $\mathfrak{su}(3)_C$, spinor fields $\bar{q}(x)$ representing  transform in the antifundamental representation, and Lorentz vector fields $G_\mu(x)$ representing gluons transform in the adjoint representation.
With $N_f$ quark flavors explicitly represented and $N_c=3$ colors, $q(x)$ is a $4 N_c N_f$ component vector with three color states, four spin states, and $N_f$ flavor states, while each of the four spin components of $G_\mu(x)$ is as a $N_c\times N_c$ anti-Hermitian matrix.
In Minkowski spacetime,
QCD states are represented by vectors in a Hilbert space
and time evolution is described by a unitary operator $e^{-i H_{QCD} t}$, where $H_{QCD}$ is the QCD Hamiltonian and $t$ is the duration of the time evolution.
The probability that an initial quantum state $\ket{I}$ prepared at time $t=0$ dynamically evolves into a final quantum state $\ket{F}$ at time $t$ is the squared magnitude of the amplitude $\mbraket{F}{e^{-iH_{QCD}t}}{I}$.
Denoting the initial state field configurations by $q_I(x),\;\bar{q}_I(x)$ and $G_I(x)$ and the final state field configurations by $q_F(x)$, $\bar{q}_F(x)$, and $G_F(x)$, this generic amplitude can be represented by the path integral
\begin{equation}
  \begin{split}
    \mbraket{F}{e^{-iH_{QCD}t}}{I} = \int_{q=q_I,\bar{q}=\bar{q}_I,G=G_I}^{q=q_F,\bar{q}=\bar{q}_F,G=G_F} \mathcal{D}G\mathcal{D}q\mathcal{D}\bar{q} e^{iS_{QCD}[q,\bar{q},G]},
  \end{split}\label{eq:pathintegraldef}
\end{equation}
where $S_{QCD}$ is the Minkowski-space QCD action
\begin{equation}
  \begin{split}
    S_{QCD}(q,\bar{q},G) = \int d^4x \left[ \frac{1}{2g^2}\tr\left(G_{\mu\nu}(x)G^{\mu\nu}(x)\right)  + \bar{q}(x)\left(\fs{D} - m_q\right)q(x)\right].
  \end{split}\label{eq:minkowskiactiondef}
\end{equation}
The QCD action involves the covariant derivative
\begin{equation}
  D_\mu = \partial_\mu + G_\mu,
  \label{eq:covariantderivativedef}
\end{equation}
responsible for parallel transport of quark color vectors, as well as the gluon field strength tensor
\begin{equation}
  \begin{split}
    G_{\mu\nu} &= [D_\mu, D_\nu] = \partial_\mu G_\nu - \partial_\nu G_\mu + [G_\mu, G_\nu],\\
\end{split}
  \label{eq:fieldstrengthdef}
\end{equation}
The Dirac operator is $\fs{D} = \gamma^\mu D_\mu$ where $\gamma^\mu$ represent a Lorentz vector of $4\times 4$ spin matrices satisfying $\{\gamma_\mu,\gamma_\mu\} = 2g_{\mu\nu}$ with $g_{\mu\nu} = \text{diag}(-1,1,1,1)$. 
Quarks are taken to be in mass eigenstates where the quark mass matrix $m_q$ is diagonal in spin, flavor, and color. 
The mass of each quark flavor is an input parameter of QCD. 
$g$ denotes the bare gauge coupling, an input parameter whose value, as discussed below, sets the scale of hadronic and nuclear masses and energies in physical units.
Formally defining the path integral measure and divergent-looking oscillatory integrals in Eq.~\eqref{eq:pathintegraldef} requires non-perturbative regularization with the methods of lattice field theory and is deferred to Sec.~\ref{sec:lattice}.

Real-time QCD path integrals cannot be calculated exactly or numerically with known methods because of the sign problem,  
but perturbation theory and other approximations can be used to understand semi-classical fluctuations that illuminate the structure of QCD~\cite{Hooft:1972fi}.
The zero-coupling limit of QCD is a free field theory that can be solved exactly.
Expanding the integrands of path integrals about the zero field configuration (or another saddle point of the action) in powers of the QCD coupling constant provides a method of deriving asymptotic expansions to path integrals valid when the QCD coupling is weak~\cite{tHooft:1977xjm}.
The utility of a weak-coupling expansion for so-called strong interactions may not be obvious upon first glance, but
at very short distances such as the interaction region of a high energy collision, the QCD interactions of quarks become perturbatively weak~\cite{Politzer:1973fx,Gross:1973id}.

\begin{figure}[!ht]
  \begin{center}
  \includegraphics[width=.7\columnwidth]{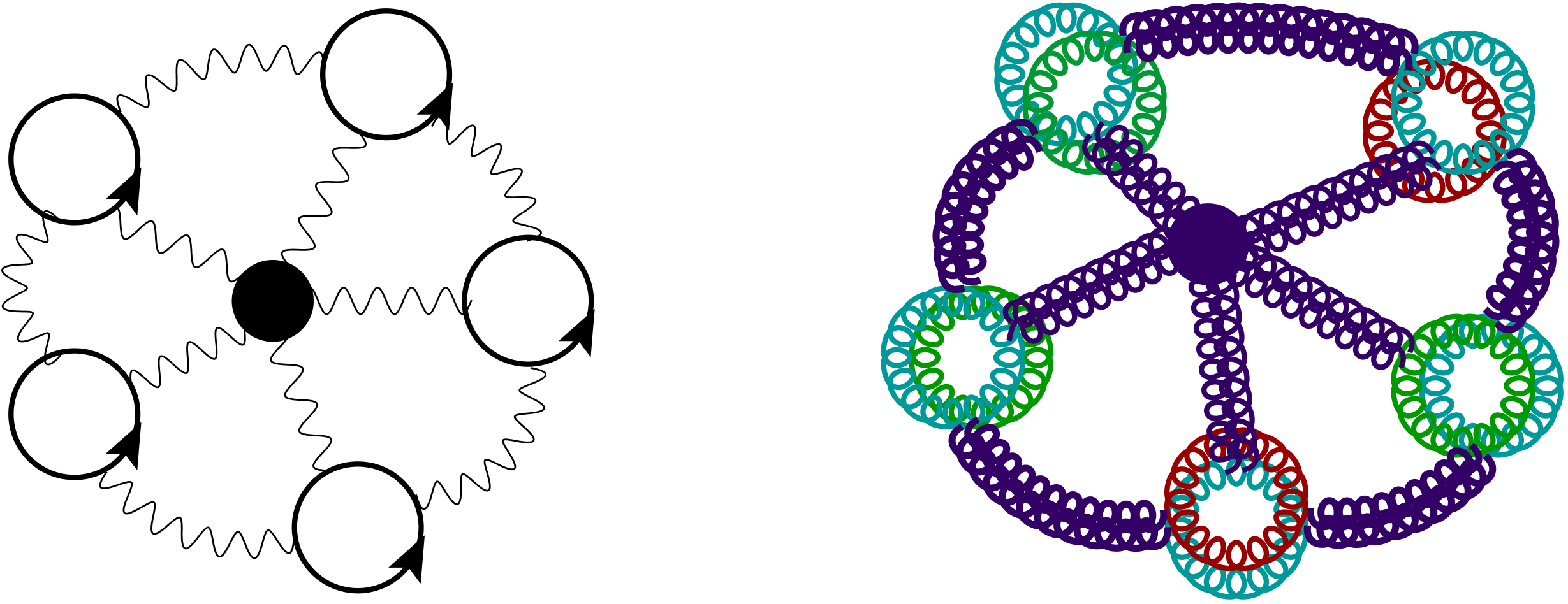}
  \end{center}
  \caption{
    Perturbative fluctuations of the QED vacuum, left, and QCD vacuum, right, about a source of charge in the vacuum.
    Fluctuations of electron-positron pairs in QED are more likely to be oriented with the opposite charge closer to the test charge at center,
    and the QED vacuum accordingly acts as a dielectric where charge is slightly screened at large distances compared to classical expectations.
    In QCD, quark-antiquark pairs act similarly, but fluctuations of gluon fields also contribute.
    A gluon source, right, induces vacuum fluctuations from gluons in orthogonal color orientations that behave like charged vector particle loops.
    Both the QED and QCD configurations shown act as paramagnets as well as dielectrics if the spins of the fluctuations are aligned.
    Pauli blocking raises the energy of each spin-aligned fermion, but no such effect occurs for bosons.
    Explicit calculation shows that paramagnetic effects lead to an overall decrease in the vacuum energy if color charge is effectively enhanced, rather than screened, at large distances~\cite{Appelquist:1977tw,opac-b1131978}.}
  \label{fig:qcdmagnet}
\end{figure}

This property, known as asymptotic freedom, arises from perturbative quantum fluctuations of the gluon field.
Perturbative quantum fluctuations give rise to vacuum polarization effects that modify the effective color charge appearing for example in the chromoelectric Coulomb's law, see Fig.~\ref{fig:qcdmagnet} and for further discussion Refs.~\cite{Appelquist:1977tw,opac-b1131978}.
The modifications are simplest in momentum space, where a well-known calculation shows that the effective coupling $\alpha_s(\mu^\prime) = g(\mu^\prime)^2/(4\pi)$ evaluated at an energy scale $\mu^\prime$ is related to the effective coupling at a different energy scale $\mu$ by~\cite{Politzer:1973fx,Gross:1973id}
\begin{equation}
  \begin{split}
    \alpha_s(\mu^\prime) = \frac{\alpha_s(\mu)}{1 + \frac{\alpha_s(\mu)}{4\pi}(\frac{11}{3}N_c - \frac{2}{3}N_f )\ln(\mu^2/\mu^{\prime 2})},
  \end{split}\label{eq:alphas}
\end{equation}
where $N_c=3$ is the number of colors and $N_f$ is the number of quark fields explicitly included in the theory.
Comparison of the evolution of $\alpha_s(\mu)$ to Eq.~\eqref{eq:alphas} and it's higher-order corrections to results from experiments obtained at a variety of scales provides strong experimental support for QCD at high energies, as summarized in Ref.~\cite{Bethke:2009jm}.
The renormalization group provides useful tools for relating theories that describe different length scales.\footnote{Wen changing renormalization scales from $\mu$ to $\mu^\prime$, the renormalization group can be used to re-sum logarithms such as the one appearing in Eq.~\eqref{eq:alphas} that become large when $\mu/\mu^\prime$ becomes sufficiently large (or sufficiently small). Without this re-summation, perturbation theory fails~\cite{Collins:105730}.}

The one-loop running coupling diverges at a scale
\begin{equation}
  \begin{split}
  \Lambda_{QCD} = \mu \exp\left[ \frac{-1}{\alpha_s(\mu)}\left(\frac{2\pi}{11N_c/3 - 2N_f/3}\right) \right].
  \end{split}\label{eq:LambdaQCDdef}
\end{equation}
Before $\mu$ reaches $\Lambda_{QCD}$ in a limit from above, $\alpha_s(\mu)$ becomes large, neglected two-loop corrections to Eq.~\eqref{eq:alphas} become as important as one-loop corrections, and Eq.\eqref{eq:alphas} becomes unreliable.
The divergence in the one-loop result is not a physical divergence; however, Eq.~\eqref{eq:LambdaQCDdef} allows a physical length scale to be defined from the dimensionless running coupling evaluated at a given scale.
By construction, $\frac{d}{d\mu}\Lambda_{QCD}=0$, so $\Lambda_{QCD}$ is renormalization scale invariant.
It is also independent of the renormalization scheme used to relate $\alpha_s(\mu)$ to a physical observable at the level of perturbative accuracy considered.
This emergence of a physical, dimensionful scale that only depends on the dimensionless gauge coupling, $N_c$, and $N_f$ is known as dimensional transmutation.
This is a non-perturbative phenomenon, signaled by the fact that a perturbative expansion in $\alpha_s(\mu)$ of the RHS of Eq.~\eqref{eq:LambdaQCDdef} vanishes to all orders in $\alpha_s(\mu)$.
For energies and momenta much larger than $\Lambda_{QCD}$, or distances and times much smaller than $\Lambda_{QCD}^{-1}$, the QCD running coupling is small and weak-coupling perturbation theory applies.
The predictive accuracy of weak-coupling expansions for high-energy QCD is demonstrated by comparing the best-fit $\alpha_s(\mu)$ determined from comparing perturbative QCD predictions to the results of various experiments and various $\mu$~\cite{Bethke:2009jm}.
For low-energies and large distances, the QCD running coupling is large and perturbation theory is unreliable.

At larger distances than $\Lambda_{QCD}^{-1}$, the effective potential between static color sources rises approximately linearly rather than falling according to Coulomb's law.
This adds an infinite energy cost to isolating a static color charge, and the static charge is said to be confined.
Confinement of static color charges can be analytically demonstrated for the strong-coupling limit of LQCD~\cite{Wilson:1974sk}.
In the strong coupling limit a linearly rising potential between static color charges arises from gluon field configurations resembling tubes of color flux joining the charges.
There is currently no analytic proof that QCD is confining outside the strong-coupling limit, but a wealth of non-perturbative LQCD results demonstrate that the spectrum of QCD describes bound and scattering states of color-singlet particles.
This is in accordance with experimental non-observation of isolated quarks or gluons.

\begin{figure}[ht!]
  \begin{center}
  \includegraphics[width=\columnwidth]{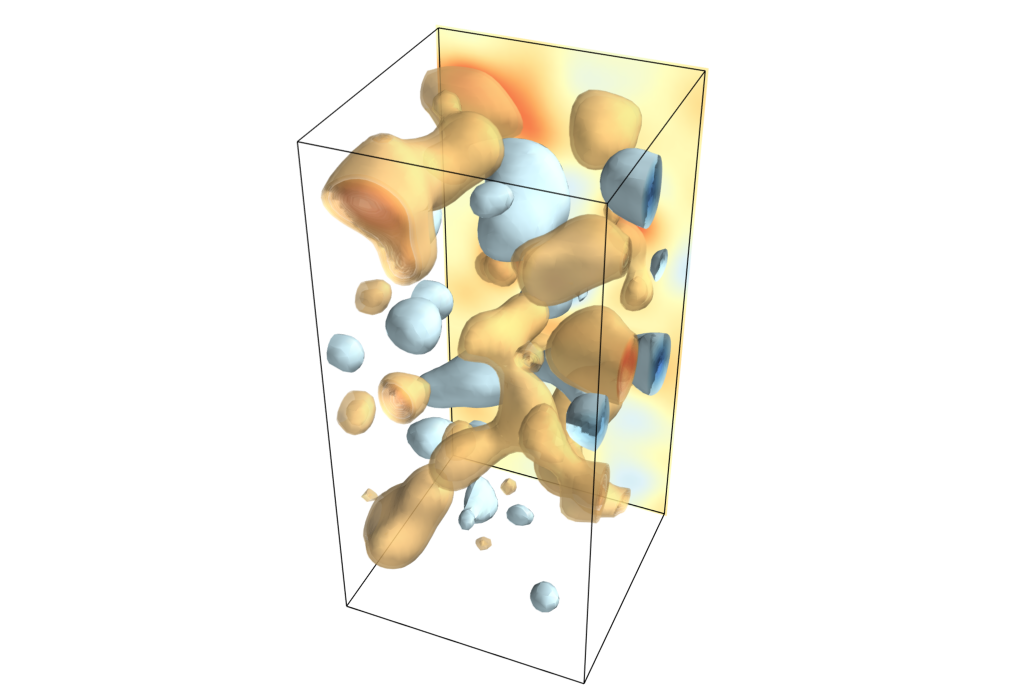}
  \end{center}
  \caption{
    A visualization of the topological charge density of a generic LQCD vacuum gluon configuration, courtesy Daniel Trewartha.
    Cooling has been applied to the gauge field configuration to average over ultraviolet fluctuations.
    The remaining three-dimensional structures that are apparent are localized regions of large positive, orange, or large negative, blue, topological charge density~\cite{Trewartha:2015ida}.}
  \label{fig:topcharge}
\end{figure}

It is noteworthy that Eq.~\eqref{eq:minkowskiactiondef} is not the most general local action comprised of quark and gluon fields that is consistent with Poincar{\'e} invariance and $SU(3)_C$ gauge invariance.
First, there are an infinite number of higher dimension operators that could be included in the action.
The renormalizability of the SM shows that these higher dimensional interactions will not be generated by perturbative fluctuations of SM fields.
Higher dimensional operators can arise when the SM is considered as an EFT for BSM physics, but
renormalization group arguments and EFT suggest that interactions involving higher dimensional operators are suppressed by $(p/\Lambda_{BSM})^n$ where $p$ is a relevant physical momentum scale, $\Lambda_{BSM} \gtrsim 10$ TeV is the cutoff scale where BSM physics could give rise to such interactions, and $n$ is the dimension of the operator.
Second, QCD is not the most general action containing dimension four operators.
Since the full SM does not respect the discrete symmetry of $CP$, an additional $CP$ violating term is expected to appear in the QCD action,
\begin{equation}
  \begin{split}
    S_{\theta} &= -  \bar{\theta}  \frac{N_f}{32\pi^2} \int d^4 x\; \varepsilon^{\mu\nu\alpha\beta}\tr\left[ G_{\mu\nu}(x)G_{\alpha\beta}(x) \right]
  \end{split}\label{eq:thetaactiondef}
\end{equation}
where $\bar{\theta}$ is a free parameter.\footnote{Strictly, we assume a basis where the determinant of the quark mass matrix is real and positive. An anomalous $U(1)_A$ transformation can be used to remove an overall phase from the quark matrix while introducing a shift in the vacuum angle $\theta$. The notation $\bar{\theta}$ is reserved for the vacuum angle in a basis where all $CP$ violation has been shifted from the quark mass matrix to the vacuum angle~\cite{coleman:1988aspects}.}
Spatial integrals of the topological charge density appearing in Eq.~\eqref{eq:thetaactiondef} are constrained to be integers, and count non-perturbative excitations of the gluon field localized in both space and time called instantons.
Single instanton field configurations are associated with saddle points of the (Euclidean) action and can be analyzed semi-classically~\cite{Belavin:1975fg,tHooft:1976fv}.
The thermal vacuum of QCD resembles a gas of instantons at high temperatures~\cite{Gross:1980br}, and it is expected that non-perturbative features of QCD are associated with instantons and multi-instanton gluon field configurations~\cite{Schafer:1996wv}.
Random samples of the topological charge density generated from smoothed LQCD gauge field ensembles suggest the non-perturbative QCD vacuum can be intuitively described as a fluctuating medium with regions of large positive or negative topological charge density localized in spacetime reminiscent of instantons and anti-instantons~\cite{Chu:1994vi,Gattringer:2001ia}.
While local fluctuations in the topological charge density appear commonplace in the QCD vacuum, the non-observation of $CP$ violating strong interactions shows that $\bar{\theta}$ in Eq.~\eqref{eq:thetaactiondef} must be very small or vanish.
Non-zero $\bar{\theta}$ would introduce $CP$ violating interactions such as a neutron electric dipole moment, and experimental constraints place stringent bounds $\bar{\theta} \lesssim 10^{-9}$ ~\cite{He:1990qa,Baker:2006ts}.
LQCD calculations needed to relate experimental observation or bounds of electric dipole moments to rigorous constraints on particular $CP$ violating BSM theory are underway~\cite{Shintani:2005xg,Guo:2015tla,Shintani:2015vsx,Bhattacharya:2016zcn,Gupta:2017anz}.
LQCD calculations can also directly probe theoretically how $\bar{\theta}\neq 0$ affects the physics of QCD.
However, such LQCD calculations are difficult because setting $\bar{\theta}\neq 0$ introduces a sign problem, see Ref.~\cite{Cai:2016eot} for a recent discussion.

\section{Lattice Field Theory}\label{sec:lattice}

In continuous spacetime, any field can occupy an infinite number of momentum states. 
For free fields, each of these momentum modes acts as a harmonic oscillator and contributes to the vacuum energy. 
When quantum fluctuations of this infinite number of modes are considered in perturbative expansions about the free field vacuum,
divergences arise in quantities such as the vacuum energy density.
There are many regularization schemes that can be used to control these divergences and verify that perturbative relationships between physical observables in QFT are finite.
Renormalization group arguments and universality suggest that different regularizations of the same QFT approach the same continuum limit when the regularization scale cutting off high-momentum, ultraviolet (UV) fluctuations is is taken to infinity.

To describe the confinement of quarks into hadrons with QCD, non-perturbative quantum effects beside fluctuations about the free field vacuum must be computed. 
To include non-perturbative quantum fluctuations in path integrals, the measure $\mathcal{D}U\mathcal{D}\bar{q}\mathcal{D}q$ schematically presented in Eq.~\eqref{eq:pathintegraldef} must be concretely defined. 
Lattice regularization is the only known non-perturbative regulator for UV divergences in QCD path integrals.\footnote{Construction of a lattice regularization of chiral gauge theories necessary for a non-perturbative regulator for electroweak interactions is a longstanding challenge that has seen exciting recent development~\cite{Grabowska:2015qpk,Grabowska:2016bis}.}
In lattice field theory, continuous spacetime is replaced by a discrete lattice of points.\footnote{Textbook introductions to lattice field theory can be found in Refs.\cite{creutz:1983quarks,Gupta:1997nd,smit:2002introduction,montvay:1997quantum}. The scalar field theory construction briefly sketched in this section most closely follows the presentation of Montvay and M{\"u}nster~\cite{montvay:1997quantum}.} 
Each spacetime point is associated with a quantum mechanical Hilbert space.
For a theory involving only a single complex scalar field $\varphi(x)$, this Hilbert space can be defined in the coordinate basis where field operators $\hat{\varphi}(x)$ act by
\begin{equation}
  \begin{split}
    \hat{\varphi}(x)\ket{\varphi(x)} = \ket{\varphi(x)}\varphi(x),
  \end{split}\label{eq:scalarstatedef}
\end{equation}
and states are normalized as
\begin{equation}
  \begin{split}
    \braket{\varphi(x)}{\varphi(x^\prime)} = \delta(\text{Re}\varphi(x) - \text{Re}\varphi(x^\prime))\delta(\text{Im}\varphi(x) - \text{Im}\varphi(x^\prime)).
  \end{split}\label{eq:scalarnormdef}
\end{equation}
The overall state of the system at time $t$ is specified the full Hilbert space, a tensor product of the Hilbert space at each spatial point with states 
\begin{equation}
  \begin{split}
    \ket{\varphi_t} &= \prod_\v{x}\ket{\varphi(\v{x},t)}. 
  \end{split}\label{eq:scalarproduct}
\end{equation}
The dynamics of the system are fully specified by the form of the Hamiltonian $H_\varphi$, a Hermitian operator defined on the full Hilbert space.
The path integral formulation of the theory is instead specified by an action $S_\varphi$ and a path integral measure
\begin{equation}
  \begin{split}
    \mathcal{D}\varphi^\dagger\mathcal{D}\varphi = \prod_x d\text{Re}\varphi(x)d\text{Im}\varphi(x).
  \end{split}\label{eq:scalarmeasuredef}
\end{equation}
Hilbert space matrix elements are related to path integrals by
\begin{equation}
  \begin{split}
    \mbraket{\varphi_t}{e^{-i\hat{H}_\varphi t}}{\varphi_0} = \int_{\varphi_0}^{\varphi_t} \mathcal{D}\varphi^\dagger \mathcal{D}\varphi\;e^{iS_\varphi}.
  \end{split}\label{eq:scalarpathintegraldef}
\end{equation}
To make the path integral finite, the lattice field theory can be defined on a spacetime volume of spatial volume $L^3$ and time extent $\beta$.
Eq.~\eqref{eq:scalarpathintegraldef} then requires a large but finite number $2L^3\beta$ of integrals over real variables. 

Stochastic integration methods broadly called Monte Carlo techniques are often useful for evaluating high-dimensional integrals.
These techniques rely on sampling field configurations rather than enumerating and evaluating the contributions of all field configurations.
The error of estimating a well-behaved integral from a random sample of $N$ field configurations with Monte Carlo techniques scales as $1/\sqrt{N}$ independently of the dimension of the integrals.
Minkowksi space path integrals are not well-behaved, and the oscillatory integrand $e^{iS_\varphi}$ leads to a sign problem obstructing Monte Carlo evaluation of real-time path integrals.
The action is extensive, and $e^{iS_\varphi}$ is $O(1)$ for generic points in field configuration space.
UV fluctuations that do not contribute significantly to the average path integral make contributions of the same magnitude as semi-classical field configurations expected to dominate the path integral.
In a finite statistical sample, it is difficult to accurately reproduce these cancellations in the average path integral relying on interference between the amplitudes of different field configurations.

Monte Carlo simulations can preference field configurations making dominant contributions to path integrals by using importance sampling,
in which field configurations are sampled from a distribution that weights configurations with larger contributions to the integral with higher probability than configurations that make smaller contributions.
Importance sampling relies on being able to treat the integrand under consideration as a probability distribution.
To compute the average real part, field configurations could be drawn from an importance sampled probability distribution approximately proportional to $|\cos(iS_\varphi)|$ separately in regions of positive and  negative $\cos(iS_\varphi)$.
This will efficiently sample field configurations from the regions giving dominant positive and dominant negative contributions.
However, without knowing the nodal structure of $e^{iS_\varphi}$ or the relative proportion of positive-valued and negative-valued configurations in field space, it is impossible to know the relative degree of cancellation between separately importance sampled positive and negative contributions.
Importance sampling cannot be applied to integrals of the pure phase $e^{iS_\varphi}$.
The sign problem similarly obstructs real-time simulation of hadrons in QCD.

Instead, standard practice is to consider analytically continuing the path integral to imaginary time $t = i x^0$.
This changes Minkowski spacetime to Euclidean spacetime with metric $(++++)$, and correspondingly changes the isometry group of flat spacetime to $SO(4)$.
The latter affects fields with spin non-trivially, but for the case of a scalar field the only charge to the action is the relative sign of the kinetic and potential terms.
For simplicity, we only consider isotropic lattices where the lattice spacing is identical in all space and time directions and unless otherwise specified will work in units where the lattice spacing is set to unity.
There is no unique way to define a lattice derivative, and one simple choice of the free scalar field action is
\begin{equation}
  \begin{split}
  S_\varphi &= \sum_{x} \varphi^\dagger(x) \left( \sum_\mu \partial_\mu^\dagger\partial_\mu + m^2 \right)\varphi(x)\\
  &= \sum_{x;\mu} |\varphi(x+\hat{\mu}) -  \varphi(x)|^2 + \sum_x m^2|\varphi(x)|^2.
  \label{eq:scalareuclideanaction}
\end{split}
\end{equation}
Here and below the same notation is used for the Minkowski and Euclidean action.
The only free parameter in Eq.~\eqref{eq:scalareuclideanaction} is the bare mass in lattice units $m$.
In particular the value of the lattice spacing is not an independent parameter.
Results from a Monte Carlo calculation are dimensionless numbers that can be interpreted as results in lattice units
and cannot be immediately converted to physical units.
Agreement between a prediction of the lattice field theory and an experimental observable, or some other specified constraint, must assumed in order to set the lattice scale in physical units.
Other observables can then be predicted in physical units.
In gauge theories, this plays the role of coupling constant renormalization or equivalently of determining $\Lambda_{QCD}$ (appropriately defined non-perturbatively) in lattice units.

After transforming to Euclidean spacetime, the path integral weight changes from $e^{iS_\varphi}$ to $e^{-S_\varphi}$.
Minima of the Euclidean action are associated with solutions of the classical field equations of motion, and the weight $e^{-S_\varphi}$ exponentially damps UV fluctuations with larger action than classical field configurations of minimum action.
The path integral weight is real and non-negative for a scalar field in Euclidean spacetime, and it can therefore be interpreted as a probability distribution.
In a finite spacetime volume with $L^3 \beta$ lattice points and field boundary conditions specified, configurations can sampled using a Markov process of sequential random updates that is constructed to have a non-zero probability of reaching any configuration from any other configuration with enough random update steps.
The desired scalar field vacuum distribution can be generated from this Markov process using for instance the Metropolis algorithm where updates from field configuration $\varphi_i$ to $\varphi_{i+1}$ that decrease the action are accepted and included in a statistical ensemble, but updates that increase the action are only accepted with a probability equal to $e^{-S_\varphi(\varphi_{i+1})+S_\varphi(\varphi_i)}$.
If periodic boundary conditions are chosen in the time direction (anti-periodic for fermions), then $e^{-S_\varphi}$ is precisely the Boltzmann distribution describing the equilibrium state of the QFT in the language of statistical mechanics.
The partition function of statistical mechanics can be identified with the analytic continuation of the vacuum-to-vacuum amplitude to Euclidean spacetime,
\begin{equation}
  \begin{split}
    Z = \int_{\varphi_\beta=\varphi_0}\mathcal{D}\varphi^\dagger \mathcal{D}\varphi e^{-S_\varphi}.
  \end{split}\label{eq:scalarZdef}
\end{equation}
By analogy with statistical mechanics, the length of the time direction $\beta$ can be identified with the inverse physical temperature of the system.
One can intuitively think that the system is immersed in a non-zero temperature heatbath made up of its periodic images separated by distance $\beta$.

There is no distinction between space and time directions in Euclidean spacetime, and including periodic boundary conditions in spatial directions can be described as coupling the system to a heatbath of its periodic images to form a cubic lattice with the original volume acting as the unit cell.
For a system with a finite correlation length, $m_{\pi}^{-1}$ for a system of isolated hadrons in QCD, effects of interactions with these periodic images will be exponentially small in the ratio of the box size to the correlation length, $e^{-m_\pi L}/L$~\cite{Luscher:1985dn}.
Multi-particle scattering states of strongly interacting hadrons receive additional power law finite volume corrections that encode the probability of particles with finite-range interactions scattering within the finite volume~\cite{Huang:1957im,Luscher:1986pf}.
Infinite volume Euclidean correlation functions describing scattering states are independent of the scattering phase shift away from kinematic thresholds~\cite{Maiani:1990ca}, but
L{\"u}scher showed that power law finite volume corrections in QFT depend on the specific form of the interaction and can be used to extract the phase shifts of hadronic scattering states from finite volume correlation functions~\cite{Luscher:1986pf}.
The original work of L{\"u}scher has been extended to increasingly more complex systems, and the finite volume effects of strong interactions in two, three, and more hadron systems~\cite{Luscher:1990ux,Luscher:1990ck,Rummukainen:1995vs,Lellouch:2000pv,Beane:2003da,He:2005ey,Kim:2005gf,Christ:2005gi,Lage:2009zv,Luu:2011ep,Briceno:2012yi,Briceno:2012rv,Hansen:2012bj,Hansen:2012tf,Gockeler:2012yj,Briceno:2013bda,Briceno:2013lba,Hansen:2014eka,Briceno:2016xwb,Hansen:2017mnd}.
Loosely bound systems, which for these purposes includes light nuclei, also receive additional finite volume corrections that are exponentially small in the binding momentum of the system~\cite{Beane:2003da}.
In the deuteron, a light nucleus with the quantum numbers of a spin-triplet neutron-proton bound state, the binding momentum is much smaller than $m_\pi$ and these near-threshold scattering effects are the much larger than $e^{-m_\pi L}$ effects.

The correspondence Eq.~\eqref{eq:scalarZdef} between statistical mechanics and Euclidean QFT for a free scalar field
can be established through the construction of a transfer matrix, a bounded, self-adjoint operator $\hat{T}_\varphi$ defined on the full Hilbert space by
\begin{equation}
  Z = \tr(\hat{T}_\varphi^{\beta}) = \int \mathcal{D}\varphi^\dagger\mathcal{D}\varphi \mbraket{\varphi_0}{\hat{T}_\varphi}{\varphi_{\beta-1}}\mbraket{\varphi_{\beta-1}}{\cdots\vphantom{\hat{T}_\varphi}}{\varphi_1}\mbraket{\varphi_1}{\hat{T}_\varphi}{\varphi_0},
  \label{eq:scalarZsplit}
\end{equation}
where the second equality follows by inserting complete sets of coordinate basis states with the normalization of Eq.~\eqref{eq:scalarstatedef} at each time-slice.
The path integral representation of the free-field partition function follows if a bounded, self-adjoint operator $\hat{T}_\varphi$ can be defined that has matrix elements
\begin{equation}
  \mbraket{\varphi_{t+1}}{\hat{T}}{\varphi_t} =  e^{-L[\varphi_{t+1},\varphi_t]},
  \label{eq:scalarT}
\end{equation}
where 
\begin{equation}
  L[\varphi_{t+1},\varphi_t] = \sum_\v{x} |\varphi(\v{x},t+1) - \varphi(\v{x})|^2 + \frac{1}{2}\sum_\v{x} V[\varphi_{t+1}] + \frac{1}{2}\sum_\v{x} V[\varphi_t],
  \label{eq:scalarL}
\end{equation}
\begin{equation}
  V[\varphi_t] = \sum_k |\varphi(\v{x} + \uv{k},t) - \varphi(\v{x},t)|^2 + m^2 |\varphi(\v{x},t)|^2.
  \label{eq:scalarV}
\end{equation}
Note that locality of the action plays an essential role in allowing the path integral weight to be decomposed into a product of factors depending on time-slice pairs $\{\varphi_t,\varphi_{t+1}\}$ in Eq.~\eqref{eq:scalarZsplit}.
Textbooks on lattice field theory explicitly demonstrate the construction of the scalar field transfer matrix in terms of field operators and their conjugate momenta~\cite{creutz:1983quarks,smit:2002introduction,montvay:1997quantum}.

If the partition function trace is computed in the eigenbasis of $\hat{T}$, guaranteed to exist by self-adjointness, it can be expressed as
\begin{equation}
  \begin{split}
    Z = \tr(\hat{T}^\beta) = \sum_n e^{-E_n \beta}\sim e^{-E_0 \beta},
  \end{split}\label{Zspectral}
\end{equation}
where $e^{-E_n}$ denotes the $n$-th eigenvalue of the transfer matrix. 
If the transfer matrix is bounded, than $E_n \geq 0$.
Minus the log of the largest eigenvalue is the smallest $E_n$ and is denoted $E_0$.
Proportionality at large values of a time parameter up to exponential corrections is denoted as $\sim$ in Eq.\eqref{Zspectral} and below, and allows $E_0$ to be determined from the partition function as
\begin{equation}
  \begin{split}
    E_0 = \lim_{\beta\rightarrow \infty} -\partial_\beta \ln Z = \lim_{\beta\rightarrow\infty} \frac{1}{\beta}\ln\left[ \frac{Z(\beta)}{Z(\beta+1)} \right],
  \end{split}\label{ZEM}
\end{equation}
where the first continuous-time definition and second discrete-time definition are both independent of the overall normalization of $Z$.
$E_0$ can be identified with the free energy of the zero temperature vacuum.
The $E_n$ for $n>0$ similarly can be identified with the free energies of excited states of the vacuum that contribute to the free energy at non-zero temperature.
The spectral representation of the transfer matrix in Eq.~\eqref{Zspectral} allows Euclidean time evolution to be represented by sums of exponentials dominated by a single ground-state exponential decay at large times.
If transfer matrix $\hat{T}$ for a generic QFT is given, then a Hamiltonian can be defined through
\begin{equation}
  \begin{split}
    \hat{T} = e^{-\hat{H}}.
  \end{split}\label{THdef}
\end{equation}
It can be shown through Osterwalder-Schrader reflection positivity~\cite{Osterwalder:1973dx} that the Euclidean Hamiltonian operator coincides with the Hamiltonian operator defined in Minkowski space.

By inserting additional fields in the path integral representing $Z$, sources creating states other than the vacuum can be introduced.
If these sources carry conserved charge, then the theory will evolve according to the spectrum of the Hamiltonian in the sector of the theory with appropriate quantum numbers.
If thermal boundary conditions are employed, then then initial and final state scalar fields must have the same quantum numbers.
Euclidean correlation functions are defined as traces of the product of transfer matrices with additional operators, and can be non-vanishing if these operators have zero net charge.
The Euclidean-spacetime propagator for the scalar field is defined as the two-point correlation function
\begin{equation}
  \begin{split}
    G_\varphi(\v{p},t) = \sum_\v{x} e^{-i\v{p}\cdot \v{x}} \avg{\varphi(\v{x},t)\varphi^\dagger(0)} = \sum_\v{x} e^{-i\v{p}\cdot \v{x}} \int \mathcal{D}\varphi^\dagger\mathcal{D}\varphi \;e^{-S_\varphi}\varphi(\v{x},t)\varphi^\dagger(0).
  \end{split}\label{eq:scalar2ptdef}
\end{equation}
Existence of a transfer matrix guarantees a spectral representation for arbitrary Euclidean correlation functions comprised of products of local field operators, for example
\begin{equation}
  \begin{split}
    G_\varphi(\v{p}=0,t) &= \sum_\v{x}   \tr(\hat{T}^{\beta-t} \hat{\varphi}(\v{x})\hat{T}^t \hat{\varphi}^\dagger(0)) \\
    &= V \int \mathcal{D}\varphi^\dagger \mathcal{D}\varphi \mbraket{\varphi_0}{\hat{T}^{\beta-t}}{\varphi_t} \mbraket{\varphi_t}{\vphantom{\hat{T}^t}\hat{\varphi}(\v{x})}{\varphi_t} \mbraket{\varphi_t}{\hat{T}^t}{\varphi_0} \mbraket{\varphi_0}{\vphantom{\hat{T}^t}\hat{\varphi}^\dagger(0)}{\varphi_0}\\
    &= \sum_{n,m} |Z_{nm}^\varphi|^2 e^{-E_m(\beta-t)} e^{-E_n t} \\
    &\sim e^{-M_\varphi t}
  \end{split}
\end{equation}
where $Z_{nm}^\varphi$ describes the amplitude of the field operator to annihilate state $m$ and create state $n$, that is $Z_{nm}^\varphi=V^{1/2} \mbraket{n}{\hat{\varphi}(\v{x})}{m}$, and $M_\varphi$ is the ground-state energy in the single-particle state created by $\varphi^\dagger$.
Excited state effects are exponentially small in $t$, and thermal effects are exponentially small in $\beta$ (or often $\beta - t$).
Both can be neglected when $t \ll \beta$ is much larger than the relevant energy gap to the first excited state in the spectrum.
$M_\varphi$ can be straightforwardly determined from the large-time behavior of the Euclidean correlation function by forming an effective mass analogous to Eq.~\eqref{ZEM}.

LQCD was first constructed by Wilson~\cite{Wilson:1974sk} in the path integral formalism with $\mathfrak{su}(3)$ valued gauge fields $A_\mu$ replaced by $SU(3)$ valued gauge fields $U_\mu$ sometimes called gauge links,
\begin{equation}
  U_\mu(x) = \mathcal{P}\exp\left( \int_x^{x+\hat{\mu}} A_\mu(x)dx\mu \right),
\end{equation}
where $A_\mu$ again is anti-Hermitian.
Introducing plaquettes
\begin{equation}
  U_{\mu\nu}(x) = U_\mu(x)U_\nu(x+\hat{\mu})U_\mu^\dagger(x+\hat{\nu})U_\nu^\dagger(x) = U_{\nu\mu}^\dagger(x),
\end{equation}
for $\mu\neq\nu$ and $U_{\mu\mu}(x)=0$, the Wilson action for pure Yang-Mills theory is defined by
\begin{equation}
  S_G(U)  = \frac{1}{g^2}\sum_{x;\mu,\nu}\text{Tr} \left[ 1 - U_{\mu\nu}(x) \right].
\end{equation}
Defining the path integral measure $\mathcal{D}U$ to be the $SU(3)$ Haar measure, gauge invariance of the Wilson action $S_G$ guarantees that only color-singlet states contribute to the QCD partition function.
At each spacetime point in the lattice, coordinate basis states can be defined for the spatial components of the gauge field.
The path integral formulation includes time-like gauge fields $A_0$ that naively give rise to negative-norm states and cannot appear in the full Hilbert space defined from the tensor product as in the scalar example.
Time-like gauge links arise in the Hilbert space formulation of LQCD as projectors that restrict the action of the transfer matrix to the gauge singlet sector of the full Hilbert space.
The gluon part of the physical Hilbert space of LQCD is the gauge singlet projection of the coordinate basis eigenstates of spatial link operators $\ket{U_k(x)}$.

Defining quarks in LQCD is more complicated because the naive discretized action for a Dirac fermion in $D$ spacetime dimensions actually describes $2^D$ degenerate fermions.
This fermion doubling problem was first solved by Wilson, who demonstrated that adding higher dimensional operators to the action that are irrelevant in the continuum limit is enough to break this degeneracy and give all but one fermion a large mass on the order of the inverse lattice spacing.
The Wilson quark action is given by
\begin{equation}
  \begin{split}
    S_F(q,\bar{q},U) &= \sum_{x;\mu} \bar{q}(x)\left[\frac{1}{2}(D_\mu -D_\mu^\dagger)\gamma_\mu + \frac{1}{2} D_\mu^\dagger D_\mu\right]q(x) + \sum_x \bar{q}(x)m_q q(x), \\
\end{split}
\label{eq:wilsonquarkdef}
\end{equation}
where
\begin{equation}
  \begin{split}
    D_\mu q(x) &= U_\mu(x)q(x+\hat{\mu}) - q(x),\\
    D_\mu^\dagger q(x) &= U_\mu^\dagger(x-\hat{\mu})q(x-\hat{\mu}) - q(x),\\
    \sum_\mu D_\mu^\dagger D_\mu q(x) &= 8\,q(x) - \sum_\mu\left[ U_\mu(x)q(x+\hat{\mu}) + U_\mu^\dagger(x-\hat{\mu})q(x-\hat{\mu}) \right].
\end{split}
\label{eq:quarkgaugecovariant}
\end{equation}
Because the naive fermion action represented by the first two terms in  Eq.~\eqref{eq:wilsonquarkdef} is linear in derivatives, its discrete Fourier transform will involve a sine of the momentum.
Sine vanishes at both zero and $\pi$, and so the naive fermion propagator includes poles corresponding to isolated particles of mass $m_q$ both at momentum zero and $\pi$.
These spurious poles with non-zero momentum represent the $2^D-1$ fermion doublers, and are removed when the last term in Eq.~\eqref{eq:wilsonquarkdef} is included in the fermion propagator.
An explicit construction of the LQCD transfer matrix with Wilson's quark and gluon actions was given by L{\"u}scher~\cite{Luscher:1977} and shows that the Wilson term can be incorporated through a gluon field dependent renormalization of the quark states.

The Wilson action demonstrates that it can be helpful to add higher dimensional terms to the LQCD action.
These terms are irrelevant in the continuum limit and so leave the physics under study unchanged, but they can be chosen to reduce the size of systematic errors associated with non-zero lattice spacing.
Lattice artifacts can be studied in EFT by considering the Symanzik action, the most general action including higher dimensional operators consistent with the symmetries of the discretized lattice field theory~\cite{Symanzik:1983dc}.
Interactions terms in the Symanzik action are organized in a derivative expansion so that lattice artifacts can be computed as a power series in ratios of low-energy scales to the lattice cutoff.
Adding higher-dimensional terms to the LQCD action will effectively shift the renormalized values of couplings in the Symanzik action.
Tuning the coefficients of higher-dimensional operators explicitly included in the action can be used to cancel the effects of quantum fluctuations and drive the renormalized couplings associated with particular lattice artifacts to zero or another specified value.
By tuning all the renormalized low-energy constants parametrizing lattice artifacts at a given order in the derivative expansion to zero, the parametric scaling of lattice artifacts in low-energy observables can be systematically improved from linear to quadratic or better.

Many different LQCD groups use many different discretized QCD actions, and in particular difficulties related to chiral symmetry and additive mass renormalization\footnote{See Ref.~\cite{Kaplan:2009yg} for a review of lattice fermions and chiral symmetry.} motivated the construction of staggered~\cite{Kogut:1974ag}, domain wall~\cite{Kaplan:1992bt,Shamir:1993zy,Furman:1994ky}, and overlap~\cite{Narayanan:1993sk,Narayanan:1994gw,Neuberger:1997fp} quark actions.
Some improved actions can be shown to not possess a positive-definite transfer matrix~\cite{Luscher:1984is}.
It is expected, and assumed below, that all discretized LQCD actions are in the same universality class and approach the same continuum limit fixed point, and that
results from LQCD discretizations such as domain wall fermions that do not possess a positive-definite transfer matrix only differ by perturbatively small amounts from the well-defined results of the Wilson action.
Performing calculations at multiple lattice spacings and preferably with multiple different discretized actions allows systematic uncertainties related to these and other lattice artifacts to be quantified.

Exploratory LQCD calculations discussed in Chapters~\ref{chap:statistics}-\ref{chap:PR} were generated by the NPLQCD collaboration and use 
the improved L{\"u}scher-Weize gauge action action~\cite{Luscher:1984xn} and
the clover-improved quark action including a term of the form $\bar{q}\sigma_{\mu\nu}G_{\mu\nu}q$ constructed by Sheikholeslami and Wohlert~\cite{Sheikholeslami:1985ij}.
See Ref.~\cite{Orginos:2015aya} for further details.


\section{Meson and Baryon Correlation Functions}\label{sec:corr}

The formal construction of LQCD sketched in Sec.~\ref{sec:lattice} provides a finite-dimensional path integral representation of the QCD partition function
\begin{equation}
  \begin{split}
    Z_{QCD} &= \int \mathcal{D}U\mathcal{D}\bar{q}\mathcal{D}q e^{-S_G(U) - \sum_x \bar{q}(x)D(U;x,x)q(x)} \\
    &= \int \mathcal{D}U e^{-S_G(U)}\det[D(U)].
  \end{split}\label{eq:qcdZ}
\end{equation}
The free energy of the QCD vacuum can be defined straightforwardly as $F_{QCD} = -\partial_\beta \ln Z_{QCD}$, and in principle computed by performing the integral over gauge field configurations.
As discussed above, high-dimensional integrals such as QFT path integrals in large spacetime volumes are most efficiently performed using Monte Carlo techniques where the error after sampling $N$ field configurations decreases as $1/\sqrt{N}$ regardless of the dimensionality of the integral.
Monte Carlo integration relies on sampling the integrand as a probability distribution.
The Wilson Dirac operator, as well as other variants including the clover-improved Dirac operator, obeys the $\gamma_5$-Hermiticity property
\begin{equation}
  \begin{split}
    D(U;x,y) = \gamma_5 D(U;y,x)^\dagger \gamma_5,
  \end{split}\label{eq:g5Hermiticity}
\end{equation}
which guarantees that the determinant inside the path integral weight is real in the presence of an arbitrary gauge field configuration $U$
\begin{equation}
  \begin{split}
    \det[D(U)] = \det[\gamma_5 D(U)^\dagger \gamma_5] = \det[\gamma_5]^2 \det[D(U)^\dagger] = \det[D(U)]^*. 
  \end{split}\label{eq:positivity}
\end{equation}
In the continuum, chiral symmetry guarantees that $\fs{D}(U)$ has eigenstates of opposite chirality with complex conjugate eigenvalues, and therefore that $\det[D(U)]$ is a positive definite product of paired eigenvalues,~\cite{Kaplan:2009yg}
\begin{equation}
  \begin{split}
    \mathcal{P}(U) = e^{-S_G(U)}\det[D(U)] \geq 0.
  \end{split}\label{eq:probdef}
\end{equation}
It is therefore possible to interpret $\mathcal{P}(U)$ as a probability distribution.
At finite lattice spacing, there is no exact chiral symmetry and configurations with $\det D < 0$ are possible.
In $N_f = 2$, that is calculations of two degenerate flavors, the determinant factorizes into a product of equal determinants for the two flavors and is positive semi-definite.
Heavier quarks are less likely to see fluctuations of negative $\det D$, and for the physical strange quark the average determinant phase factor $\avg{\det D/|\det D|}$ is close to one~\cite{Beane:2010em}.
Reweighting and other methods have allowed determinants of light, non-degenerate quarks to be effectively included in Monte Carlo sampling techniques, see Refs.~\cite{Hasenfratz:2008fg,Finkenrath:2013soa} for further discussions.
With a non-negative $\mathcal{P}(U)$ with the same continuum limit as Eq.~\eqref{eq:probdef} constructed, averages of generic observables $\mathcal{O}$ can be identified with the sample means of functions of an ensemble of $N$ random gauge fields $U_i$ distributed according to $\mathcal{P}(U)$,
\begin{equation}
  \begin{split}
    \avg{\mathcal{O}(U)} &= \int \mathcal{D}U \;e^{-S_G(U)}\det[D(U)]\mathcal{O}(U) \\
    &= \int \mathcal{D}U \;\mathcal{P}(U)\;\mathcal{O}(U) \\
    & = \frac{1}{N}\sum_{i=1}^N \mathcal{O}_i + O(N^{-1/2}),
  \end{split}\label{eq:MC}
\end{equation}
where $\mathcal{O}_i = \mathcal{O}(U_i)$ represents the observable calculated in the presence of gauge field configuration $U_i$.

\begin{figure}[!ht]
  \begin{center}
  \includegraphics[width=.7\columnwidth]{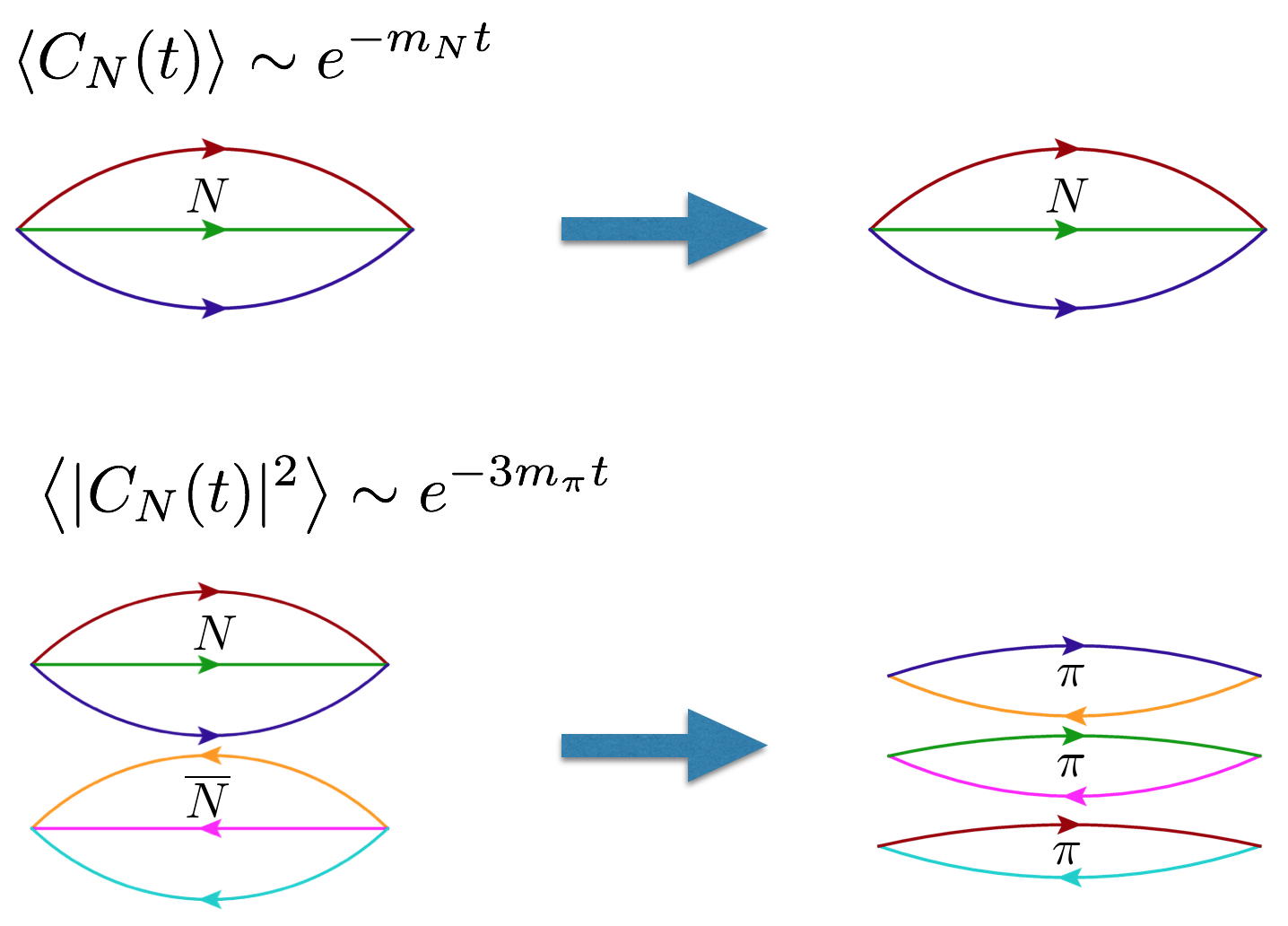}
\end{center}
  \caption{A schematic illustration of Parisi-Lepage scaling.
    At times close to the source, left, the nucleon correlation function, top, decays at a rate set by $M_N$, and it's mean-square, bottom, decays at a rate set by $2M_N$ minus an interaction energy shift.
    Baryon number conservation guarantees that the nucleon is the lightest state contributing to $C_N(t)$, but no conservation law prevents the quarks and antiquarks in $|C_N(t)|^2$ from rearranging to form pions.
    The lowest energy state in $|C_N(t)|^2$ preserving all quark lines explicitly inserted as propagators includes three pions.
  }
  \label{fig:lepage}
\end{figure}

Higher moments of observables dictate statistical properties of Monte Carlo calculations such as their variance.
Parisi first noted that it can be helpful to analyze these moments from a QFT point of view in order to understand the statistical variation of Monte Carlo results~\cite{Parisi:1983ae}.
The variance corresponding to the observable in Eq.~\eqref{eq:MC}, for example, is
\begin{equation}
  \begin{split}
    \text{Var}(\mathcal{O}) &= \avg{\mathcal{O}(U)^2} - \avg{\mathcal{O}(U)}^2 \\
    &= \int \mathcal{D}U\;e^{-S_G(U)}\det[D(U)] \left(\mathcal{O}(U)^2 - \avg{\mathcal{O}(U)}^2\right).
  \end{split}\label{eq:MCvar}
\end{equation}
The Monte Carlo variance is therefore controlled by the QFT operator $(\mathcal{O} - \avg{\mathcal{O}})^2$ encoding quantum fluctuations about the expectation value, and the StN ratio that will be seen in a stochastic calculation can be analytically calculated in the infinite statistics limit,
\begin{equation}
  \begin{split}
    \text{StN}(\mathcal{O}) &= \frac{\avg{\mathcal{O}}}{\sqrt{\text{Var}(\mathcal{O})}}.
  \end{split}\label{eq:MCstn}
\end{equation}
If $\mathcal{O}$ is a complex operator with $\avg{\text{Im}\mathcal{O}} = \avg{\text{Re}\mathcal{O}\text{Im}\mathcal{O}}=0$, which holds for many observables by $C$, then the variance of the real part of a complex random variable can be written
\begin{equation}
  \begin{split}
    \text{Var}(\text{Re}\mathcal{O}) &= \avg{(\text{Re}\mathcal{O})^2} - \avg{\text{Re}\mathcal{O}}^2\\
    &= \frac{1}{2}\avg{|\mathcal{O}|^2} + \frac{1}{2}\avg{\mathcal{O}^2} - \avg{\mathcal{O}}^2.
  \end{split}\label{eq:ComplexVar}
\end{equation}
Lepage first considered the statistics of baryon correlation functions, and by considering the scaling of terms in Eq.~\eqref{eq:ComplexVar} argued there must be exponential degradation of the baryon correlation function StN ratio with Eulcidean time~\cite{Lepage:1989hd}.
Baryon correlation functions analogous to the scalar field correlation functions in Eq.~\eqref{eq:scalar2ptdef}, describe the propagation of a baryon between points separated by a fixed Euclidean spacetime extent.
They also receive additional contributions from excited baryon states that decay exponentially faster than the ground state and can be neglected at large separation.
The conjugate of a baryon correlation function represents propagation of an anti-baryon, and the magnitude-squared appearing in the StN ratio represents propagation of three quarks and three anti-quarks.
The appropriate ground state contains three pions, as shown in Fig.~\ref{fig:lepage}.
At very large Euclidean separations, the root-mean-square magnitude of a baryon correlation function will decay exponentially with a rate $\frac{3}{2}m_\pi$, while the mean decays exponentially at a rate $M_N$.
The StN problem describes the resulting issue that the StN ratio decays at a rate $M_N - \frac{3}{2}m_\pi$.
With physical quark masses, $M_N - \frac{3}{2}m_\pi \sim 0.78\; M_N$.
High-statistics LQCD calculations~\cite{Beane:2009kya,Beane:2010em,Beane:2014oea,Detmold:2014rfa,Detmold:2014hla,Wagman:2016bam,Wagman:2017xfh}, have confirmed that numerical Monte Carlo path integral calculations in LQCD have StN ratios consistent with Parisi-Lepage scaling. 

Before discussing Parisi-Lepage scaling and statistics of nucleon correlation functions further, we briefly consider the intimately related scalings of alternative approaches to calculating the properties of dense matter in QCD.
Thermodynamic properties of nuclei and nuclear matter can be determined from free energy calculations if a quark-number chemical potential is introduced.
Large artifacts of the lattice discretization can appear with some definitions of the chemical potential, but can be avoided by recognizing that the chemical potential acts like the time-like component of an imaginary background gauge field that couples to $U(1)_B$~\cite{Hasenfratz:1983ba}.
Inclusion of a quark chemical potential breaks $\gamma_5$-Hermiticity of the Dirac operator.
This breaks reality of the quark determinant in an arbitrary gauge field configuration and introduces a sign problem.
Reweighting methods~\cite{Ferrenberg:1988yz} exploit the freedom to redefine the factorization between a probability distribution and an observable,
and can be used to (inefficiently) avoid sign problems.
To determine the free energy in the presence of a baryon chemical potential $F_{QCD}(\mu_B)$ with reweighting methods, one can sample from the standard $\mu_B=0$ vacuum probability distribution
and include the ratio of determinants with and without chemical potentials as an observable,
\begin{equation}
  \begin{split}
    Z(\mu_B) = e^{-F(\mu_B)V\beta} &= \avg{\frac{\det(D[U,\mu_B])^{N_f}}{\det(D[U,\mu_B=0])^{N_f}}} \\
    &= \sum_{i=1}^N \frac{\det(D[U_i,\mu_B])^{N_f}}{\det(D[U_i,\mu_B=0])^{N_f}}\\
    &\equiv \sum_{i=1}^N Z_i(\mu_B),
  \end{split}\label{eq:ZmuBdef}
\end{equation}
The free energy and hence $Z(\mu_B)$ are real, but $Z(\mu_B)$ is the real part of the average of a random complex function $Z_i(\mu_B)$ sampled for each gauge field configuration $U_i$.
Higher moments of random complex quark determinants have been studied, starting from the observation by Gibbs~\cite{Gibbs:1986ut} that reweighting determinants with baryon chemical potential has an exponentially bad StN problem for $\mu_B > m_\pi / 2$ and continuing with analysis in EFT and random matrix theory~\cite{Cohen:2003kd,Cohen:2003ut,Splittorff:2006fu,Splittorff:2006vj,Splittorff:2007ck,deForcrand:2010ys,Alexandru:2014hga}.
The StN ratio of a Monte Carlo calculation of $Z(\mu_B)$ employing reweighting of complex random functions $Z_i$ can be expressed by Eq.~\eqref{eq:ComplexVar} as~\cite{Alexandru:2014hga}
\begin{equation}
  \begin{split}
    \text{StN}(Z(\mu_B)) &= \frac{\avg{Z_i}}{\sqrt{\text{Var}(Z_i)}}\\
    &= \left( \frac{\avg{|Z_i|^2}}{2\avg{Z_i}^2} + \frac{\avg{Z_i^2}}{2\avg{Z_i}^2} - 1 \right)^{-1/2}.
  \end{split}\label{eq:ZStN}
\end{equation}
The mean-square determinant $\avg{Z_i^2}$ represents $2N_f$ quarks experiencing the same chemical potential as the quarks contributing to $\avg{Z_i}$, and will have a free energy that differs from $2F(\mu_B)$ by energy shifts arising from interactions between the two baryon species,
\begin{equation}
  \begin{split}
    \avg{Z_i^2} &= \int \mathcal{D}U e^{-S_G(U)} \det{D[U,\mu_B]^{2N_f}} =  e^{-\left(2F(\mu_B) - 2\delta F(\mu_B)\right)V\beta},
  \end{split}\label{eq:deltaFdef}
\end{equation}
where the second equality defines the interaction energy shift $\delta F(\mu_B)$.
For two-flavor QCD with up and down quarks, the magnitude of the baryon number determinant is equal to $\det[D(U,\mu_B)]\det[D(U,\mu_B)^*] = \det[D(U,\mu_B)]\det[D(U,-\mu_B)]$, which is equivalent to the determinant describing an isospin chemical potential acting opposite on up and down quarks~\cite{Son:2000xc},
\begin{equation}
  \begin{split}
    \avg{|Z_i|^2} &= \int \mathcal{D}U\;e^{-S_G(U)} \det(D[U,\mu_I])\det(D[U,-\mu_I]) \equiv e^{-2F_{PQ}(\mu_B)V\beta}.
  \end{split}\label{eq:FPQdef}
\end{equation}
The StN ratio of the fermion determinant is therefore given by
\begin{equation}
  \begin{split}
    \text{StN}(Z_i) = \left( \frac{1}{2}e^{[F(\mu_B)-F_{PQ}(\mu_B)]V\beta} + \frac{1}{2}e^{\delta F V \beta} - 1 \right)^{-1},
  \end{split}\label{eq:ZStNexp}
\end{equation}
Due to the bosonic nature of pions compared with the fermionic nature of baryons, the free energy of the pion system is lower than the free energy of the associated baryon system, $F_{PQ}(\mu_B) < F(\mu_B)$~\cite{Cohen:2003ut}.
This implies that reweighting faces an exponentially hard StN problem where the precision of Monte Carlo calculations degrades exponentially with increasing spacetime volume.
The baryon-baryon interaction term is negligible for repulsive interactions, $\delta F <0$, and for attractive interactions, $\delta F > 0$, adds a second source of exponential StN degradation.
For low-density, low-temperature systems in QCD, the phase-quenched term dominates the term arising from baryon-baryon interactions. 
Using $\chi PT$, it can be shown that at large $\beta$ and small $\mu_B$ the phase-quenched free energy difference provides the dominant effect and is given by~\cite{Splittorff:2006fu}
\begin{equation}
  \begin{split}
    F(\mu_B) - F_{PQ}(\mu_B) \simeq  \frac{f_\pi^2 \mu_B^2}{9}\left( 1 - \frac{9m_\pi^2}{4\mu_B^2} \right)^2\theta\left(\mu_B - \frac{3 m_\pi}{2}\right).
  \end{split}\label{eq:ZStNChPT}
\end{equation}
Once $\mu_B$ is large enough to start producing pions in the phase-quenched theory, the StN problem associated with reweighting  $\mu_B \neq 0$ determinants becomes exponentially hard in spacetime volume.
It is interesting to note that in this analysis $\mu_B > \frac{3}{2} m_\pi$ emerges as a natural scale from the phase-quenched theory even though $\mu_B > M_N$ sets the threshold of particle production~\cite{Gibbs:1986ut,Gocksch:1987ha,Cohen:2003kd}.

It is also possible in principle to determine thermodynamic properties of nuclei and nuclear matter using LQCD calculations at fixed baryon number rather than fixed chemical potential, that is working with the canonical ensemble of statistical mechanics instead of the grand canonical ensemble~\cite{Roberge:1986mm,Hasenfratz:1991ax,Kratochvila:2005mk,Alexandru:2005ix,Gattringer:2009wi,Danzer:2012vw,Alexandru:2014hga}.
The partition function at fixed baryon number is the Fourier transform of the partition function with imaginary baryon chemical potential.
Imaginary baryon chemical potential does not lead to a sign problem; however the Fourier transform requires linear combinations with complex coefficients and therefore introduces a sign problem.
This sign problem can be inefficiently avoided by reweighting as above and calculating canonical ensemble quark determinants averaged across a statistical ensemble of gauge fields importance sampled with the QCD vacuum distribution.
A calculation of the StN ratio of canonical ensemble determinants using $\chi$PT and the hadron resonance gas model for baryons shows that the difference in free energy density  between QCD at fixed baryon number and the phase-quenched theory at fixed isospin charge is given by~\cite{Alexandru:2014hga}
\begin{equation}
  \begin{split}
    F(B) - F_{PQ}(B) \simeq \frac{B}{V}\left( M_N - \frac{3}{2}m_\pi \right),
  \end{split}\label{ZStNCanChPT}
\end{equation}
where $B$ is the total baryon number of the system.
The difficulty of solving the canonical ensemble sign problem with reweighting is seen similarly to Eq.~\eqref{eq:ZStNexp} to be exponentially hard in the baryon number density times the spacetime volume, or equivalently exponentially hard in the total baryon number times the inverse temperature $B(M_N - \frac{3}{2}m_\pi)\beta$.
The scale setting this exponential difficulty is again $M_N - \frac{3}{2}m_\pi$, 
the difference between quark contributions to the ground state energy in the full theory containing baryons and phase-quenched theory containing pions.

It is difficult to compute quark determinants with large spacetime volumes,
especially a statistical sample of cold, dense quark determinants whose size is exponentially large in $B(M_N - \frac{3}{2}m_\pi)\beta$.
To compute the low-temperature physics of hadrons, 
it is standard to calculate ensembles of gauge fields distributed according to the QCD thermal vacuum probability distribution $e^{-S_G(U)}\det[D(U)]$
and then calculate hadronic correlation functions as observables, as in Eq.~\eqref{eq:MC}.
By Parisi-Lepage scaling, the StN problem associated with the correlation function of a single baryon is exponentially hard in $(M_N - \frac{3}{2}m_\pi)t$.
The NPLQCD collaboration verified Parisi-Lepage scaling in LQCD calculations and extended it to nuclei where up to corrections from interactions it is seen that the StN problem associated with correlation functions of baryon number $B$ is $B(M_N - \frac{3}{2}m_\pi)t$~\cite{Beane:2009kya}.
Using hadronic correlation functions, it is possible to probe low-temperature hadronic physics where $B(M_N - \frac{3}{2}m_\pi)\beta \gg 1$ and partition function methods would face a severe StN problem,
while only including non-zero baryon charge on a fraction $t/\beta \ll 1$ of the spacetime volume such that $B(M_N - \frac{3}{2}m_\pi)t \lesssim 1$ and the Parisi-Lepage StN problem is manageable.
It is also helpful that statistical ensembles of $N$ correlation functions for a number of different hadrons can all be computed on the same ensemble of gauge field configurations and correlated differences can be computed between them.

Once a gauge field ensemble distributed according to the vacuum distribution $e^{-S_G(U)}\det[D(U)]$ is generated, correlation functions must be defined and computed in the representation of QCD where the quark path integral has been performed analytically.
This can be readily accomplished if the interpolating operator is a product of local quark fields.
Since the QCD action in a fixed gluon field configuration is quadratic in the action of the quark fields,
correlation functions of products of quark fields in a fixed gauge field background
can be expressed a products of quark propagators in the same gauge field background
by performing free-fermion contractions.
Quark propagators in a given gauge field configuration are given by the inverse of the Dirac operator in that gauge field configuration,
\begin{equation}
  \begin{split}
    S(U_i;x,0) &= \int \mathcal{D}\bar{q}\mathcal{D}q e^{-\sum_y \bar{q}(y)D(U;y,y)q(y)} \bar{q}(x)q(0) \\
    &= D(U_i;x,0)^{-1}, \\
  \end{split}\label{propdef}
\end{equation}
Constructing the inverse of a large sparse numerical matrix can be performed with iterative Krylov solvers such as conjugate gradient and optimized using methods such as deflation~\cite{Stathopoulos:2007zi}.

The average quark propagator vanishes by Elitzur's theorem because it is not gauge invariant.
Non-vanishing expectation values only arise for color-singlet functions of quark propagators.
Such two-point functions describe for instance the propagation of mesons built with quark-antiquark sources and baryons and nuclei built from multi-quark sources.
A simple interpolating operator for the $\pi^+$, a pseudoscalar with quantum numbers of an anti-up and a down quark, is given by $\pi^+(x) = \bar{d}^a(x) \gamma_5 u^a(x)$,
where $u$ and $d$ are up and down quark fields and $a,b,\cdots$ denote $\mathfrak{su}(3)$ fundamental indices.
The pion correlation function computed with these operators can be expressed in terms of propagators as
\begin{equation}
  \begin{split}
    G_\pi(\v{x},t) &= \avg{\pi^+(\v{x},t)\pi^-(0)}\\
    &= \avg{\left( \bar{d}^b(x)\gamma_5 u^b(x)\right) \left( \bar{d}^a(0)\gamma_5 u^a(0) \right)}\\
    &= \sum_{i=1}^N \tr_s \left[S_u^{ab}(U_i;x,0)\gamma_5 S_d^{ba}(U_i;0,x)\gamma_5\right]\\
    &= \sum_{i=1}^N \tr\left[ S_u(U_ix,0) S_d^\dagger(U_i;x,0) \right],
  \end{split}\label{picorrden}
\end{equation}
where in the second line $\tr_s$ denotes a trace over spin and we use matrix notation for propagator spin contractions and in the third line $\tr = \tr_c \tr_s$ denotes a trace over color and spin. 

Interpolating fields for baryons are slightly more complicated.
The non-relativistic quark model provides a useful guide for constructing QCD interpolating operators, and for instance proton interpolating operators can be constructed with spin-singlet diquarks as
\begin{equation}
  \begin{split}
    p(x) = \varepsilon_{ijk}(u_i^T C \gamma_5 d_j)u_k,
  \end{split}\label{eq:protoninterp}
\end{equation}
where $i,j,k,\cdots$ represent $\mathfrak{su}(3)$ fundamental indices and $u$ and $d$ are fields representing up and down quarks.
Correlation functions of spin-1/2 operators are spin matrices, and it is convenient to analyze them by projecting out particle spin and parity components.
In accordance with the axial anomaly, positive-parity baryons are lighter than negative-parity baryons.
Projectors onto spin-up and spin-down positive-parity states $\Gamma_{+\uparrow}$ and $\Gamma_{+\downarrow}$ are useful for isolating the lowest-energy states in baryon propagators.
In a chiral basis with
\begin{equation}
  \begin{split}
    \gamma_i = -\sigma_2\otimes \sigma_i, \hspace{20pt}     \gamma_4 = \gamma_1\otimes 1, \hspace{20pt} \gamma_5 = \gamma_1\gamma_2\gamma_3\gamma_4 = \sigma_3 \otimes 1, \hspace{20pt}C = \gamma_2\gamma_4 = \sigma_3\otimes i\sigma_2,
  \end{split}\label{gammadef}
\end{equation}
these projectors are given by
\begin{equation}
  \begin{split}
    \Gamma_{+\uparrow/\downarrow} &= \frac{1}{2}(1 \pm i \gamma_5\gamma_3\gamma_4)(1 + \gamma_4).
  \end{split}\label{spinprojdef}
\end{equation}
Defining spin-0 color-\textbf{6} diquarks as
\begin{equation}
  \begin{split}
    \text{diq}_{\alpha\beta}^{c^\prime c}(A,B) = \varepsilon^{abc}\varepsilon^{a^\prime b^\prime c^\prime} A_{\gamma \alpha}^{ab}B_{\gamma \beta}^{a^\prime b^\prime},
  \end{split}\label{eq:diquarkdef}
\end{equation}
Zero-momentum, positive parity baryon correlation functions are explicitly defined as
\begin{equation}
  \begin{split}
    G_{p}(t) &= \sum_\v{x} \tr\left\lbrace \Gamma S_u^{ab}(U_i;\v{x},t;0) \tr_s\text{diq}^{ba}\left[ C\gamma_5 S_u(U_i;\v{x},t;0), S_d(U_i;\v{x},t;0)C\gamma_5 \right]\right\rbrace\\
    &\hspace{20pt} +\tr\left\lbrace \Gamma S_u^{ab}(U_i;\v{x},t;0) \text{diq}^{ba}\left[ S_u(U_i;\v{x},t;0) C\gamma_5, C\gamma_5 S_d(U_i;\v{x},t;0)  \right] \right\rbrace.
  \end{split}\label{eq:protoncorr}
\end{equation}

The spectrum of hadronic masses and binding energies can be extracted from two-point correlation functions with the same techniques as the scalar filed two-point functions of Sec.~\ref{sec:lattice}.
Multi-particle bound states can be constructed from interpolating operators that are products of the single-particle interpolating operators projected to the correct quantum numbers.
Binding energies can be directly measured from the large-time behavior of Euclidean correlation functions.
As discussed in Sec.~\ref{sec:lattice}, the phase shifts of scattering states are related to the energy levels of multi-particle Euclidean correlation functions in a finite volume.
Basic ingredients to models of nuclear forces such as the deuteron binding energy and neutron-neutron scattering length can be computed directly from QCD using the large-time behavior of finite volume LQCD multi-baryon correlation functions~\cite{Luscher:1985dn,Luscher:1986pf,Luscher:1990ux,Aoki:2002in,Beane:2003da}.

\begin{figure}
  \begin{center}
  \includegraphics[width=\columnwidth]{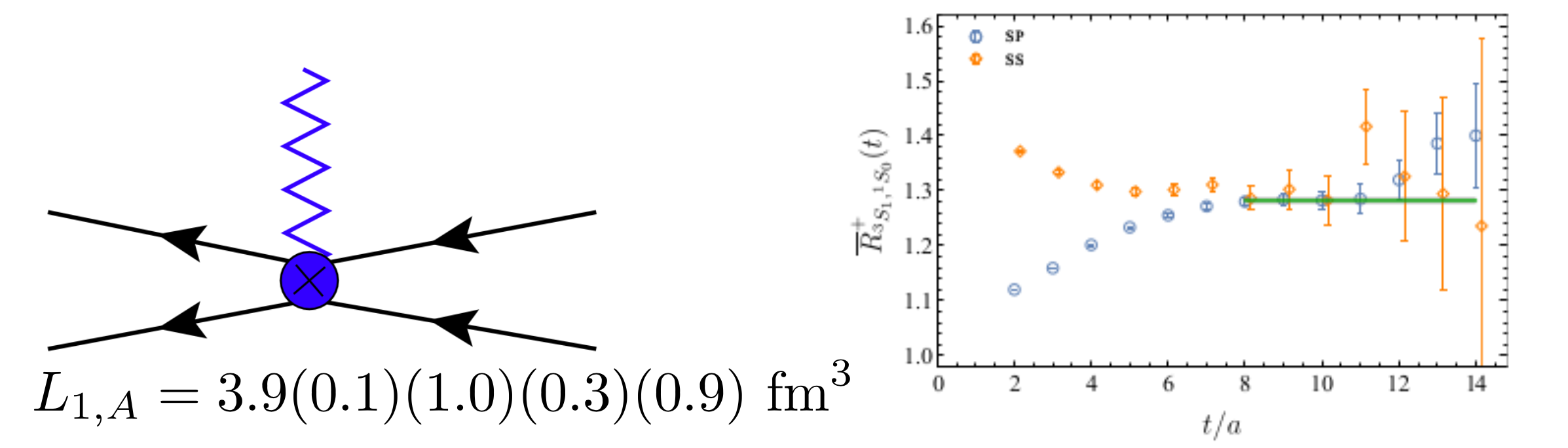}
\end{center}
  \caption{
    Results from the LQCD calculation of proton-proton fusion in Ref.~\cite{Savage:2016kon}. 
    EFT($\fs{\pi}$) is used to separate long-distance axial interactions that can be described as an isolated nucleon interacting with a $W^\pm$ boson from short-distance axial interactions that require more detailed QCD input about the structure of nuclear interactions, for example include a $W^\pm$ boson interacting with a pion exchanged between the nucleons.
    Both the proton-proton fusion cross-section and the finite-volume energy shift of a two-nucleon system in a background axial field depend on a poorly experimentally constrained low-energy constant in EFT($\fs{\pi}$) called $L_{1A}$.
    A calculation of the two-nucleon finite volume energy shift in LQCD, right, therefore allows an extraction of $L_{1A}$ and in turn the proton-proton fusion cross-section.
    Results for $L_{1A}$ shown are consistent with phenomenological determinations within uncertainties, shown as statistical, fitting systematic, scale-setting systematic, and quark mass extrapolation uncertainties.
    The quark mass extrapolation uncertainty in particular requires refinement with additional calculations at lighter values of the quark masses than the $m_\pi \sim 800$ MeV ensembles used here.
    Further details about the LQCD ensembles used for this production are given in Ref.~\cite{Orginos:2015aya}.
  }
  \label{fig:plateau}
\end{figure}

Following early calculations of pion scattering~\cite{Gupta:1993rn,Fiebig:1999hs,Liu:2001ss,Yamazaki:2004qb,Beane:2005rj,Aoki:2005uf}, nucleon-nucleon scattering was computed from Euclidean correlation functions in the quenched approximation~\cite{Gupta:1993rn,Fukugita:1994na}, and then in fully dynamical $N_f=2+1$ LQCD by the NPLQCD collaboration~\cite{Beane:2006mx}.
Calculations initially have been performed at heavy quark masses where the StN problem is less severe.
A silver lining from this is that early calculations have now begun to explore and understand the dependence of seemingly fine-tuned parameters in nuclear physics on the underlying parameters of the SM.
Early studies of two- and three-baryon systems determined that light nuclei exist with heavy quark masses and that binding energies are generally larger than with physical quark masses~\cite{Beane:2009gs,Beane:2010em,Yamazaki:2009ua,Beane:2009py,Beane:2009kya,Beane:2011pc,Beane:2011iw,Beane:2011zpa,Beane:2012vq,Yamazaki:2012hi,Beane:2013br,Yamazaki:2015asa,Yamazaki:2015vjn,Orginos:2015aya}.
In particular, the dineutron is unbound in nature but bound at heavier quark masses~\cite{Beane:2011iw}.
Detailed studies of nuclear correlation functions for nuclei with atomic number $A=2-5$ performed by the NPLQCD collaboration using ensembles tuned to heavy quark masses where $m_\pi \sim 800$ MeV~\cite{Beane:2012vq,Beane:2013br},
the PACS-CS collaboration in the quenched approximation~\cite{Yamazaki:2009ua,Yamazaki:2011nd}, with $m_\pi \sim 500$ MeV~\cite{Yamazaki:2012hi}, and $m_\pi \sim 300$ MeV~\cite{Yamazaki:2015asa,Yamazaki:2015vjn},
and HALQCD~\cite{Doi:2011gq,Inoue:2011nq}, but see Ref.~\cite{Savage:2016egr} for a critique of the HALQCD potential method.

Studies of nuclear structure and reactions can also be performed using LQCD calculation of multi-baryon correlation functions.
The magnetic moments of light nuclei at $m_\pi \sim 800$ MeV and $m_\pi \sim 450$ MeV obey shell-model like relations, suggesting that nuclei are describable as collections of interacting nucleons rather than structureless blobs of quarks and gluons~\cite{Chang:2015qxa,Detmold:2015daa,Parreno:2016fwu}.
Electromagnetic background field calculations were also used to postdict the measured cross-section of the electromagnetic radiative capture reaction $np\rightarrow d\gamma$~\cite{Beane:2015yha} by relating finite volume energy shifts to poorly known parameters in EFT($\fs{\pi}$) with the formalism of Detmold and Savage~\cite{Detmold:2004qn}.
Weak nuclear reactions have also recently been studied in lattice QCD, and the rates of proton-proton fusion and tritium $\beta$-decay have been determined at $m_\pi \sim 800$ MeV by the NPLQCD collaboration using LQCD in conjunction with EFT($\fs{\pi}$), see Fig.~\ref{fig:plateau}~\cite{Savage:2016kon}.
Calculations of the kinematically forbidden doubly-weak reaction $nn\rightarrow pp$~\cite{Shanahan:2017bgi,Tiburzi:2017iux} have identified and made preliminary LQCD explorations of previously overlooked isotensor polarizability effects that add significant theoretical uncertainties to interpretation of experimental searches for neutrinoless double-$\beta$ decay.
Understanding the emergence of nuclear structure and reactions from quarks and gluons will require studies similarly combining LQCD with EFT in order to incorporate known infrared nuclear physics and quantify the effects of short-distance physics specific to QCD.

As the number of quark fields included in a correlation function is increased, the number of contractions needed to express the correlation function in terms of propagators increases factorially.
Faster contraction algorithms exploiting symmetries of nuclear interpolating operators and the Grassmannian nature of quark fields lead to dramatic accelerations in the efficiency of multi-baryon contraction algorithms~\cite{Detmold:2012eu,Doi:2012xd}.
Calculations of up to $A=28$ have been performed~\cite{Detmold:2012eu}, but exponential decrease in statistical precision from the StN problem and other challenges have so far kept detailed studies of all but the lightest nuclei out of reach.

Methods have been developed that are able to exponentially reduce the severity of the StN problem in simple theories, and recently in LQCD.
In particular, the hierarchical integration approach of Ref.~\cite{Luscher:2001up} exploits the locality of QCD to express Wilson loops and other observables as products of factors that only depend on the gauge field on a subset of the lattice.
By calculating the average observable as a product of sub-lattice averages, the overall variance is reduced.
In a two-level hierarchical integration scheme, for instance, taking the product of averages instead of the average of products leads to  effectively $N^{-1}$ instead of $N^{-1/2}$ error scaling, or equivalently reduces the exponential scale of StN degradation for nuclei from $B(M_N - \frac{3}{2}m_\pi)$ to $\frac{B}{2}(M_N - \frac{3}{2}m_\pi)$.
Gluonic observables have been calculated using hierarchical integration~\cite{Meyer:2002cd,DellaMorte:2007zz,DellaMorte:2008jd,DellaMorte:2010yp,Vera:2016xpp}, and recently
C{\`e}, Giusti, and Schaefer have applied hierarchical integration to baryon correlation functions in quenched QCD~\cite{Ce:2016idq}
and unquenched fermion determinants~\cite{Ce:2016ajy}. 
The success of hierarchical integration schemes can be physically understood with the ideas that fluctuations in fields at widely separated spacetime points are approximately uncorrelated and averaging over uncorrelated fluctuations add little signal but a lot of noise.

Other investigations of the StN problem have focused on understanding the probability distributions of noisy observables.
It is lucidly argued by Hamber, Marinari, Parisi and Rebbi in Appendix B of Ref.~\cite{Hamber:1983vu} and further explained by Guagnelli, Marinari, and Parisi~\cite{Guagnelli:1990jb} that probably distributions of single-particle correlation functions are sensitive to the effects of multi-particle interactions.
Assuming a model of two-body forces, the $n$-th moment of a correlation functions receives a contribution proportional to the mass of the particle that scales with $n$, and a contribution arising from two-body interactions that scales like $n(n-1)/2$ times the binding energy or finite volume energy shift for a scattering state.
This pattern of moments for the average mass is satisfied for a correlation function that is log-normally distributed and whose $n$-th moment is therefore given by $e^{\mu n + \sigma^2 n^2/2}$.
Since pions experience perturbative two-body forces and even weaker multi-body forces, pion correlation functions are predicted by these arguments to have a log-normal distribution where the variance can be quantitatively related in $\chi$PT to the $\pi-\pi$ scattering length.

Endres, Kaplan, Lee, and Nicholson~\cite{Endres:2011jm} found that log-normal correlation functions are ubiquitous in Monte Carlo calculations of non-relativistic unitary fermions, and
developed a cumulant expansion that determines the average correlation function of a log-normally distribution sample more precisely than a calculation of the sample mean.
They presented statistical and mean-field arguments suggesting log-normal distributions are generic for QFT correlation functions,
and the cumulant expansion is used in further studies of unitary fermions~\cite{Endres:2011er,Endres:2011mm,Lee:2011sm,Endres:2012cw}.
DeGrand~\cite{DeGrand:2012ik} observed that meson, baryon, and gauge-field correlation functions in 
$SU(N_c)$ gauge theories with a range of $N_c$ are also approximately log-normal at early times where 
imaginary parts of correlation functions can be neglected. 

At large time separations, a nucleon correlation function calculated in a generic gauge configuration is complex,
and a log-normal distribution provides a poor fit to the real part.
Parisi-Lepage analysis was extended to higher moments of the correlation function distribution by Savage~\cite{Savage:2010misc},
who showed that all odd moments are exponentially suppressed compared to even moments at late times.
Kaplan noted that a stable distribution provides a reasonable fit to the real part of the late time nucleon distribution~\cite{davidkaplanLuschertalk}.
Correlation function distributions have been studied analytically in the Nambu-Jona-Lasinio 
model~\cite{Grabowska:2012ik,Nicholson:2012xt}, where it was found that real correlation functions were approximately log-normal but complex correlation functions in a physically equivalent formulation of the theory were broad and symmetric at late times with qualitative similarities to the QCD nucleon distribution. 

Chapter~\ref{chap:statistics} describes the statistics of baryon correlation functions at large times, where they must be treated as complex.
Complex correlation functions for the nucleon and other hadrons are found to be well-described by an approximately uncorrelated product of a log-normal magnitude and a wrapped normal phase factor.
From the behavior of the phase, the nucleon StN problem described by Parisi-Lepage scaling is found to follow directly from the sign problem associated with the nucleon correlation function phase.
The time evolution of the log-magnitude and the phase is further shown to resemble a heavy-tailed random walk where steps are only correlated over hadronic timescales.
Building on these observations, a new estimator based on correlation function ratios is proposed for which StN degradation only appears in a time paramter that can be treated independently from the source-sink separation time.

A variant on this technique callled phase reweighting is discussed in Chapter~\ref{chap:PR}.
Exploratory studies are conducted for meson, baryon, and two-baryon systems.
Phase-reweighted results with large time separations inacessible to standard analysis techniques 
are found to give consistent results with comparable precision to standrd techniques,
and possibilities for expanding the scope and improving the precision of phase reweighting are highlighted.

Chapter~\ref{chap:mesons} discusses isovector meson correlation functions.
These meson correlation functions are real but non-positive definite and face a ``sign'' rather than a ``phase'' problem.
Similar statistical distributions are found to describe the real parts of meson and baryon correlation functions,
and the asymptotic time dependence of moments of this distribution is shownw to possess generic features.
Applications of phase reweighting and ratio-based estimators are discussed.

 
\chapter{Statistical Distributions of Baryon Correlation Functions}\label{chap:statistics}

Noise in numerical calculations is generally considered to be a nuisance.
In Monte Carlo calculations of path integrals, statistical noise can also encode quantum fluctuations that play an essential role in asymptotic freedom, confinement, and other physics of strongly coupled QFTs.
Understanding the statistical distributions of quantum fluctuations in Monte Carlo path integral calculations can provide physical understanding of fluctuation-driven phenomena, as well as improved statistical estimators to extract physical results from noisy calculations more precisely.
To better understand the statistical distributions of quantum fluctuations giving rise to the StN problem,
this chapter focuses on a detailed study of quantum fluctuations in the simplest system experiencing thee baryon StN problem: one baryon at rest.

This statistical study is performed on a high-statistics analysis of 500,000 nucleon correlation functions 
generated on a single ensemble of gauge-field configurations by the NPLQCD collaboration~\cite{Orginos:2015aya} with LQCD.
This ensemble has a pion mass of $m_\pi \sim 450\text{ MeV}$, approximately physical strange quark mass, 
lattice spacing $\sim 0.12$ fm, and spacetime volume $32^3\times 96$. 
The L{\"u}scher-Weisz gauge action~\cite{Luscher:1984xn} and $N_f = 2+1$ clover-improved 
Wilson quark actions~\cite{Sheikholeslami:1985ij} were used to generate these ensembles,
details of which can be found in Ref.~\cite{Orginos:2015aya}.
In physical units, these calculations describe a proton at rest inside a cubic box with length $L \sim 3.8$ fm discretized on a $\sim 0.12$ fm grid.
The box is in thermal equilibrium with an ensemble of other boxes, and kept at a temperature of $\beta^{-1} \sim 17$ MeV.
Periodic boundary conditions are placed at the spatial boundaries of the box, and the proton experiences strong interactions with its periodic images suppressed by $e^{-m_\pi L}/L$.
For reference the longest QCD correlation length in the vacuum is set by $m_\pi^{-1}$, which is $m_\pi^{-1} \sim 0.43\text{ fm} \sim 3.6$ lattice units corresponding to $m_\pi L \sim 8.5$.
The LQCD calculations described are performed in part using the Chroma software suite~\cite{Edwards:2004sx}.

In this chapter, $G$ denotes the average nucleon correlation function and $C_i$ denotes the zero-momentum nucleon correlation function sampled from a Monte Carlo ensemble of $i=1,\cdots,N$ correlation functions,
\begin{equation}
  \begin{split}
    G(t) = \avg{C_i(t)} = \frac{1}{N}\sum_{i=1}^N C_i(t).
  \end{split}\label{GCdef}
\end{equation}
The average correlation function has a spectral representation 
\begin{equation}
  \begin{split}
    G(t) &= \sum_\v{x} \avg{N(\v{x},t) \bar{N}(0)} \\
    &= \sum_\v{x} \tr\left[ \hat{T}^{\beta - t}\hat{N}(x)\hat{T}^t \hat{\bar{N}}(0) \right]\\
    &= \sum_{m,n} |\sqrt{V}\mbraket{m}{\hat{N}(0)}{n}|^2 \; e^{-E_n t} e^{-E_m(\beta -t)}\\
    &\sim e^{-M_N t}
  \end{split}\label{eq:qcd2pt}
\end{equation}
The nucleon can be extracted from the late-time behavior of the correlation function by defining the effective mass
\begin{equation}
  \begin{split}
    M(t) = - \partial_t \ln G(t) = \ln\left( \frac{G(t)}{G(t+1)} \right),
  \end{split}\label{eq:qcdem}
\end{equation}
and extrapolating to the limit $t\rightarrow \infty$.
As described in Sec.~\ref{sec:corr},
the nucleon correlation function at large times has an exponentially degrading StN ratio 
\begin{equation}
  \begin{split}
    \text{StN}(G(t)) \sim \frac{\avg{C_i(t)}}{\sqrt{\avg{|C_i(t)|^2}}} \sim e^{-\left( M_N - \frac{3}{2}m_\pi \right)t}
    \ \ \  .
  \end{split}\label{Lepage}
\end{equation}
A phase convention for creation and annihilation operators is assumed so that $C_i(0)$ is real for all correlation functions in a statistical ensemble. At early times $C_i(t)$ is then approximately real, but at late times it must be treated as a complex quantity.

A generalization of the log-normal distribution for complex random variables that approximately 
describes the QCD nucleon correlation function at late times is discussed in this chapter. 
To study the logarithm of a complex correlation function, it is useful to introduce the magnitude-phase decomposition 
\begin{equation}
  \begin{split}
    C_i(t) = |C_i(t)|e^{i\theta_i(t)} = e^{R_i(t) + i\theta_i(t)}
    \ \ \ \ .
  \end{split}\label{RThdef}
\end{equation}
At early times where the imaginary part of $C_i(t)$ is negligible, previous observations of log-normal correlation 
functions~\cite{DeGrand:2012ik} demonstrate that $R_i(t)$ is approximately normally distributed. It is shown below that 
$R_i(t)$ is approximately normal at all times, and that $\theta_i(t)$ is approximately normal at early times. 
Statistical analysis of $\theta_i(t)$ is complicated by the fact that it is defined modulo $2\pi$. 
In particular, the sample mean of a phase defined on $-\pi < \theta_i(t) \leq \pi$ does not necessarily provide a faithful 
description of the intuitive average phase (consider a symmetric distribution peaked around 
$\pm\pi$ with a sample mean close to zero). 
Suitable statistical tools for analyzing $\theta_i(t)$ are found in the theory of circular statistics and as will 
be seen below that $\theta_i(t)$ is described by an approximately wrapped normal distribution.~\footnote{
See Refs.~\cite{Fisher:1995,Borradaile:2003,Mardia:2009} for 
textbook introductions to circular statistics.
} 

Before discussing of the magnitude-phase decomposition in Sec.~\ref{sec:decomposition},
it is worthwhile to briefly review relevant aspects of standard analysis methods of correlation functions.
Typically, in calculations of meson and baryon masses and  their interactions, 
 correlation functions are generated  from combinations of quark- and gluon-level  
sources and sinks with the appropriate hadron-level quantum numbers.  
Linear combinations of these correlation functions are formed, 
either using Variational-Method type techniques~\cite{Luscher:1990ck}, 
the Matrix-Prony technique~\cite{Beane:2009kya},
or other less automated methods,
in order to optimize overlap onto the lowest lying states in the spectrum 
and establish extended plateaus in 
relevant effective mass plots (EMPs).
In the limit of an infinite number of independent measurements, 
the expectation value of the correlation function is a real number at all times, and 
the imaginary part can be discarded as it is known to average to zero. 
The large-time behavior of such correlation functions 
becomes a single exponential (for an infinite time-direction) with an argument determined by the ground-state energy associated with the particular quantum numbers, or more generally the energy of the lowest-lying state with non-negligible overlap.

The structure of the source and sink play a crucial role in determining the utility of sets of correlation functions.
For many observables of interest, it is desirable to optimize the overlap onto the ground state of the system, and to minimize the overlap onto the correlation function dictating the variance of the ground state.
In the case of the single nucleon, the sources and sinks, ${\cal O}$,  
are tuned in an effort to have maximal overlap onto the ground-state nucleon, while minimizing overlap 
of ${\cal O}{\cal O}^\dagger$ onto the three-pion ground state of the variance~\cite{Detmold:2014rfa}.
NPLQCD uses  momentum projected hadronic blocks~\cite{Beane:2006mx} generated from quark propagators 
originating from localized smeared sources to suppress the overlap into the three-pion ground state of the variance by a factor of 
$1/\sqrt{V}$ where $V$ is the lattice volume, e.g. Ref.~\cite{Beane:2009kya}.  
For such constructions, the variance of the average scales as
$\sim  e^{- 3 m_\pi t}/(V N)$ at large times, 
where $N$ is the number of statistically independent correlation functions,
while the nucleon correlation function scales as $\sim e^{- M_N t}$.
For this set up, the StN ratio scales as $\sim \sqrt{V N} e^{ - (M_N - 3 m_\pi/2) t}$, from which it is clear that 
exponentially large numbers of correlation functions or volumes are required to overcome the StN problem at large times.
The situation is quite different at small and intermediate times in which the variance correlation function is dominated, not by the three-pion ground state, but by the ``connected'' nucleon-antinucleon excited state, which provides a variance contribution that scales as $\sim  e^{- 2 M_N t}/N$.  

This time interval where the nucleon correlation function is in its ground state and the variance correlation function is in a nucleon-antinucleon excited state has been called the ``golden window''~\cite{Beane:2009kya} (GW).
The variance in the GW is generated, in part, by the distribution of overlaps of the source and sink onto the ground state, that differs at each lattice site due to variations in the gluon fields.
In the work of NPLQCD, correlation functions arising from Gaussian-smeared quark-propagator sources and point-like or Gaussian-smeared sinks
that have been used to form single-baryon hadronic blocks.  
Linear combinations of these blocks are 
combined with coefficients 
(determined using the Matrix-Prony technique of Ref.~\cite{Beane:2009kya} or simply by minimizing the 
$\chi^2$/dof in fitting a constant to an extended plateau region)
that
extend the single-baryon plateau region to smaller times, eliminating the contribution from the first excited state of the baryon
and providing access to smaller time-slices of the correlation functions where StN degradation is less severe.
High-statistics analyses of these optimized correlation functions have shown that GW results are exponentially more precise
and have a StN ratio that degrades exponentially more slowly than larger time results~\cite{Beane:2009kya,Beane:2009gs,Beane:2009py} (for a review, see Ref.~\cite{Beane:2010em}).
In particular StN growth in the GW has been shown to be consistent with an energy scale close to zero,
 as is expected from a variance correlation function dominated by baryon, as opposed to meson, states.
Despite the ongoing successes of GW analyses of few-baryon correlation functions,
the GW shrinks with increasing baryon number~\cite{Beane:2009kya,Beane:2009gs,Beane:2009py}
and calculations of larger nuclei may require different analysis strategies suitable for correlation function without a GW.

\begin{figure}[!ht]
	\includegraphics[width=\columnwidth]{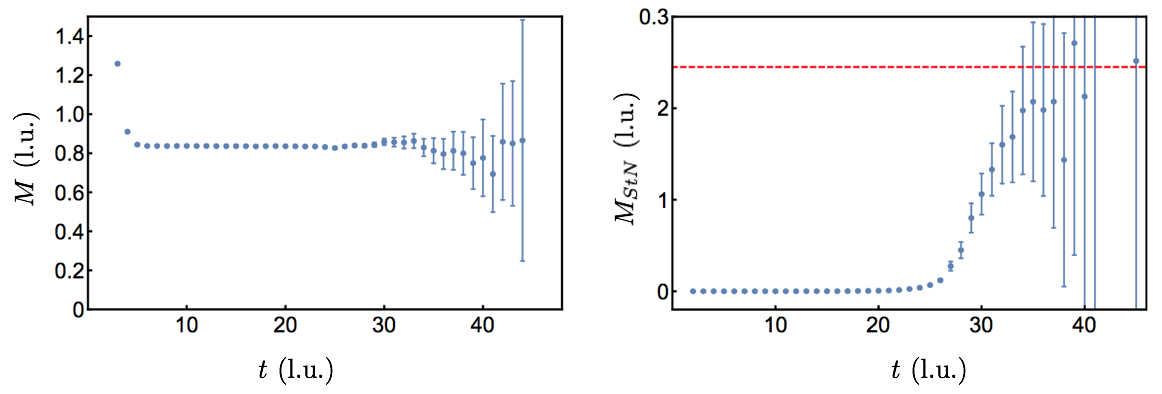}
	\caption{
	\label{fig:xiemp} 
	The EMP associated with the $\Xi$-baryon correlation function with $t_J =2$ (left panel)
	and the energy scale associated with the standard deviation of the ground state energy (right panel).
	This correlation function  is a tuned linear combination 
	of those resulting from localized smeared and 
	point sinks and from a localized smeared source at a pion mass of $m_\pi\sim 450~{\rm MeV}$
	calculated  from 96 sources per configuration on 3538 statistically independent isotropic clover gauge-field 
	configurations~\protect\cite{Orginos:2015aya}.	  
	They have been blocked together to form 100 independent samplings of the combined correlation function.
	The red dashed line in the right panel corresponds to the lowest energy contributing to the StN ratio that is expected to dominate at large times.
	}		
\end{figure}
EMPs, such as that associated with the $\Xi$-baryon shown in Fig.~\ref{fig:xiemp},
 are formed from ratios of correlation functions, which become constant when only 
a single exponential is contributing to the correlation function,
\begin{eqnarray}
  M(t) = \frac{1}{t_J}\ln\left[ {\langle C_i(t) \rangle \over \langle C_i(t+ t_J) \rangle}  \right] & \rightarrow & E_0
\label{eq:emdef}
\ \ \ ,
\end{eqnarray}
where $E_0$ is the ground state energy in the channel with appropriate quantum numbers.
The average over gauge field configurations is typically  over  correlation functions derived from multiple source points on multiple gauge-field configurations.
This is well-known technology and is a ``workhorse'' in the analysis of LQCD calculations.
Typically, $t_J$ corresponds to one temporal lattice spacing, and the jackknife and bootstrap resampling techniques are used to generate covariance matrices in the plateau interval used to extract the 
ground-state energy from a correlated $\chi^2$-minimization~\cite{DeGrand:1990ss,Beane:2010em,Beane:2014oea}.~\footnote{
For pedagogical introductions to LQCD uncertainty quantification with resampling methods, 
see Refs.~\cite{Young,DeGrand:1990,Luscher:2010ae,Beane:2014oea}.
}
The energy can be extracted from an exponential fit to the correlation function or by a direct fit to the effective mass itself.
Because correlation functions generated from the same, and nearby, gauge-field configuration are correlated,
typically they are blocked to form one average correlation function 
per configuration, and  blocked further over multiple configurations, to create an smaller ensemble 
containing (approximately) statistically independent samplings of the correlation function.

\begin{figure}[!ht]
	\includegraphics[width=\columnwidth]{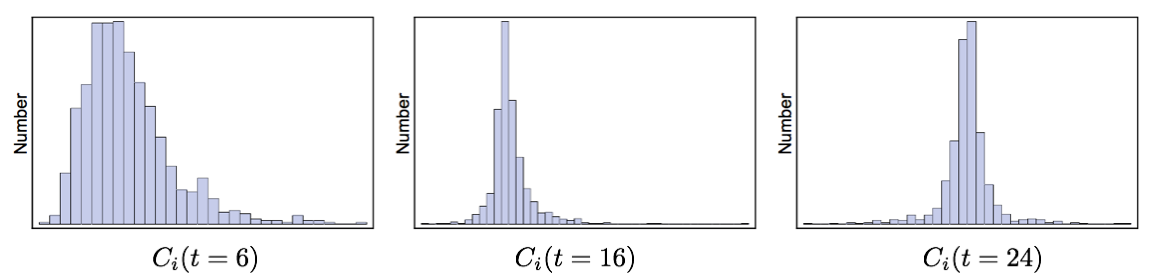}
	\caption{
	\label{fig:ReCdist} 
	The distribution of the real part of
	$10^3$ nucleon correlation functions at time slices $t=6$ (left panel), $t=16$ (middle panel) and $t=24$ (right panel). 
	}		
\end{figure}

It is known that baryon correlation functions contain strong correlations over $\sim m_\pi^{-1}$ time scales, and that these correlations are sensitive the presence of outliers.
Fig.~\ref{fig:ReCdist} shows the distribution of the real part of small-time nucleon correlation functions,
which resembles a heavy-tailed log-normal distribution~\cite{DeGrand:2012ik}.
Log-normal distributions are associated with a larger number of ``outliers'' than arise when sampling a Gaussian distribution,
and the sample mean of these small-time correlation function will be strongly affected by the presence of these outliers.
The distribution of baryon correlation functions at very large source-sink separations is also heavy-tailed;
David Kaplan has analyzed the real parts of NPLQCD baryon correlation functions and found that they resemble a stable distribution~\cite{davidkaplanLuschertalk}.
Cancellations between positive and negative outliers occur in determinations of the sample mean of this large-time distribution,
leading to different statistical issues that are explored in detail in Sec.~\ref{sec:decomposition}.

\begin{figure}[!ht]
  \includegraphics[width=.9\columnwidth]{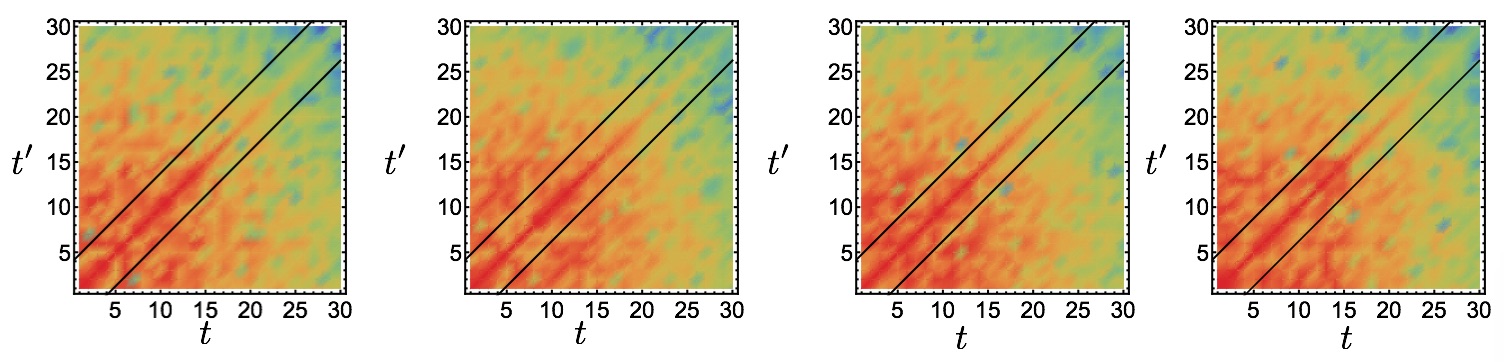}\hspace{5pt}
	\includegraphics[width=.07\columnwidth]{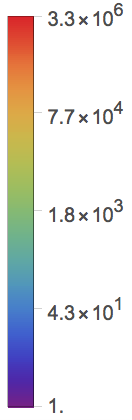}
	\caption{
	\label{fig:jugemean} 
        The logarithm of the inverse covariance matrix determined using booststrap resampling of the sample mean.  
        Lines with $t = m_\pi^{-1}$ and $t^\prime = m_\pi^{-1}$ are shown to demonstrate expected hadronic correlation lengths.
	The correlation function is the same as that described in the caption of Fig.~\protect\ref{fig:xiemp}.
        The normalization of the color scale is identical for all $t_J$.
	}		
\end{figure}

To analyze temporal correlations in baryon correlation functions in more detail,
results for inverse covariance matrices generated through bootstrap resampling of the $\Xi$ baryon effective mass are shown in Fig.~\ref{fig:jugemean}.
The size of off-diagonal elements in the inverse covariance matrix directly sets the size of contributions to the least-squares fit result 
from temporal correlations in the effective mass,
and so it is appropriate to use their magnitude to describe the strength of temporal correlations.
The inverse covariance matrix is seen to possess large off-diagonal elements associated with small time separations
that appear to decrease exponentially with increasing time separation at a rate somewhat faster than $m_\pi^{-1}$.
Mild variation in the inverse covariance matrix is seen when $t_J$ is varied.
Since correlations between $M(t)$ and $M(t^\prime)$ are seen in Fig.~\ref{fig:jugemean} to decrease rapidly as $|t-t^\prime|$ becomes large
compared to hadronic correlation lengths,
is expected that small distance correlations in the covariance matrix decrease
when $C_i(t)$ and $C_i(t-t_J)$ are separated by $t_J \gg m_\pi^{-1}$ and Fig.~\ref{fig:ReCdist},
though such an effect is not clearly visible in the inverse covariance matrix on the logarithmic scale shown.

\begin{figure}[!ht]
  \includegraphics[width=.9\columnwidth]{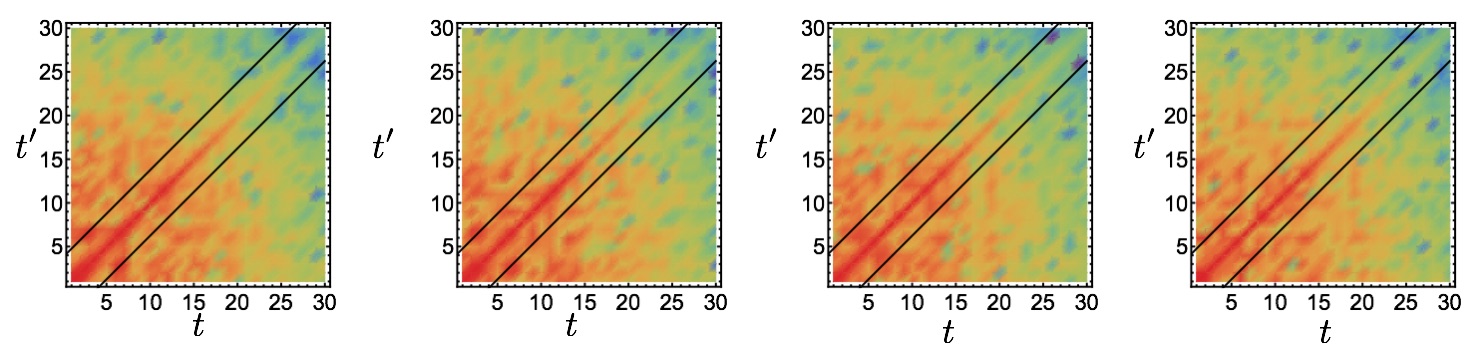} \hspace{5pt}
	\includegraphics[width=.07\columnwidth]{colorscale.png}
	\caption{
	\label{fig:jugeHL} 
	The logarithm of the inverse of the $\Xi$ baryon effective mass covariance matrix for $t_J = 1,2,3,16$ 
	determined using bootstrap resampling of the Hodges-Lehman estimator. 
        Lines with $t = m_\pi^{-1}$ and $t^\prime = m_\pi^{-1}$ are shown to demonstrate expected hadronic correlation lengths.
        The normalization of the color scale is identical for all $t_J$ and further is identical to the normalization of Fig.~\ref{fig:jugemean}.
	}		
\end{figure}

The role of outliers in temporal correlations on timescales $\lesssim m_\pi^{-1}$ is highlighted in Fig.~\ref{fig:jugeHL},
where inverse covariance matrices determined with the Hodges-Lehmann estimator are shown.
The utility of robust estimators, such as the median and the Hodges-Lehmann estimator, with reduced sensitivity to outliers, 
has been explored in Ref.~\cite{Beane:2014oea}.  
When the median and  average of a function are known to coincide,   
there are advantages to using 
the median or Hodges-Lehmann estimator
 to determine the average of a distribution. 
 The associated uncertainty can be estimated with the ``median absolute deviation'' (MAD), and be related to the 
 standard deviation with a well-known scaling factor.
Off-diagonal elements in the inverse covariance matrix associated with timescales $\lesssim m_\pi^{-1}$ are visibly
smaller on a logarithmic scale when the covariance matrix is determined with the Hodges-Lehmann estimator instead of the sample mean.
This decrease in small-time correlations when a robust estimator is employed strongly suggests that
short-time correlations on scales $\lesssim m_\pi^{-1}$ are associated with outliers.

%

\section{A Magnitude-Phase Decomposition}
\label{sec:decomposition}

In terms of the log-magnitude and  phase, the mean nucleon correlation functions is
\begin{equation}
  \begin{split}
    \avg{C_i(t)} =  \int\mathcal{D}C_i\;\mathcal{P}(C_i(t))\; e^{R_i(t) + i\theta_i(t)}
    \ \ \ .
  \end{split}
  \label{Csignproblem}
\end{equation}
In principle, $e^{R_i(t)}$ could be included in the MC probability distribution used for importance sampling. 
With this approach, $R_i(t)$ would contribute as an additional term in a new effective action. 
The presence of non-zero $\theta_i(t)$ demonstrates that this effective action would have an imaginary part. 
The resulting  weight therefore could not be interpreted as a probability and importance sampling 
could not proceed; importance sampling of $C_i(t)$ faces a sign problem. 
In either the canonical or grand canonical approach, one-baryon correlation functions are described by 
complex correlation functions that cannot be directly importance sampled without a sign problem,
but 
 it is formally permissible to importance sample according to the vacuum probability distribution, calculate the  phase resulting from the imaginary effective action on each background field configuration produced in this way, 
 and average the results on an ensemble of background fields. 
 This approach, known as reweighting, has a long history in grand canonical ensemble calculations 
 but has been generically unsuccessful because statistical averaging is impeded by large fluctuations in the complex phase that grow exponentially with increasing spacetime volume~\cite{Gibbs:1986ut,Splittorff:2006fu,Splittorff:2007ck}. 
 Canonical ensemble nucleon calculations averaging $C_i(t)$ over background fields importance sampled 
 with respect to the vacuum probability distribution are in effect solving the sign problem associated with 
 non-zero $\theta_i(t)$ by reweighting. As emphasized by Ref.~\cite{Grabowska:2012ik}, 
 similar chiral physics is responsible for the exponentially hard StN problem appearing in canonical 
 calculations and exponentially large fluctuations of the complex phase in grand canonical calculations.

Reweighting a pure phase causing a sign problem generically produces a StN problem in theories with a mass gap. 
Suppose $\avg{e^{i\theta_i(t)}}\sim e^{-M_\theta t}$ for some $M_\theta \neq 0$. 
Then because $|e^{i\theta_i(t)}|^2 = 1$ by construction, $\theta_i(t)$ has the StN ratio
\begin{equation}
  \begin{split}
    \frac{\avg{e^{i\theta_i(t)}}}{\sqrt{\avg{|e^{i\theta_i(t)}|^2}}} = \avg{e^{i\theta_i(t)}} \sim e^{-M_\theta t}
    \ \ \ ,
  \end{split}\label{ThStN}
\end{equation}
which is necessarily exponentially small at large times. 
Non-zero $M_\theta$ guarantees that statistical sampling of $e^{i\theta_i(t)}$ has a StN problem. 
Strictly, this argument applies to a pure phase but not to a generic complex observable such as 
$C_i(t)$ which might receive zero effective mass contribution from $\theta_i(t)$ and could have important 
correlations between $R_i(t)$ and $\theta_i(t)$. 
MC LQCD studies are needed to understand whether the pure phase StN problem of Eq.~\eqref{ThStN} 
captures some or all of the nucleon StN problem of Eq.~\eqref{Lepage}.

To determine the large-time behavior of correlation functions, it is useful to consider the effective-mass estimator 
commonly used in LQCD spectroscopy, a special case of eq.~(\ref{eq:emdef}), 
\begin{equation}
  \begin{split}
    M(t) = \ln \left[ \frac{\avg{C_i(t)}}{\avg{C_i(t+1)}} \right]
    \ \ \  .
  \end{split}\label{EMdef}
\end{equation}
As $t\rightarrow \infty$, the average correlation function can be described by a single exponential whose  
decay rate is set by the ground state energy, and therefore $M(t)\rightarrow M_N$. 
The uncertainties associated with  $M(t)$ can be estimated by resampling methods such as 
bootstrap. 
The variance of $M(t)$ is generically smaller than that of $\ln\avg{C_i(t)}$ due to cancellations arising from 
correlations between 
$\ln\left[\avg{C_i(t)}\right]$ 
and 
$\ln\left[\avg{C_i(t+1)}\right]$ 
across bootstrap ensembles. 
Assuming that these correlations do not affect the asymptotic scaling of the variance of $M(t)$, 
propagation of uncertainties for bootstrap estimates of the variance of 
$\ln\left[\avg{C_i(t)}\right]$ 
shows that the variance of $M(t)$ scales as
\begin{equation}
  \begin{split}
    \text{Var}\left( M(t) \right) \sim \frac{\text{Var}\left( C_i(t) \right)}{\avg{C_i(t)}^2} \sim e^{2\left( M_N - \frac{3}{2}m_\pi \right)t}
    \ \ \ .
  \end{split}
  \label{EMLepage}
\end{equation}
An analogous effective-mass estimator for the large-time exponential decay of the magnitude is
\begin{equation}
  \begin{split}
    M_R(t) = \ln \left[ \frac{\avg{e^{R_i(t)}}}{\avg{e^{R_i(t+1)}}} \right]
    \ \ \ ,
  \end{split}
  \label{EMRdef}
\end{equation}
and an effective-mass estimator for the  phase is
\begin{equation}
  \begin{split}
    M_\theta(t) = \ln \left[  \frac{\avg{e^{i\theta_i(t)}}}{\avg{e^{i\theta_i(t+1)}}} \right] 
    = 
    \ln \left[ \frac{\avg{\cos(\theta_i(t))}}{\avg{\cos(\theta_i(t+1))}} \right],
  \end{split}
  \label{EMThdef}
\end{equation}
where the reality of the average correlation function has been used.  

\begin{figure}
  \centering
  \includegraphics[width=\columnwidth]{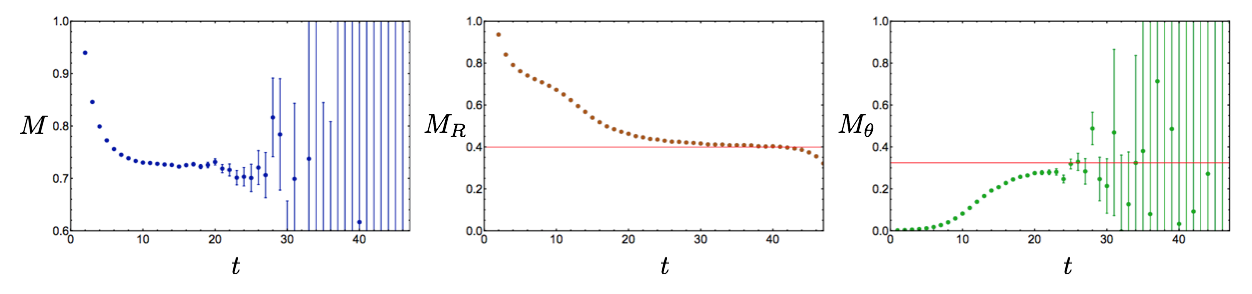}
  \caption{
  The left panel shows the nucleon effective mass $M(t)$ as a function of Euclidean time in lattice units. 
  The middle and right panels show the effective masses $M_R(t)$ and $M_\theta(t)$ of the magnitude and  
  phase respectively. 
  The asymptotic values of $M_R(t)$ and $M_\theta(t)$ are close to $\frac{3}{2}m_\pi$ and $M_N - \frac{3}{2}m_\pi$ respectively, whose values are indicated for comparison with horizontal red lines. 
  The uncertainties are calculated  using bootstrap methods.
  Past $t\gtrsim 30$ the imaginary parts of $\avg{C_i(t)}$ and $\avg{\cos\theta_i(t)}$ are not negligible compared to the real part.
  Here and below we display the real part of the complex log in Eq.~\eqref{EMdef}-\eqref{EMThdef}; taking the real part of the average correlation functions before taking the log or some other prescription would modify the results after $t\gtrsim 30$ in the left and right panels.
  All definitions are equivalent in the infinite statistics limit where $\avg{C_i(t)}$ is real.
  }
  \label{RTh_EM}
\end{figure}
\begin{figure}
  \centering
  \includegraphics[width=\columnwidth]{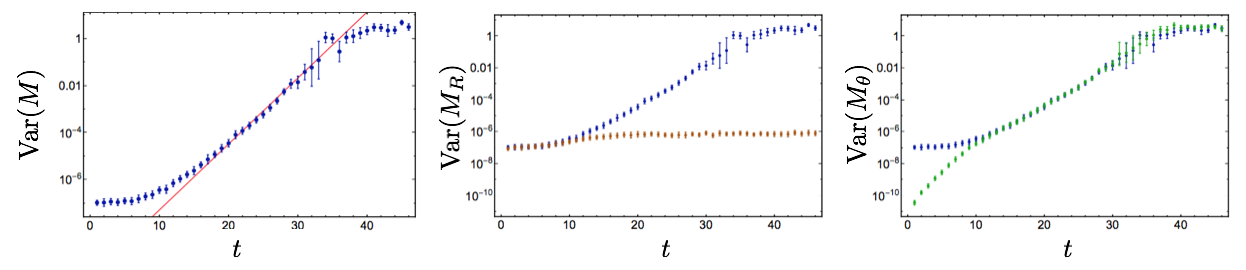}
  \caption{
  Variances of the effective mass estimates shown in Fig.~\ref{RTh_EM}. 
  The blue points common to all panels show the variance of $M(t)$. 
  The red line in the left plot shows a fit to $e^{2(M_N - \frac{3}{2}m_\pi)t}$ variance growth, where the normalization has been fixed to reproduce the observed variance at $t=22$. 
  The orange points in the middle panel show the variance associated with $M_R(t)$.
   The green points in the right panel show the variance associated with $M_\theta(t)$.
   }
  \label{RTh_EMErrors}
\end{figure}

Figure~\ref{RTh_EM} shows EMPs for $M(t)$, $M_R(t)$, and $M_\theta(t)$ calculated from the LQCD ensemble described previously. 
The mass of the nucleon, 
determined from a constant fit in the shaded plateau region $15 \leq t \leq 25$ indicated in Fig.~\ref{RTh_EM},
is found  to be 
$M_N = 0.7253(11)(22)$,
 in agreement with the mass obtained from the golden window in previous studies~\cite{Orginos:2015aya} of 
 $M_N = 0.72546(47)(31)$.
$M_R(t)$ and $M_\theta(t)$ do not visually plateau until much larger times. 
For the magnitude, a constant fit in the shaded region $30 \leq t \leq 40$ gives an effective mass 
$M_R(t) \rightarrow M_R = 0.4085(2)(13)$ which is  close to the value 
$\frac{3}{2}m_\pi = 0.39911(35)(14)$~\cite{Orginos:2015aya} 
indicated by the red line. 
For the  phase, a constant fit to the shaded region $25\leq t \leq 29$ gives an effective mass 
$M_\theta(t) \rightarrow M_\theta = 0.296(20)(12)$, which is  consistent with the value 
$M_N - \frac{3}{2} m_\pi = 0.32636(58)(34)$~\cite{Orginos:2015aya} 
indicated by the red line. 
It is unlikely that the  phase has reached its asymptotic value by this time, but a signal cannot be established at larger times. 
For $t\geq 30$, 
large fluctuations lead to complex effective mass estimates for $M(t)$ and $M_\theta(t)$ and unreliable estimates and uncertainties.  
$M_R(t) + M_\theta(t)$ agrees with $M(t)$ up to $\lesssim 5\%$ corrections at all times, demonstrating that the magnitude and 
cosine of the complex phase are approximately uncorrelated at the few percent level. 
This suggests the asymptotic scaling of the nucleon correlation function can be approximately decomposed as
\begin{equation}
  \begin{split}
    \avg{C_i(t)} \approx \avg{e^{R_i(t)}}\avg{e^{i\theta_i(t)}} \sim \left(  e^{-\frac{3}{2}m_\pi t}\right)\left(  e^{-\left( M_N - \frac{3}{2}m_\pi \right)t}\right)
    \ \ \ \ .
  \end{split}\label{RThScaling}
\end{equation}

For small times $t \lesssim 10$, the means and variances of $M(t)$ and $M_R(t)$ agree up to 
a small contribution from  $M_\theta(t)$. 
This indicates that the real part of the correlation function is nearly equal to its magnitude at small times. 
At intermediate times $10 \lesssim t \lesssim 25$, the contribution of $M_\theta(t)$ grows relative to 
$M_R(t)$, and for $t\gtrsim 15$ the variance of the full effective mass is nearly saturated by the variance of 
$M_\theta(t)$, as shown in Fig.~\ref{RTh_EMErrors}. 
At intermediate times a linear fit normalized to $\text{Var}(M(t=22))$ with slope $e^{2(M_N - \frac{3}{2}m_\pi)t}$ 
provides an excellent fit to bootstrap estimates of $\text{Var}(M(t))$, 
in agreement with the scaling of Eq.~\eqref{EMLepage}. $\text{Var}(M_\theta(t))$ is indistinguishable from 
$\text{Var}(M(t))$ in this region, and $m_\theta(t)$ has an identical StN problem. 
$\text{Var}(M_R(t))$ has 
much more mild time variation, and $M_R(t)$ can be reliably estimated at all times without almost no  StN problem. 
At intermediate times, the presence of non-zero $\theta_i(t)$ signaling a sign problem in importance sampling of $C_i(t)$ appears responsible for the entire nucleon StN problem.

$M(t)$ approaches its asymptotic value much sooner than $M_R(t)$ or $M_\theta(t)$. 
This indicates that the overlap of $\bar{N}(0)N(0)$ onto the three-pion ground state in the variance correlation function is greatly 
suppressed compared to the overlap of $\bar{N}(0)$ onto the one-nucleon signal ground state. 
Optimization of the interpolating operators for high signal overlap contributes to this. 
Another contribution arises from momentum projection, which suppresses the variance overlap factor by 
$\sim 1/(m_\pi^3 V)$~\cite{Beane:2009gs}. 
A large hierarchy between the signal and noise overlap factors provides a golden window visible 
at intermediate times $10 \lesssim t \lesssim 25$. 
In the golden window, $M(t)$ approaches it's asymptotic value 
but $\text{Var}(M(t))$  begins to grow exponentially and 
$M_\theta(t)$ is suppressed compared to $M_R(t)$. Reliable extractions of $M(t)$ are possible in the golden window.

\begin{figure}[!ht]
  \centering
  \includegraphics[width=\columnwidth]{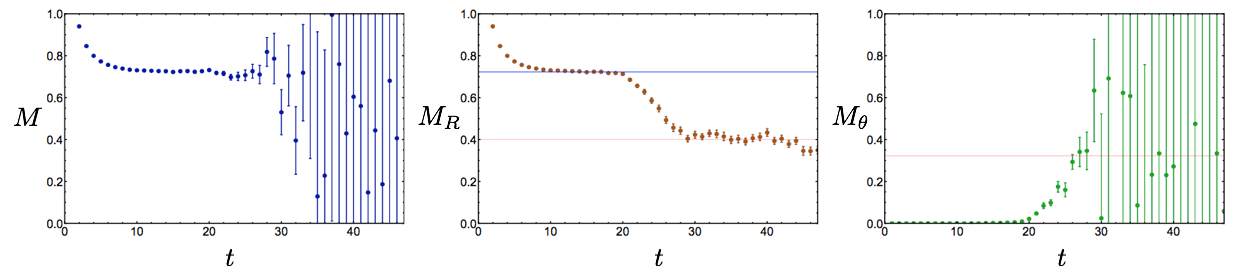}
  \caption{
  EMPs from an ensemble of 500 blocked correlation functions,
  each of which is equal to the sample mean of 1000 nucleon correlation functions. 
  The left panel shows the effective mass $M(t)$ of the blocked correlation functions. 
  The middle panel shows the magnitude contribution $m_R(t)$ and, for reference, 
  a red line at $\frac{3}{2}m_\pi$ and a blue line at $M_N$ are shown. 
  The right panel shows the phase mass $m_\theta(t)$ of the blocked correlation functions 
  along with a red line at $M_N - \frac{3}{2}m_\pi$.
  }
  \label{BlockedRTh_EM}
\end{figure}
\begin{figure}[!ht]
  \centering
  \includegraphics[width=\columnwidth]{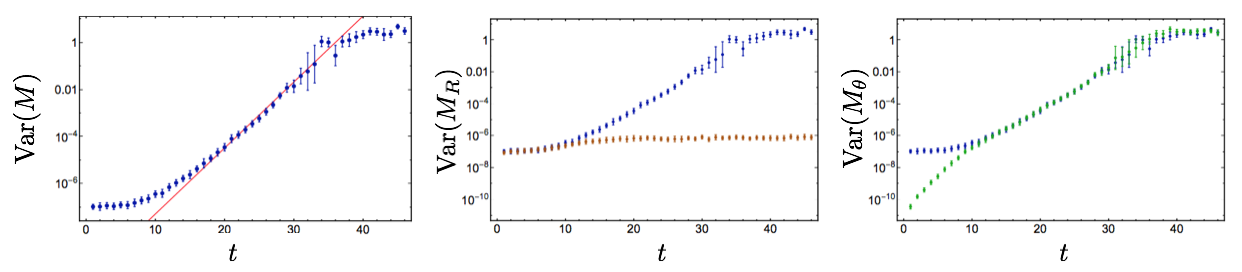}
  \caption{
  Bootstrap estimates of the variance of the effective mass using blocked correlation functions. 
The left panel shows the variance of $M(t)$ for blocked data in blue and the almost indistinguishable variance of $M(t)$ for unblocked data in gray. The middle panel shows the variance of blocked estimates of $m_R(t)$ in orange and the right panel shows the variance of blocked estimates of $m_\theta(t)$ in green.}
  \label{BlockedRTh_EMErrors}
\end{figure}

The effects of blocking, that is averaging subsets of correlation functions and analyzing the distribution of the averages, 
are shown in Fig.~\ref{BlockedRTh_EM}. 
$M_\theta(t)$ is suppressed compared to $M_R(t)$ for larger times in the blocked ensemble, and the log-magnitude 
saturates the average and variance of $M(t)$ through intermediate times $t\lesssim 25$. 
Blocking does not actually reduce the total variance of $M(t)$. 
Variance in $M(t)$ is merely shifted from the phase to the log-magnitude at intermediate times. 
This is reasonable, since the imaginary part of $C_i(t)$ vanishes on average and so blocked correlation functions 
will have smaller imaginary parts. Still, blocking does not affect $\avg{C(t)}$ and only affects bootstrap 
estimates of $\text{Var}(M(t))$ at the level of correlations between correlation functions in the ensemble. 
Blocking also does not delay the onset of a large-time noise region $t\gtrsim 35$ where $M(t)$ and 
$m_\theta(t)$ cannot be reliably estimated.

Eventually the scaling of $\text{Var}(M(t))$ begins to deviate from Eq.~\eqref{EMLepage}, 
and in the noise region $t\gtrsim 35$ the observed variance remains approximately  constant (up to large fluctuations). 
This is inconsistent with Parisi-Lepage scaling. While the onset of the noise region is close to the mid-point of the time direction $t=48$, a qualitatively similar onset occurs at smaller times in smaller statistical ensembles.
Standard statistical estimators therefore do not reproduce the scaling required by basic principles of quantum field theory 
in the noise region. 
This suggests systematic deficiencies leading to unreliable results for standard statistical estimation of correlation 
functions in the noise region. 
The emergence of a noise region where standard statistical tools are unreliable can be understood in terms of the 
circular statistics describing $\theta(t)$ and is explained in Sec.~\ref{sec:phase}. 
A more straightforward analysis of the distribution of $R_i(t)$ is first presented below.

\subsection{The Magnitude}
\label{sec:magnitude}

\begin{figure}[!ht]
  \centering
  \includegraphics[width=\columnwidth]{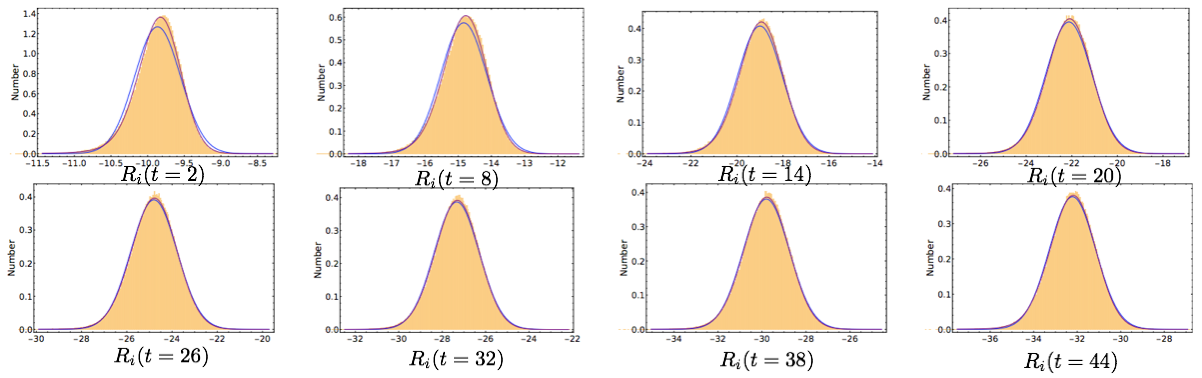}
  \caption{
  Normalized histograms of $R_i(t)$ derived from the LQCD results. 
  The blue curves correspond to  best fit normal distributions  determined from the sample mean and variance,
  while the purple curves correspond to 
  maximum likelihood fits to generic stable distributions. 
  See the main text for more details.
  }
  \label{RHistograms}
\end{figure}

\begin{figure}[!ht]
  \centering
  \includegraphics[width=\columnwidth]{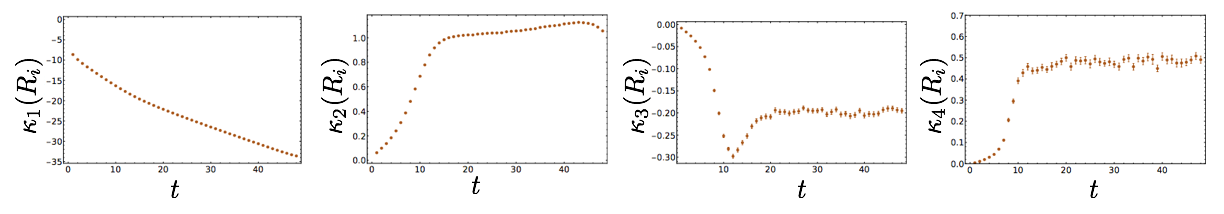}
  \caption{
  The first four cumulants of $R(t)$ as functions of $t$. Cumulants are calculated from sample moments using Eq.~\eqref{cumulantdef} 
  and the associated uncertainties are estimated by bootstrap methods. 
  From left to right, the panels show the cumulants 
   $\kappa_1(R(t))$ (mean),  $\kappa_2(R(t))$  (variance),   $\kappa_3(R(t))$  (characterizing skewness) and $\kappa_4$ (characterizing kurtosis).
  }
  \label{RCumulants}
\end{figure}

Histograms of the nucleon log-magnitude are shown in Fig.~\ref{RHistograms}. 
Particularly at large times, the distribution of $R_i(t)$ is approximately described by a normal distribution. 
Fits to a normal distribution are qualitatively good but not exact, and  deviations between normal distribution fits 
and $R_i(t)$ results are  visible in Fig.~\ref{RHistograms}. 
Cumulants of $R_i(t)$ can be used to quantify these deviations, which can be  recursively calculated  from its moments by
\begin{equation}
  \begin{split}
    \kappa_n\left( R_i(t) \right) = \avg{R_i(t)^n} - \sum_{m=1}^{n-1} {{n-1}\choose{m-1}} \kappa_m\left( R_i(t) \right)\avg{R_i(t)^{n-m}}
    \ \ \  .
  \end{split}\label{cumulantdef}
\end{equation}
The first four cumulants of a probability distribution characterize its mean, variance, skewness, and kurtosis respectively. 
If $|C_i(t)|$ were exactly log-normal, the first and second cumulants of $R_i(t)$, its mean and variance, would fully describe the distribution. 
Third and higher cumulants of $R_i(t)$ would all vanish for exactly log-normal $|C_i(t)|$. 
Fig.~\ref{RCumulants} shows the first four cumulants of $R_i(t)$. 
Estimates of higher cumulants of $R_i(t)$ become successively noisier.

The cumulant expansion of Ref.~\cite{Endres:2011jm} relates the effective mass of a correlation function to the cumulants 
of the logarithm of the correlation function. The derivation of Ref.~\cite{Endres:2011jm} is directly applicable to $M_R(t)$. 
The characteristic function $\Phi_{R(t)}(k)$, defined as the Fourier transform of the probability distribution function of $R_i(t)$, 
can be described by a Taylor series for $\ln[\Phi_{R(t)}(k)]$ whose coefficients are precisely the cumulants of $R_i(t)$,
\begin{equation}
  \begin{split}
    \Phi_{R(t)}(k) = \avg{e^{i k R_i(t)}} = \exp\left[\sum_{n=1}^\infty \frac{(ik)^n}{n!}\kappa_n(R_i(t))\right].
  \end{split}
  \label{cumulants}
\end{equation}
The average magnitude of $C_i(t)$ is given in terms of this characteristic function by
\begin{equation}
  \begin{split}
    \avg{e^{R_i(t)}} = \Phi_{R(t)}(-i) = \exp\left[\sum_{n=1}^\infty \frac{\kappa_n(R_i(t))}{n!}\right].
  \end{split}
  \label{cumulantR}
\end{equation}
This allows application of the cumulant expansion in Ref.~\cite{Endres:2011jm} 
to the effective mass in Eq.~(\ref{EMRdef}) to give,
\begin{equation}
  \begin{split}
    M_R(t) = \sum_{n=1}^\infty \frac{1}{n!}\left[ \kappa_n(R_i(t)) - \kappa_n(R_i(t+1)) \right].
  \end{split}
  \label{cumulantEM}
\end{equation}
Since $\kappa_n(R_i(t))$ with $n >2$ vanishes for normally distributed $R_i(t)$, the cumulant expansion provides a rapidly 
convergent series for correlation functions that are close to, but not exactly, log-normally distributed. 
Note that the right-hand-side of Eq.~\eqref{cumulantEM} is simply a discrete approximation suitable for a lattice regularized 
theory of the time derivative of the cumulants.

\begin{figure}[!ht]
  \centering
  \includegraphics[width=\columnwidth]{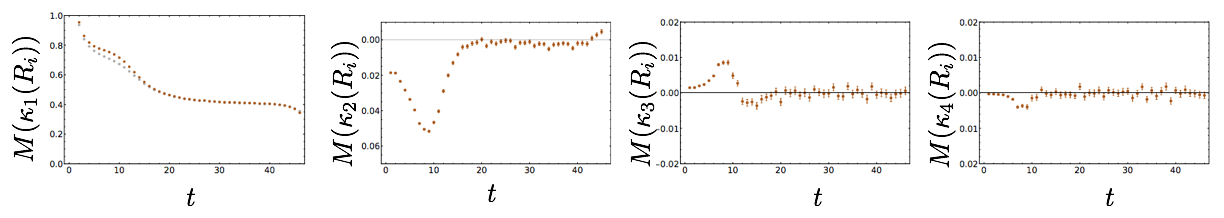}
  \caption{
  Contributions to $M_R(t)$ from the first four terms in the cumulant expansion of Ref.~\cite{Endres:2011jm} 
  given in Eq.~\eqref{cumulantEM}. In the leftmost panel, the gray points correspond to the unapproximated 
  estimate for $M_R(t)$ (that are also shown in  Fig.~\ref{RTh_EM}),
  while the orange points show the contribution from the mean $\kappa_1(R(t))$. 
  The other panels show the contributions to Eq.~\eqref{cumulantEM} associated with the 
  higher cumulants
  $\kappa_2(R_i(t))$, $\kappa_3(R(t))$, and $\kappa_4(R(t))$, respectively.
  }
  \label{RCumulantEM}
\end{figure}
Results for the effective mass contributions of the first few terms in the cumulant expansion of Eq.~\eqref{cumulantEM} 
are shown in Fig.~\ref{RCumulantEM}. 
The contribution $\kappa_1(R_i(t)) - \kappa_1(R_i(t+1))$, 
representing the time derivative of the mean, provides an excellent approximation to $M_R(t)$ after small times. 
$(\kappa_2(R_i(t)) - \kappa_2(R_i(t+1)))/2$ provides a very small negative contribution to $M_R(t)$, and contributions  
from $\kappa_3(R_i(t))$ and $\kappa_4(R_i(t))$ are statistically consistent with zero. 
As $M_R(t)$ approaches its asymptotic value, the log-magnitude distribution can be described to high-accuracy by a 
nearly normal distribution with very slowly increasing variance and small, approximately 
constant  $\kappa_{3,4}$. 
The slow increase of the variance of $R_i(t)$ is consistent with observations above that $|C_i(t)|$ has no severe StN problem.
It is also consistent with expectations that $|C_i(t)|^2$ describes a (partially-quenched) three-pion correlation function with a very mild StN problem, 
with a  scale  set by the attractive isoscalar pion interaction energy.

As Eq.~\eqref{cumulantEM} relates $M_R(t)$ to time derivatives of moments of $R_i(t)$, it is interesting to consider the distribution of the time derivative $\frac{dR_i}{dt}$. 
Defining generic finite differences,
\begin{equation}
  \begin{split}
    \Delta R_i(t, \Delta t) = R_i(t) - R_i(t -  \Delta t)
    \ \ \  ,
  \end{split}
  \label{DeltaRdef}
\end{equation}
the time derivative of lattice regularized results can be defined as the finite difference,
\begin{equation}
  \begin{split}
    \frac{dR_i}{dt} = \Delta R_i(t, 1)
    \ \ \ .
  \end{split}
  \label{dRdtdef}
\end{equation}
If $R_i(t)$ and $R_i(t-1)$ were statistically independent,  it would be straightforward to extract the time derivatives of the 
moments of $R_i(t)$ from the moments of $\frac{dR_i}{dt}$. 
The presence of correlations in time, arising from non-trivial QCD dynamics, obstructs a naive extraction of 
$M_R(t)$ from moments of $\frac{dR)i}{dt}$. 
For instance, without knowledge of $\avg{R_i(t)R_i(t-1)}$ it is impossible to extract the time derivative of the variance of 
$R_i(t)$ from the variance of $\frac{dR_i}{dt}$. 
While the time derivative of the mean of $R_i(t)$ is simply the mean of $\frac{dR_i}{dt}$, 
 time derivatives of the higher cumulants of $R_i(t)$ cannot be extracted from the cumulants of $\frac{dR_i}{dt}$ 
without knowledge of dynamical correlations.

\begin{figure}[!ht]
  \centering
  \includegraphics[width=\columnwidth]{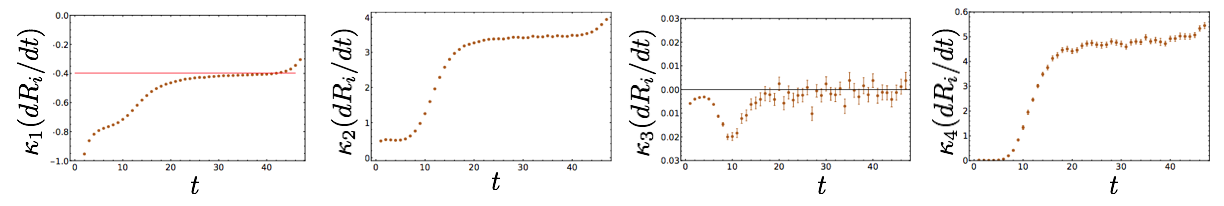}
  \caption{
    The first four cumulants of $\frac{dR_i}{dt}$, determined analogously to the cumulants in Fig.~\ref{RCumulants}.
  }
  \label{dRdtCumulants}
\end{figure}
\begin{figure}[!ht]
  \centering
  \includegraphics[width=\columnwidth]{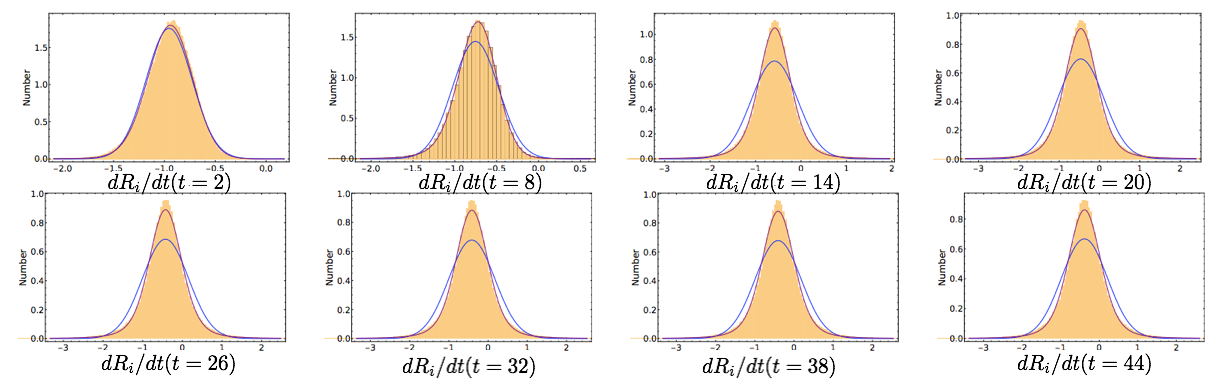}
  \caption{
  Histograms of $\frac{dR}{dt}$, defined as the finite difference $\Delta R(t, 1)$ given in Eq.~(\ref{DeltaRdef}). 
  The blue curves in each panel correspond to the best-fit normal distribution, while the purple curves correspond to the best-fit 
  stable distribution.
  }
  \label{dRdtHistograms}
\end{figure}

The cumulants of $\frac{dR_i}{dt}$ are shown in Fig.~\ref{dRdtCumulants}.  
As expected, the mean of $\frac{dR_i}{dt}$  approaches $\frac{3}{2}m_\pi$ at large times. 
The variance of $\frac{dR_i}{dt}$ is tending to a plateau which is approximately one-third of the variance of $R_i(t)$. 
This implies there are  correlations between $R_i(t)$ and $R_i(t-1)$ that are on the same order of the individual variances 
of $R_i(t)$ and $R_i(t-1)$. 
This is not surprising, given that the QCD correlation length is larger than the lattice spacing. 
No statistically significant $\kappa_3$ is seen for $\frac{dR_i}{dt}$ at large times, but a statistically significant positive $\kappa_4$ is found. 
Normal distribution fits to $\frac{dR_i}{dt}$ are found to be poor, as  shown in Fig.~\ref{dRdtHistograms}, as they 
underestimate both the peak probability and the probability of finding ``outliers'' in the tails of the distribution. 
Interestingly, 
Fig.~\ref{dRdtCumulants}, and histograms of $\frac{dR_i}{dt}$ shown in Fig.~\ref{dRdtHistograms},
suggest that the distribution of $\frac{dR_i}{dt}$ becomes approximately time-independent at large times.

Stable distributions are found to provide a much better description of $\frac{dR_i}{dt}$, and are  consistent with the heuristic arguments 
for log-normal correlation functions given in Ref.~\cite{Endres:2011jm}. 
Generic correlation functions can be viewed as products of creation and annihilation operators with many transfer matrix factors 
describing Euclidean time evolution. 
It is difficult to understand the distribution of products of transfer matrices in quantum field theories, 
but following Ref.~\cite{Endres:2011jm} insight can be gained by considering products of random positive numbers.
As a further simplification, one can consider a product of independent, identically distributed positive numbers, each schematically representing a product of many transfer matrices describing time evolution over a period much larger than all temporal correlation lengths.
Application of the central limit theorem to the logarithm of a product of many independent, identically distributed 
random numbers shows that the logarithm of the product tends to become normally distributed as the number of 
factors becomes large. 
The central limit theorem in particular assumes that the random variables in question 
originate from distributions that have a finite variance.
A generalized central limit theorem proves that sums of heavy-tailed random 
variables tend to become distributed according to stable distributions (that 
include the normal distribution as a special case), 
suggesting that
 stable distributions arise naturally in the logs of products of random variables.

Stable distributions are named as such because their shape is stable under averaging of independent copies of a random variable. 
Formally, stable distributions form a manifold of fixed points in a Wilsonian space of probability distributions where averaging 
independent random variables from the distribution plays the role of renormalization group evolution. 
A parameter $\alpha$, called the index of stability, dictates the shape of a stable distribution and remains fixed under averaging transformations. 
All probability distributions with finite variance evolve under averaging towards the normal distribution, 
a special case of the stable distribution with $\alpha = 2$. 
Heavy-tailed distributions with ill-defined variance evolve towards generic stable distributions with $0 < \alpha \leq 2$. 
In particular, stable distributions with $\alpha < 2$ have power-law tails; 
for a stable random variable $X$ the tails decay as $X^{-(\alpha + 1)}$. 
The heavy-tailed Cauchy, Levy, and Holtsmark distributions are special cases of stable distributions with $\alpha = 1,\;1/2,$ and $3/2$ respectively,
that arise in physical applications.~\footnote{
Further details can be found in  textbooks and reviews on stable distributions and their applications in physics. 
See, for instance, Refs.~\cite{Chandrasekhar:1943,Bouchaud:1990,Bardou:2000,Voit:2005,Nolan:2015} and references within.
}

Stable distributions for a real random variable $X$ are defined via Fourier transform,
\begin{equation}
  \begin{split}
    \mathcal{P}_S(X;\alpha,\beta,\mu,\gamma) &= \int \frac{dk}{2\pi}e^{-i k X}\Phi_X(k;\alpha,\beta,\mu,\gamma)
    \ \ \ ,
  \end{split}
  \label{PSdef}
\end{equation}
of their characteristic functions
\begin{equation}
  \begin{split}
    \Phi_X(k;\alpha,\beta,\mu,\gamma) = \exp\left( i \mu k - |\gamma k|^\alpha\left[1 - i\beta\frac{k}{|k|}\tan(\pi\alpha/2)\right]\right)
    \ \ \  ,
  \end{split}
  \label{PhiSdef}
\end{equation}
where $0<\alpha\leq 2$ is the index of stability, 
$-1\leq \beta \leq 1$ determines the skewness of the distribution, 
$\mu$ is the location of peak probability, 
$\gamma$ sets the width.
For $\alpha=1$,  the above parametrization does not hold and $\tan(\pi\alpha/2)$ should be replaced by $-\frac{2}{\pi}\ln|k|$.
For $\alpha > 1$ the mean is $\mu$, and for $\alpha \leq 1$ the mean is ill-defined. 
For $\alpha =2$ the variance is $\sigma^2 = \gamma^2/2$ and Eq.~\eqref{PhiSdef} implies the distribution is independent of $\beta$,
while
for $\alpha < 2$ the variance is ill-defined.

\begin{figure}[!ht]
  \centering
  \includegraphics[width=\columnwidth]{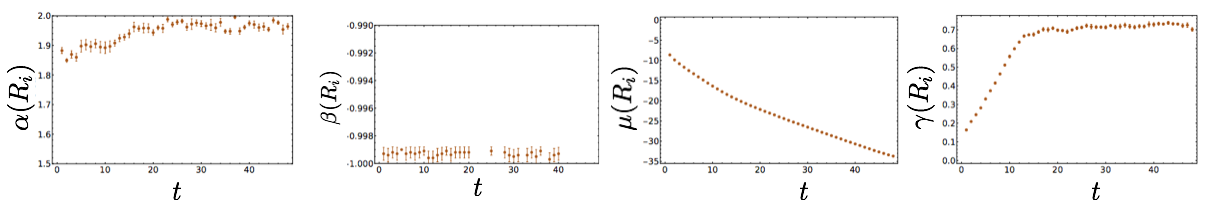}
  \caption{
  Maximum likelihood estimates for stable distribution fits of $R_i(t)$ in terms of the parameters of 
  Eq.~\eqref{PSdef}-\eqref{PhiSdef}. 
  $\alpha=2$ corresponds to a normal distribution. 
  The associated uncertainties are estimated by bootstrap methods.
  Changes in $\beta$ do not affect the likelihood when $\alpha=2$, and reliable estimates of $\beta(R_i(t))$ are not obtained at all times.
  }
  \label{RStable}
\end{figure}
The distributions of $R_i(t)$ obtained from the LQCD calculations
can be fit to stable distributions through maximum likelihood estimation of the stable parameters 
$\alpha,\;\beta,\;\mu,$ and $\gamma$, 
obtaining the results that are shown in Fig.~\ref{RStable}. 
Estimates of $\alpha(R_i)$ are consistent with $2$, corresponding to a normal distribution. 
This is not surprising, because higher moments of $|C_i(t)|$ would be ill-defined and diverge in the infinite statistics 
limit if $R_i(t)$ were literally described by a heavy-tailed distribution. 
$\beta(R_i)$ is strictly ill-defined when $\alpha(R_i)=2$, but results consistent with $\beta(R_i) = -1$ indicate negative 
skewness in agreement with observations above. 
Estimates of $\mu(R_i)$ and $\gamma(R_i)$ are consistent with the cumulant results above if a normal distribution ($\alpha(R_i) = 2$) is assumed. 
Fits of $R(t)$ to generic stable distributions are shown in Fig.~\ref{RHistograms},  and are roughly consistent with  fits to a normal distribution, though some skewness is captured by the stable fits.

\begin{figure}[!ht]
  \centering
  \includegraphics[width=\columnwidth]{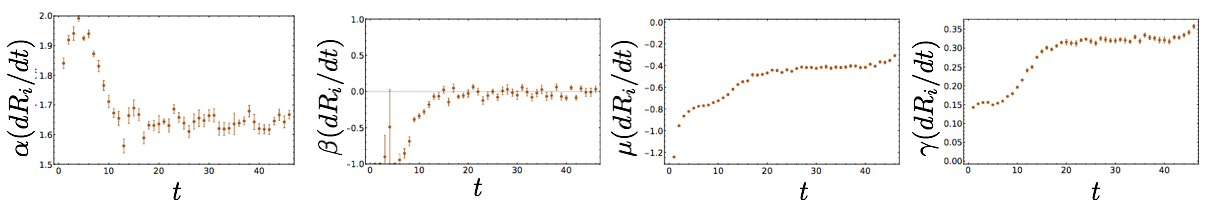}
  \caption{
  Maximum likelihood estimates for stable distribution fits of $\frac{dR_i}{dt}$ similar to Fig.~\ref{RStable}.
  The associated uncertainties are estimated by bootstrap methods.
}
  \label{dRdtStable}
\end{figure}
Stable distribution fits to $\frac{dR_i}{dt}$ indicate statistically significant deviations from a normal distribution ($\alpha=2$), 
as seen in Fig.~\ref{dRdtStable}.
 The large-time distribution of $\frac{dR_i}{dt}$ appears time independent, and fitting $\alpha\left( \frac{dR_i}{dt} \right)$ in the large-time plateau region gives an estimate of the large-time index of stability. Recalling $\frac{dR_i}{dt}$ describes a finite difference over a physical time interval of one lattice spacing, the estimated index of stability is
\begin{equation}
  \begin{split}
    \alpha\left( \Delta R(t\rightarrow \infty, \Delta t \sim 0.12\text{ fm}) \right) \rightarrow 1.639(4)(1).
  \end{split}
  \label{alphaDeltaR}
\end{equation}
Maximum likelihood estimates for $\mu\left( \frac{dR_i}{dt} \right)$ are consistent with the sample mean, and $\beta\left( \frac{dR_i}{dt} \right)$ is consistent with zero in agreement with observations of vanishing skewness.
Therefore,  the distribution of $ \frac{dR_i}{dt}$ is symmetric, as observed in Fig.~\ref{dRdtHistograms}, with power-law tails scaling as 
$\sim \left(\Delta R_i\right)^{-2.65}$ over this time interval of $\Delta t\sim 0.12~{\rm fm}$.

\begin{figure}[!ht]
  \centering
  \includegraphics[width=\columnwidth]{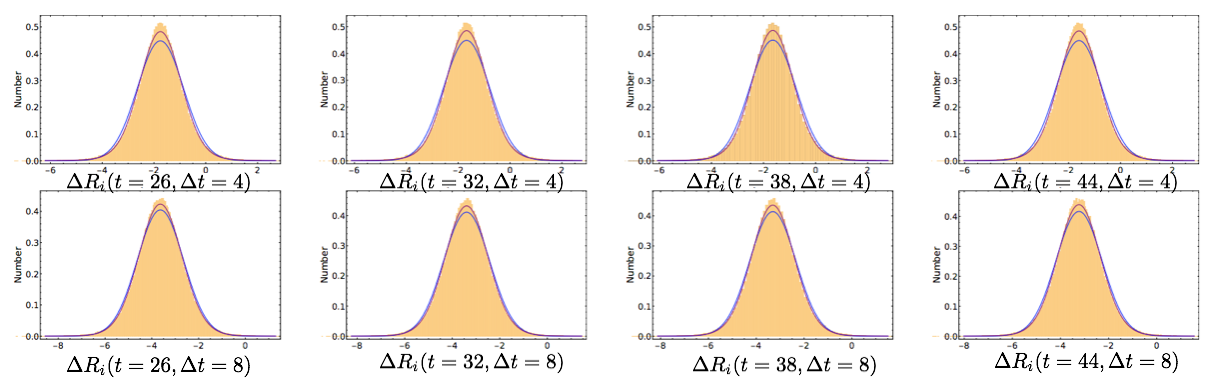}
  \caption{
  Histograms of $\Delta R_i(t, \Delta t)$ for selected large-time values of $t$. 
  The top row shows results for  $\Delta t = 4$, 
  the bottom row shows results for  $\Delta t = 8$,
  and   Fig.~\ref{dRdtHistograms} shows the results for  $\Delta t = 1$. 
  The blue curves represent fits to a normal distribution, while the purple curves represent fits to a stable distribution.
  }
  \label{DeltaRHistograms}
\end{figure}
\begin{figure}[!ht]
  \centering
  \includegraphics[width=\columnwidth]{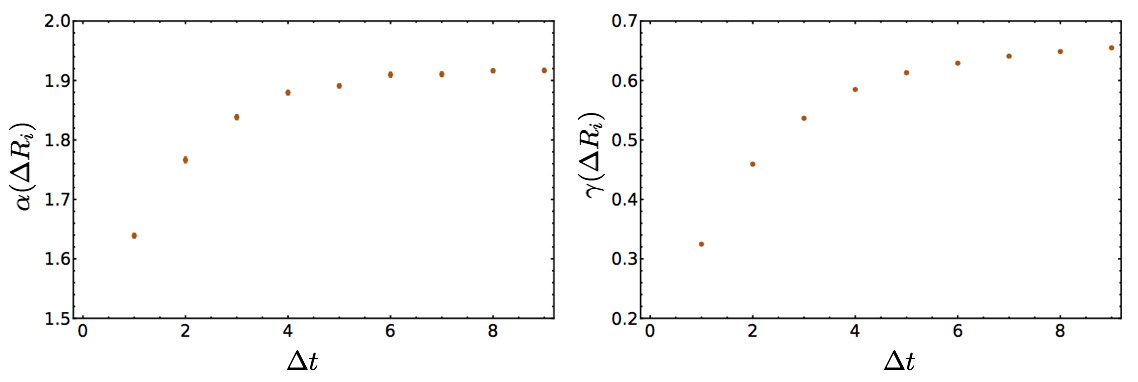}
  \caption{
  Maximum likelihood estimates for the index of stability, $\alpha\left(\Delta R_i(t, \Delta t)\right)$ and width $\gamma\left( \Delta R_i(t, \Delta t) \right)$, in the large-time 
  plateau region as a function of $\Delta t$.  Associated uncertainties are estimated with bootstrap methods.
  }
  \label{DeltaRStable}
\end{figure}

The value of $\alpha\left( \frac{dR_i}{dt} \right)$ depends on the physical time separation used in the finite difference 
definition Eq.~\eqref{DeltaRdef}, and 
stable distribution fits can be performed for generic finite differences $\Delta R_i(t, \Delta t)$. 
For all $\Delta t$, the distribution of $\Delta R_i$ becomes time independent at large times. 
Histograms of the large-time distributions $\Delta R$ for $\Delta t = 4,\;8$ are shown in Fig.~\ref{DeltaRHistograms}, and 
the best fit large-time values for $\alpha\left( \Delta R_i\right)$ and $\gamma\left( \Delta R_i \right)$ are shown in Fig.~\ref{DeltaRStable}. 
Since QCD has a finite correlation length,  $\Delta R_i(t, \Delta t)$ can be described as the difference of 
approximately normally distributed variables at large $\Delta t$. 
In the large $\Delta t$ limit, $\Delta R_i$ is therefore necessarily almost normally distributed,
and correspondingly, $\alpha(\Delta R_i)$, shown in Fig.~\ref{DeltaRStable}, 
increases with  $\Delta t$ and begins to approach the 
normal distribution value $\alpha(\Delta R_i)\rightarrow 2$ for large $\Delta t$. 
A large $\Delta t$ plateau in $\alpha(\Delta R_i)$ is observed that demonstrates small but statistically significant departures from $\alpha(\Delta R_i)< 2$.
This deviation is consistent with the appearance of small but statistically significant measures of non-Gaussianity in $R_i(t)$ seen in Fig.~\ref{RCumulants}.
Heavy-tailed distributions are 
found to be needed only to describe the distribution of $\Delta R_i$ when $\Delta t$ is small enough such 
that $R_i(t)$ and $R_i(t-\Delta t)$ are physically correlated. 
In some sense, the deviations from normally distributed differences, i.e. $\alpha(\Delta R_i) < 2$, are a measure 
the strength of dynamical QCD correlations on the scale $\Delta t$.

\begin{figure}[!ht]
  \centering
  \includegraphics[width=16cm]{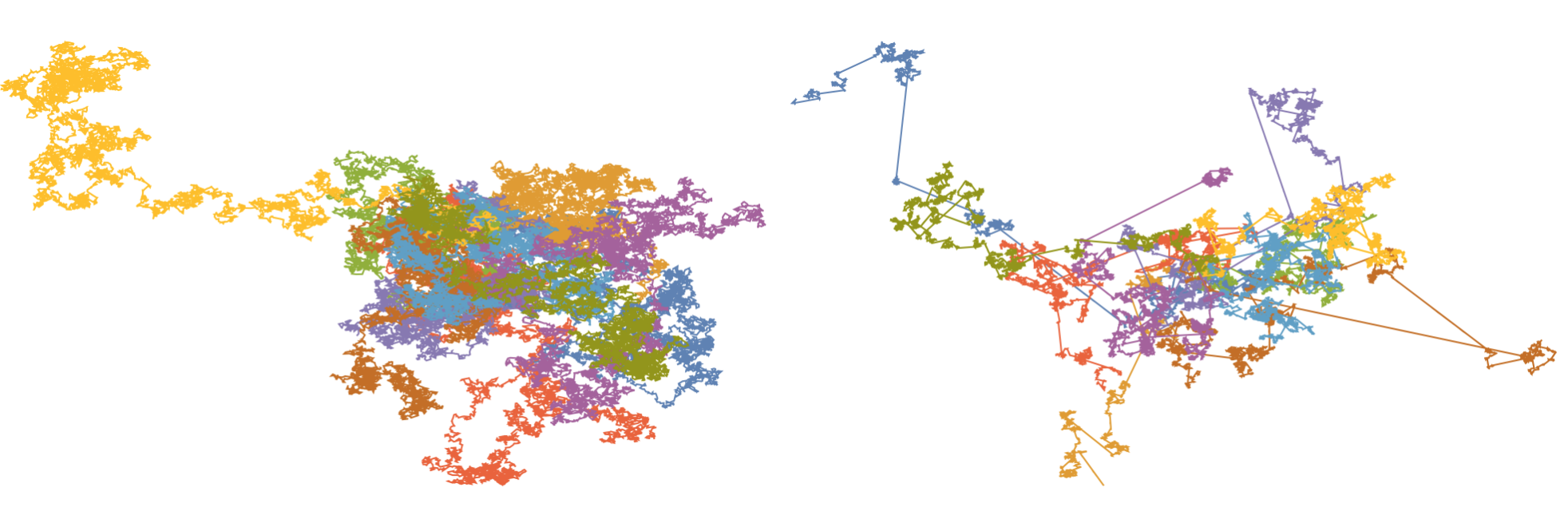}
  \caption{
  The two-dimensional motion of tests particles with their random motion taken from symmetric Stable Distributions.
  At each time step, the angle of the outgoing velocity is chosen randomly with respect to the incident velocity 
  while the magnitude of the velocity is chosen from a symmetric Stable Distribution with $\alpha=2$ 
  corresponding to Brownian motion (left panel), 
  and $\alpha=1.5$ corresponding to a Holtsmark distribution (right panel).
  In the right panel, the large separations between clusters achieved during one time interval correspond to L{\'e}vy flights.
    }
  \label{fig:levyflights}
\end{figure}

The heavy-tailed distributions of $\Delta R_i$ for dynamically correlated time separations correspond to time evolution $\frac{dR_i}{dt}$ 
that is quite  different to that of diffusive Brownian motion  describing the   quantum mechanical motion of free point particles. 
Rather than Brownian motion, heavy-tailed jumps in $R_i(t)$ correspond to a superdiffusive random walk or L{\'e}vy flight. 
Power-law, rather than exponentially suppressed, large jumps give L{\'e}vy flights a qualitatively different character than 
diffusive random walks, including fractal self-similarity, as can be seen in Fig.~\ref{fig:levyflights}.
The dynamical features of QCD that give rise to superdiffusive time evolution are presently unknown, however, 
we conjecture that instantons play a  role.
Instantons are associated with large, localized fluctuations in gauge fields,
and we expect that instantons may also be responsible for infrequent, large fluctuations 
in hadronic correlation functions generating the tails of the $dR_i/dt$ distribution.
It would be interesting to understand if 
 $\alpha\left( \frac{dR_i}{dt} \right)$ can be simply related to observable properties of the nucleon. 
It is also not possible to say from this single study whether $\alpha\left( \frac{dR_i}{dt} \right)$ has a well-defined 
continuum limit for infinitesimal $\Delta t$. 
Further LQCD studies are required to investigate the 
continuum limit of $\alpha\left( \frac{dR_i}{dt} \right)$. 
Lattice field theory studies of other systems and 
calculations of $\alpha\left( \frac{dR_i}{dt} \right)$ in perturbation theory, effective field theory, 
and models of QCD could  provide important insights into the dynamical origin of superdiffusive time evolution.~\footnote{
For example, 
an analysis of  pion correlation functions from the same ensemble of gauge-field configurations
shows that  $R_i$ and ${dR_i\over dt}$ are 
both approximately normally distributed, with $\alpha=1.96(1)$ and 
$\alpha=1.97(1)$,
respectively.
We conclude that the pion shows only small deviations from free particle Brownian motion.
}

One feature of LQCD $\frac{dR_i}{dt}$ results is not well described by a stable distribution. 
The variance of heavy-tailed distributions is ill-defined, and were $\frac{dR_i}{dt}$ truly described by a heavy-tailed 
distribution then the variance and higher cumulants of $\frac{dR_i}{dt}$ would increase without bound as the size of the 
statistical ensemble is increased. 
This behavior is not observed. 
While the distribution of $\frac{dR_i}{dt}$ is well-described by a stable distribution near its peak, the extreme tails of the 
distribution of $\frac{dR_i}{dt}$ decay sufficiently quickly that the variance and higher cumulants of $\frac{dR}{dt}$ 
shown in Fig.~\ref{dRdtCumulants} give statistically consistent results as the statistical ensemble size is varied. 
This suggests that $\frac{dR_i}{dt}$ is better described by a truncated stable distribution, a popular model for,
for example,
financial markets exhibiting high volatility but with a natural cutoff on trading prices, in which some form of sharp cutoff 
is added to the tails of a stable distribution~\cite{Voit:2005}. 
Note that the tails of the $\frac{dR_i}{dt}$ 
distribution describe extremely rapid changes in the correlation function and are sensitive to ultraviolet properties of the theory. 
One possibility is that $\frac{dR_i}{dt}$ describes a stable distribution in the presence of a (perhaps smooth) cutoff arising from ultraviolet regulator effects that damps the stable distribution's power-law decay at very large $\frac{dR_i}{dt}$.
Further studies at different lattice spacings will be 
needed to understand the form of the truncation and whether the truncation scale is indeed set by the lattice scale. 
It is also possible that there is a strong interaction length scale providing a modification to the distribution at large 
$\frac{dR_i}{dt}$,
and it is further possible that stable distributions only provide an approximate description at all $\frac{dR_i}{dt}$.
For now we simply observe that a truncated stable distribution with an unspecified high-scale modification provides a 
good empirical description of $\frac{dR_i}{dt}$.

Before turning to the complex phase of $C_i(t)$, we summarize the main findings about the log-magnitude:
\begin{itemize}
  \item 
  The log-magnitude of the nucleon correlation function in LQCD is approximately normally distributed with small 
  but statistically significant negative skewness and positive kurtosis.
  \item 
  The magnitude effective mass $M_R(t)$ approaches $\frac{3}{2}m_\pi$ at large times, consistent with expectations from Parisi-Lepage scaling for the nucleon variance $|C_i(t)|^2 \sim e^{-3 m_\pi t}$. 
  The plateau of $M(t)$ marks the start of the golden window where excited state systematics are negligible and 
  statistical uncertainties are increasing slowly. 
  The much larger-time plateau of $M_R(t)$ roughly coincides with the plateau of $M_\theta(t)$ to $M_N - \frac{3}{2}m_\pi$ and 
  occurs after variance growth of $M(t)$ reaches the Parisi-Lepage expectation $e^{2(M_N - \frac{3}{2}m_\pi)t}$. 
  Soon after, a noise region begins where the variance of $M(t)$ stops increasing and the effective mass cannot be reliably estimated.
  \item 
  The  log-magnitude does not have a severe StN problem, and $M_R(t)$ can be measured accurately across all 48 timesteps 
  of the present LQCD calculations.  
  The variance of the log-magnitude distribution only increases by a few percent in 20 timesteps after visibly plateauing.
  \item 
  The cumulant expansion describes $M_R(t)$ as a sum of the time derivatives of the cumulants of the log of the correlation function. 
  At large times, the time derivative of the mean of $R_i(t)$ is constant and approximately 
  equal to $M_R(t)$. Contributions to $M_R(t)$ from the variance and higher cumulants of $R_i(t)$ are barely 
  resolved in the sample of $500,000$ correlation functions. 
  \item 
  Finite differences in $R_i(t)$, $\Delta R_i(t, \Delta t)$, are described by time independent distributions at large times. 
  For large $\Delta t$ compared to the QCD correlation length, $\Delta R$ describes a difference of approximately 
  independent normal random variables and is therefore approximately normally distributed. 
  For small $\Delta t$, $\Delta R_i$ describes a difference of dynamically correlated variables. 
  The mean of $\frac{dR_i}{dt}$ is equal to the time derivative of the mean of $R_i(t)$ and therefore provides a good 
  approximation to $M_R(t)$. 
  The time derivatives of higher cumulants of $R_i(t)$ cannot be readily extracted from cumulants of $\frac{dR_i}{dt}$ 
  without knowledge of dynamical correlations.
  \item 
  At large times, $\frac{dR_i}{dt}$ is well described by a symmetric, heavy-tailed, truncated stable distribution. 
  The presence of heavy tails in $\frac{dR_i}{dt}$ indicates that $R_i(t)$ is not described by free particle Brownian motion 
  but rather by a superdiffusive L{\'e}vy flight. 
  Deviations of the index of stability of $\frac{dR_i}{dt}$ from a normal distribution quantify the amount of dynamical correlations 
  present in the nucleon system, the physics of which is yet to be understood. 
  Further studies are required to determine the continuum limit value of the index of stability associated with 
  $\frac{dR_i}{dt}$ and the dynamical origin and generality of superdiffusive L{\'e}vy flights in quantum field theory correlation functions.
\end{itemize}

\subsection{The Phase}
\label{sec:phase}

\begin{figure}[!ht]
  \centering
  \includegraphics[width=\columnwidth]{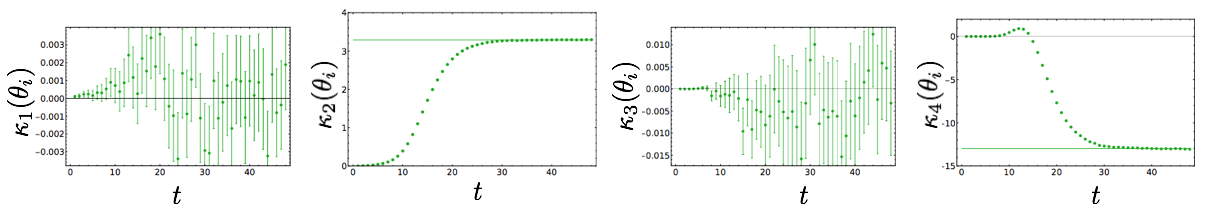}
  \caption{
  The first four cumulants of $\theta_i(t)$. 
  In these fits, no special care is given to the fact that $\theta_i(t)$ is a phase defined on $-\pi<\theta_i(t)\leq \pi$ and standard sample moments 
  are used to determine these cumulants in analogy to Eq.~\eqref{cumulantdef}. 
  Uniform distribution results of $\frac{\pi^2}{3}$ variance and $-\frac{2\pi^4}{15}$ 
  fourth cumulant are shown as green
   lines for reference.
  }
  \label{ThCumulants}
\end{figure}

\begin{figure}[!ht]
  \centering
  \includegraphics[width=\columnwidth]{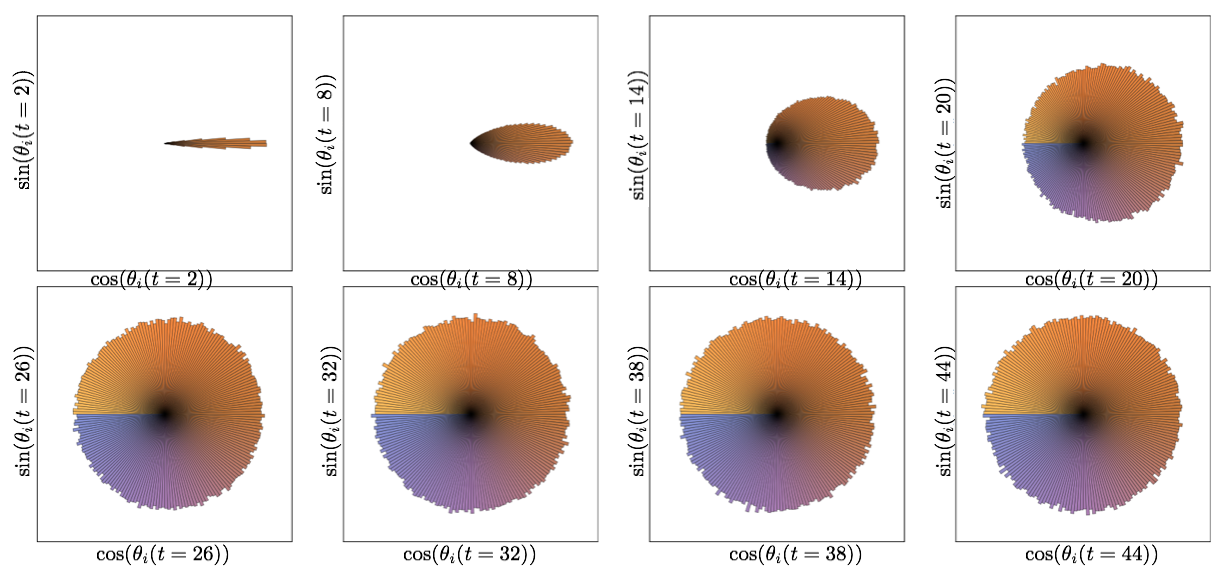}
  \caption{
  Angular histograms of $\theta_i(t)$. 
  The unit circle is split into a uniform sequence of bins, and the number of $\theta_i(t)$ samples falling in 
  each bin sets the radial length of a bar at that angle.
  Colors ranging from orange to blue also denotes angle, and is included to indicate the $\theta_i = \pi$ location of the branch cut in $\theta_i(t) = \text{arg} C_i(t)$.
  }
  \label{ThAngularHistograms}
\end{figure}

The reality of  average correlation functions requires that the distribution of $\theta_i(t)$ be symmetric under 
$\theta_i(t) \rightarrow -\theta_i(t)$. 
Cumulants of $\theta_i(t)$ calculated from sample moments in analogy to Eq.~\eqref{cumulantdef} are shown in Fig.~\ref{ThCumulants}. 
The mean and $\kappa_3$ are noisy but statistically consistent with zero as expected. 
The variance and $\kappa_4$ are small at small times since every sample of $\theta_i(t)$ is defined to vanish at $t=0$, and grow linearly 
at intermediate times $10 < t < 20$ around the golden window. 
After $t=20$, this linear growth slows and they become constant at large times, and are consistent with results from a uniform distribution. 
\begin{figure}[!ht]
  \centering
  \includegraphics[width=\columnwidth]{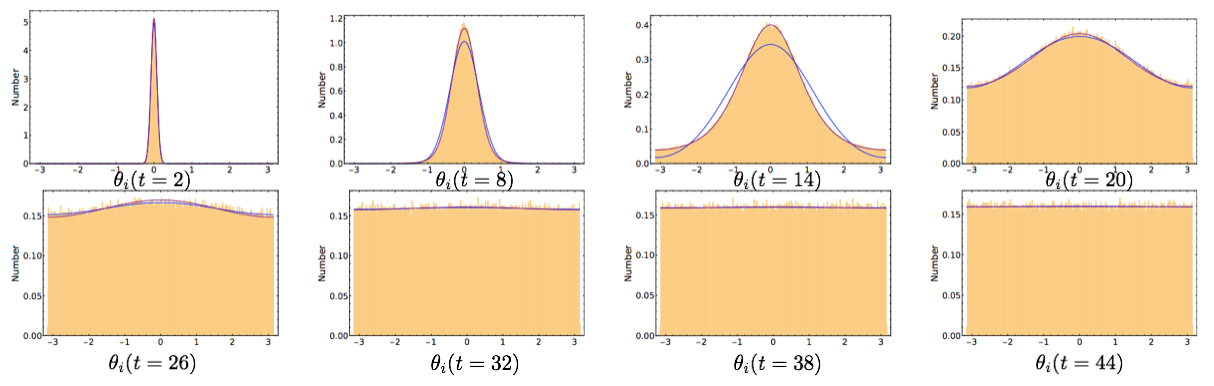}
  \caption{
    Histograms of $\theta_i(t)$ with fits to wrapped normal distributions using Eq.~\eqref{Rdef} shown in blue and fits to wrapped stable distributions using maximum likelihood estimation of the parameters of Eq.~\eqref{PWSdef} shown in purple.
  See the main text for details.
  }
  \label{ThHistograms}
\end{figure}
Histograms of $\theta_i(t)$ shown in Fig.~\ref{ThHistograms} qualitatively suggest that $\theta_i(t)$ is described by a narrow, 
approximately normal distribution at small times and an increasingly broad, approximately uniform distribution at large times. 
$\theta_i(t)$ is only defined modulo $2\pi$ and can be described as a circular variable defined on the interval 
$-\pi < \theta_i \leq \pi$. 
The distribution of $\theta_i(t)$ can therefore be described with angular histograms, as shown in Fig.~\ref{ThAngularHistograms}. 
Again, $\theta_i(t)$ resembles a uniform circular random variable at large times.

A cumulant expansion can be readily constructed for $M_\theta(t)$. 
The mean  phase is given in terms of the characteristic function and cumulants of $\theta_i(t)$ by
\begin{equation}
  \begin{split}
    \avg{e^{i\theta_i(t)}} = \Phi_{\theta(t)}(1) = \exp\left[\sum_{n=0}^\infty \frac{i^n}{n!}\kappa_n(\theta_i(t))\right]
    \ \ \ ,
  \end{split}
  \label{cumulantTh}
\end{equation}
and  the appropriate cumulant expansion for $M_\theta(t)$ is therefore, 
using Eq.~(\ref{EMThdef}),
\begin{equation}
  \begin{split}
    M_\theta(t) = \sum_{n=0}^\infty \frac{i^n}{n!}\left[ \kappa_n(\theta_i(t)) - \kappa_n(\theta_i(t+1)) \right]
    \ \ \  .
  \end{split}
  \label{cumulantThEM}
\end{equation}
Factors of $i^n$ dictate that a linearly increasing variance of $\theta_i(t)$ makes a positive contribution to $M_\theta(t)$, 
in contradistinction to the slight negative contribution to $M_R(t)$ made by linearly increasing variance of $R_i(t)$. 
Since the mean of $\theta_i(t)$ necessarily vanishes, the variance of $\theta_i(t)$ makes the dominant contribution 
to Eq.~\eqref{cumulantThEM} for approximately normally distributed $\theta_i(t)$. 
For this contribution to be positive, the variance of $\theta_i(t)$ must increase, indicating that  
$\theta_i(t)$ has a StN problem. 
For the case of approximately normally distributed  $\theta_i(t)$, non-zero $M_\theta$  requires a StN problem for the  phase.

\begin{figure}[!ht]
  \centering
  \includegraphics[width=\columnwidth]{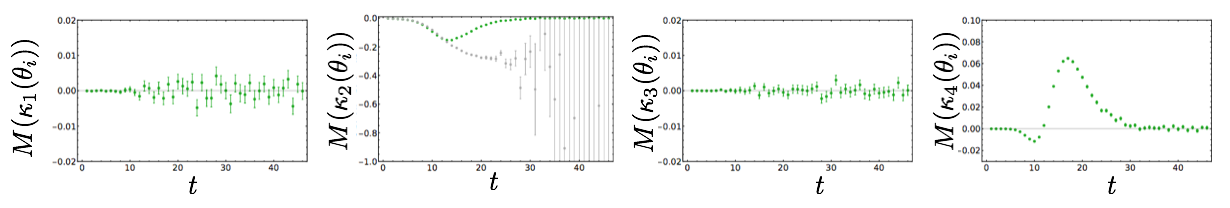}
  \caption{
  Contributions from the first four terms in  the cumulant expansion of Eq.~\eqref{cumulantEM}. 
  The variance, shown second from left, is expected to provide the dominant contribution if a truncation of 
  Eq.~\eqref{cumulantEM} is reliable. 
  Standard estimates of $M_\theta(t)$ from Eq.~\eqref{EMThdef} are shown as the gray points, alongside the cumulant contribution 
  (green points) in the second from left panel. 
  Other panels only show cumulant contributions (green points).
  }
  \label{ThCumulantEM}
\end{figure}

Contributions to Eq.~\eqref{cumulantThEM} from the first four cumulants of $\theta_i(t)$ are shown in Fig.~\ref{ThCumulantEM}. 
Contributions from odd cumulants are consistent with zero, as expected by $\theta_i(t)\rightarrow -\theta_i(t)$ symmetry. 
The variance provides the dominant contribution to $M_\theta(t)$ at small and intermediate times, and is indistinguishable 
from the total $M_\theta(t)$ calculated using the standard effective mass estimator for $t \lesssim 15$. 
Towards to end of the golden window $15 \lesssim t \lesssim 25$, the variance contribution to the effective mass 
begins to decrease. At very large times $t \gtrsim 30$ contributions to $M_\theta(t)$ from the variance are consistent with zero. 
The fourth cumulant makes smaller but statistically significant contributions to $M_\theta(t)$ at intermediate times. 
Contributions from the fourth cumulant also decrease and are consistent with zero at large times. 
The vanishing of these contributions results from the distribution becoming uniform at large times, and time independent as a
consequence.
These observations signal a breakdown in the cumulant expansion at large times $t \gtrsim 25$ where contributions 
from the variance do not approximate standard estimates of $M_\theta(t)$. Notably, the breakdown of the cumulant 
expansion at $t\gtrsim 25$ coincides with plateaus to uniform distribution cumulants in Fig.~\ref{ThCumulants} and 
with the onset of the noise region discussed in Sec.~\ref{sec:decomposition}.

Observations of these unexpected behaviors of $\theta_i(t)$ in the noise region hint at more fundamental issues with the 
statistical description of $\theta_i(t)$ used above. A sufficiently localized probability distribution of a circular random variable 
peaked far from the boundaries of $-\pi <\theta_i(t) \leq \pi$ can be reliably approximated as a standard probability distribution 
of a linear random variable defined on the real line. For broad distributions of a circular variable, the effects of a finite domain 
with periodic boundary conditions cannot be ignored. While circular random variables are not commonly encountered in 
quantum field theory, they arise in many scientific contexts, 
most notably in astronomy, biology, geography, geology, meteorology and oceanography. 
Familiarity with circular statistics is not assumed here, and a few basic results relevant for understanding the statistical 
properties of $\theta_i(t)$ will be reviewed  without proof. Further details can be found in 
Refs.~\cite{Fisher:1995,Mardia:2009,Borradaile:2003} and references therein.

A generic circular random variable $\theta_i$ can be described by two linear random variables $\cos(\theta_i)$ and $\sin(\theta_i)$ 
with support on the line interval $[-1, 1]$ where periodic boundary conditions are not imposed. 
It is the periodic identification of $\theta_i = \pm \pi$ that makes sample moments poor estimators of the distribution of $\theta_i$ and,
 in particular, allows the sample mean of a distribution symmetrically peaked about $\theta_i = \pm \pi$ to be 
 opposite the actual location of peak probability. Parameter estimation for circular distributions can be straightforwardly 
 performed using trigonometric moments of $\cos(\theta_i)$ and $\sin(\theta_i)$. 
 For an ensemble of $N$ random angles $\theta_i$, the first trigonometric moments are defined by the sample averages,
\begin{equation}
  \bar{\cal C} = \frac{1}{N}\sum_i \cos(\theta_i), \hspace{20pt} \bar{\cal S} = \frac{1}{N}\sum_i \sin(\theta_i)
  \ \ \ .
  \label{trigmeandef}
\end{equation}
Higher trigonometric moments can be defined analogously but will not be needed here. 
The average angle can be defined in terms of the mean two-dimensional vector $(\bar{\cal C},\;\bar{\cal S})$ as
\begin{equation}
  \bar{\theta} = \text{arg}\left( \bar{\cal C} + i \bar{\cal S} \right)
  \ \ \  .
  \label{thetabardef}
\end{equation}
A standard measure of a circular distribution's width is given in terms of trigonometric moments as
\begin{equation}
  \bar{\rho}^2 = \bar{\cal C}^2 + \bar{\cal S}^2
  \ \ \ 
  \label{Rdef}
\end{equation}
where $\bar{\rho}$ should be viewed as a measure of the concentration of a circular distribution. 
Smaller $\bar{\rho}$ corresponds to a broader, more uniform distribution, while larger $\bar{\rho}$ corresponds to a more localized distribution.

One way of defining statistical distributions of circular random variables is by ``wrapping'' distributions for linear random 
variables around the unit circle. The probability of a circular random variable equaling some value in $-\pi < \theta \leq \pi$ 
is equal to the sum of the probabilities of the linear random variable equaling any value that is equivalent to $\theta$ modulo $2\pi$. 
Applying this prescription to a normally distributed linear random variable gives the wrapped normal distribution
\begin{equation}
  \begin{split}
  \mathcal{P}_{WN}(\theta_i;\mu,\sigma) &= 
  \frac{1}{\sqrt{2\pi}\sigma}\sum_{k=-\infty}^\infty \exp\left[ -\frac{(\theta_i - \mu + 2\pi k)^2}{2\sigma^2} \right]
  \ =\  \frac{1}{2\pi}\sum_{n=-\infty}^\infty e^{in(\theta_i - \mu) -\sigma^2 n^2/2}
  \ \ \  ,
  \label{WNdef}
\end{split}
\end{equation}
where the second form follows from the Poisson summation formula. 
Wrapped distributions share the same characteristic functions as their unwrapped counterparts, 
and the second expression above can be derived as a discrete Fourier transform of a normal characteristic function. 
The second sum above can also be compactly represented in terms of elliptic-$\vartheta$ functions. 
For $\sigma^2 \lesssim 1$ the wrapped normal distribution qualitatively resembles a normal distribution, 
but for $\sigma^2 \gtrsim 1$ the effects of wrapping obscure the localized peak. 
As $\sigma^2 \rightarrow \infty$, the wrapped normal distribution becomes a uniform distribution on $(-\pi,\pi]$. 
Arbitrary trigonometric moments and therefore the characteristic function of the wrapped normal distribution are given by 
\begin{equation}
   \avg{e^{i n \theta_i}}_{WN} = e^{in\mu - n^2\sigma^2/2}
   \ \ \ \  .
  \label{WNmoments}
\end{equation}
Parameter estimation in fitting a wrapped normal distribution to LQCD results for $\theta_i(t)$ 
can be readily performed by relating $\bar{\theta}$ and $\bar{\rho}$ above to these trigonometric moments as
\begin{equation}
  \mu = \bar{\theta}
  \qquad {\rm and }\qquad
 e^{-\sigma^2} = \bar{\rho}^2
 \ \ \  .
  \label{WNest}
\end{equation}
Note that Eq.~\eqref{WNest} holds only in the limit of infinite statistics. 
Estimates for the average of a wrapped normal distribution 
are consistent with zero at all times, as expected. 
Wrapped normal probability distribution functions with $\sigma^2(\theta_i(t))$ determined from Eq.~\eqref{WNest} 
are shown with the histograms of Fig.~\ref{ThHistograms} and provide a good fit to the data at all times.

The appearance of a uniform distribution at large times is consistent with the heuristic argument that the logarithm of a 
correlation function should be described by a stable distribution. The uniform distribution is a stable distribution for 
circular random variables, and in fact is the only stable circular distribution~\cite{Mardia:2009}. The distribution describing 
a sum of many linear random variables broadens as the number of summands is increased, and the same is true of circular 
random variables. A theorem of Poincar{\'e} proves that as the width of any circular distribution is increased without bound, 
the distribution will approach a uniform distribution. One therefore expects that the sum of many well-localized 
circular random variables might initially tend towards a narrow wrapped normal distribution while boundary effects are negligible. 
Eventually as more terms are added to the sum this wrapped normal distribution will broaden and approach a uniform distribution. 
This intuitive picture appears consistent with the time evolution of $\theta_i(t)$  shown in 
Figs.~\ref{ThHistograms},~\ref{ThAngularHistograms}.

\begin{figure}[!ht]
  \centering
  \includegraphics[width=\columnwidth]{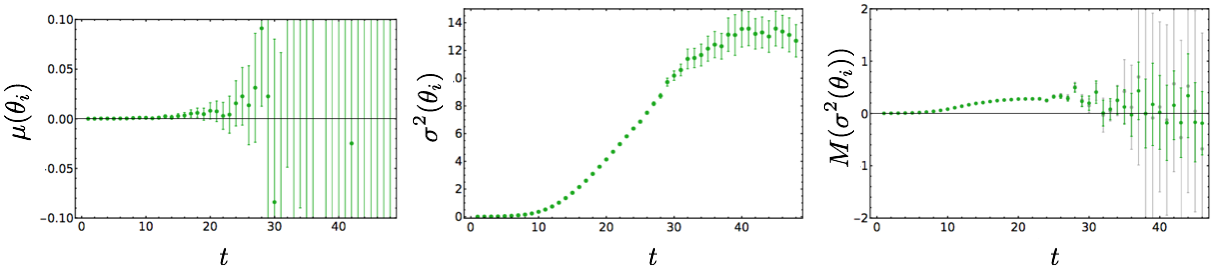}
  \caption{
    The left panel shows estimates of the wrapped normal mean $\mu(\theta_i(t))$ calculated from Eq.~\eqref{WNest} as a function of time.
  The center panel shows analagous estimates of the wrapped normal variance, $\sigma^2(\theta_i(t))$.
  The right panel shows the wrapped normal effective mass, $M_\theta^{WN}(t)$, 
  defined in Eq.~\eqref{mthetaWN}  (green points) 
  along with the standard complex phase effective mass $M_\theta(t)$ 
 defined in Eq.~\eqref{cumulantThEM}  (gray points).
 }
  \label{ThWN}
\end{figure}
The wrapped normal variance estimates for $\theta_i(t)$ that are shown in Fig.~\ref{ThWN} require further discussion. 
At intermediate times, the wrapped normal variance calculated from Eq.~\eqref{WNest} rises linearly with a slope consistent with $M_N - \frac{3}{2}m_\pi$. This is not surprising because assuming an exactly wrapped normal $\theta_i(t)$, $M_\theta(t)$ becomes
\begin{equation}
  \begin{split}
    M_\theta^{WN}(t) = \ln\left[\frac{\avg{e^{i\theta_i(t)}}_{WN}}{\avg{e^{i\theta_i(t+1)}}_{WN}}\right] = -\frac{1}{2}\left[\sigma^2(\theta_i(t)) - \sigma^2(\theta_i(t+1))\right]
    \ \ \  .
  \end{split}
  \label{mthetaWN}
\end{equation}
Eq.~\eqref{mthetaWN} resembles the first non-zero term in the cumulant expansion given in Eq.~\eqref{cumulantThEM} 
adapted for circular random variables. 
Results for $M_\theta^{WN}(t)$ are also shown in Fig.~\ref{ThWN}, where it is seen that $M_\theta^{WN}(t)$ is indistinguishable 
from $M_\theta(t)$ at small and intermediate times. In the noise region, both $M_\theta^{WN}(t)$ and standard estimates for 
$M_\theta(t)$ are consistent with zero. $M_\theta(t)$ has smaller variance than $M_\theta(t)$ in the noise region, 
but this large-time noise is the only visible signal of deviation between the two. This is not surprising, because 
$M_\theta^{WN}(t)$ is actually identical to $M_\theta(t)$ when $\bar{\cal S}(\theta(t)) = 0$. Since $\bar{\cal S}(\theta_i(t))$ 
vanishes in the infinite statistics limit by $\theta_i(t) \rightarrow -\theta_i(t)$ symmetry, $M_\theta^{WN}(t)$ must 
agree with $M_\theta(t)$ up to statistical noise. At large times $t\gtrsim 30$, the wrapped normal variance shown in 
Fig.~\ref{ThWN}  becomes roughly constant up to sizable fluctuations. 
The region where $\sigma^2(\theta_i(t))$ stops increasing coincides with the noise region previously identified.

\begin{figure}[!ht]
  \centering
  \includegraphics[width=.5\columnwidth]{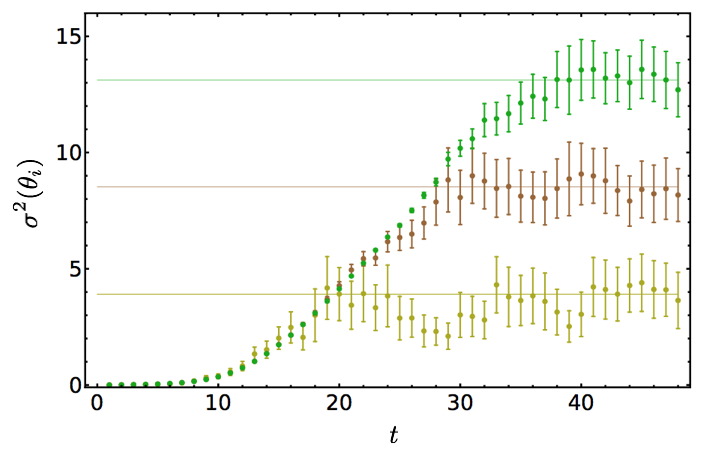}
  \caption{
  Wrapped normal variance of the phase $\sigma^2(\theta_i(t))$ for statistical ensembles of various sizes. 
  Results for an ensemble of $N=50$ nucleon correlation functions are shown in yellow, 
  $N=5,000$ in brown, and $N=500,000$ in green. 
  Lines of each color are also shown at $\sigma^2(\theta_i(t)) = \ln(N)$. 
  Above the relevant line, Eq.~\eqref{Nscaling} is violated for each ensemble and measurements of $\sigma^2(\theta_i(t))$ 
  are expected to be roughly equal to $\ln(N)$ instead of the underlying physical value of $\sigma^2(\theta_i(t))$. 
  Estimates of $\sigma^2(\theta_i(t))$ reaching these lines marks the beginning of the noise region defined by 
  violations of Eq.~\eqref{Nscaling} for each ensemble.}
  \label{ThWNAll}
\end{figure}

The time at which  the noise region begins depends on the size of the statistical ensemble $N$. 
Figure~\ref{ThWNAll} shows estimates of $\sigma^2(\theta_i(t))$ from Eq.~\eqref{WNest} for statistical ensemble sizes $N=50,\;5,000,\;500,000$ varying across four orders of magnitude. 
The time of the onset of the noise region  varies logarithmically as $t \sim 20,\;27,\;35$. 
The constant noise region value of $\sigma^2(\theta_i(t))$ is also seen to vary logarithmically with $N$. 
Equality of $M_\theta^{WN}(t)$ and $M_\theta(t)$ up to statistical fluctuations shows that $M_\theta(t)$ must be 
consistent with zero in the noise region. 
Since corrections to $M(t) \approx M_\theta(t) + M_R(t)$ from magnitude-phase correlations appear small at all times, 
it is reasonable to conclude that standard estimators for the nucleon effective mass are systematically biased in the noise 
region and that exponentially large increases in statistics are required to delay the onset of the noise region.

Besides these empirical observations, the inevitable existence and exponential cost of delaying the noise region 
can be understood from general arguments of circular statistics. 
The expected value of the sample concentration $\bar{\rho}^2$ can be calculated by 
applying Eq.~\eqref{WNmoments} to an ensemble of independent wrapped normal random variables 
$\theta_i$ in Eq.~\eqref{Rdef}. The result shows that $\bar{\rho}^2$ is a biased estimate of $e^{-\sigma^2}$, 
and that the appropriate unbiased estimator is~\cite{Fisher:1995,Mardia:2009}
\begin{equation}
  e^{-\sigma^2} = \frac{N}{N-1}\left( \bar{\rho}^2 - \frac{1}{N} \right).
  \label{Redef}
\end{equation}
For $\bar{\rho}^2 < 1/N$, Eq.~\eqref{Redef} would lead to an imaginary estimate for $\sigma^2$ and therefore 
no reliable unbiased estimate can be extracted. 
A similar calculation shows that the expected variance of $\bar{\rho}^2$ is
\begin{equation}
  \begin{split}
    \text{Var}(\bar{\rho}^2) 
    = 
    \frac{N-1}{N^3}
    \left(1 - e^{-\sigma^2} \right)^2
    \left[\  \left(1 - e^{-\sigma^2} \right)^2 + 2 N e^{-\sigma^2} \ \right]
    \ \ \ .
  \end{split}\label{VarRbar}
\end{equation}
In the limit of an infinitely broad distribution, all circular distributions tend towards uniform and the variance 
of $\bar{\rho}^2$ is set by the $\sigma^2\rightarrow\infty$ limit of Eq.~\eqref{VarRbar} regardless of the form of 
the true underlying distribution. When analyzing any very broad circular distribution, measurements of $\bar{\rho}^2$ 
will therefore include fluctuations on the order of $1/N$. For $e^{-\sigma^2} < 1/N$, the expected error from finite 
sample size effects in statistical inference based on $\bar{\rho}^2$ is therefore larger than the signal to be measured. 
In this regime $\bar{\rho}^2$ has both systematic bias and expected statistical errors that are larger than the 
value $e^{-\sigma^2}$ that $\bar{\rho}^2$ is supposed to estimate. $\bar{\rho}^2$ cannot provide accurate 
estimates of $e^{-\sigma^2}$ in this regime.

Inability to perform statistical inference in the regime $e^{-\sigma^2} < 1/N$ matters for the nucleon correlation 
function because $e^{-\sigma^2(\theta_i(t))} = \bar{\rho}^2(\theta_i(t))= \avg{\cos(\theta_i(t))}^2$ and therefore $e^{-\sigma^2(\theta_i(t))}$ 
decreases exponentially with time. 
At large times there will necessarily be a noise region where $e^{-\sigma^2(\theta_i(t))} < 1/N$ is reached and $\bar{\rho}^2(\theta_i(t))$ 
is not a reliable estimator.
Keeping $e^{-\sigma^2(\theta_i(t))}$ larger than the bias and expected fluctuations of $\bar{\rho}^2(\theta_i(t))$ requires
\begin{equation}
  N > e^{\sigma^2(\theta_i(t))} \sim e^{2\left( M_N - \frac{3}{2}m_\pi \right)t}
  \ \ \ .
  \label{Nscaling}
\end{equation}
Eq.~\eqref{Nscaling} demonstrates that exponential increases in statistics are required to delay the time where statistical uncertainties 
and systematic bias dominate physical results estimated from $\bar{\rho}^2(\theta_i(t))$. 
Formally, the noise region can be defined as the region where Eq.~\eqref{Nscaling} is violated. 
Lines at $\sigma^2(\theta_i(t)) = \ln N$ are shown on Fig.~\ref{ThWNAll} for the ensembles with $N=50,\;5,000,\;500,000$ shown. 
By this definition, the noise region formally begins once $\sigma^2(\theta_i(t))$
 (extrapolated from reliable estimates in the golden window) crosses above the appropriate line. 
 Excellent agreement can be seen between this definition and the above empirical characterizations of the 
 noise region based on constant $\sigma^2(\theta_i(t))$ and unreliable effective mass estimates with constant errors.

Breakdown of statistical inference for sufficiently broad distributions is a general feature of circular distributions. 
Fisher notes that circular distributions are distinct from more familiar linear distributions in that 
``formal statistical analysis cannot proceed'' for sufficiently broad distributions~\cite{Fisher:1995}. 
The arguments above do not rely on the particular form of the wrapped normal model assumed for $\theta_i(t)$, 
and the basic cause for the onset of the noise region for broad $\theta_i(t)$ is that $\bar{\rho}^2$ has an uncertainty of order 
$1/N$ for any broad circular distribution that begins approaching a uniform distribution.~\footnote{
One may wonder whether there is a more optimal estimator than $\bar{\rho}^2$ that could reliably calculate 
the width of broad circular distributions with smaller variance. 
While this possibility cannot be discarded in general, it is interesting to note that it can be in one model. 
The most studied distribution in one-dimensional circular statistics is the von Mises distribution, which has a 
simpler analytic form than the wrapped normal distribution. 
The von Mises distribution is also normally distributed in the limit of a narrow distribution, uniform in the limit of a broad distribution, 
and in general a close approximation but not identical to the wrapped normal distribution. 
Von Mises distributions provide fits of comparable qualitative quality to $\theta_i(t)$ as wrapped normal distributions. 
For the von Mises distribution, $\frac{N}{N-1}\left(\bar{\rho}^2 - \frac{1}{N}\right)$ is an unbiased maximum likelihood 
estimator related to the width. By the Cram{\'e}r-Rao inequality, a lower mean-squared error cannot be achieved 
if $\theta_i(t)$ is von Mises. 
Particularly in the limit of a broad distribution where all circular distributions tend towards uniform, 
it would be very surprising if an estimator could be found that satisfied this bound for the von Mises 
case but could reliably estimate the width of $\theta_i(t)$ in the noise region if a different underlying 
distribution is assumed.
}
Analogs of Eq.~\eqref{Nscaling} can be expected to apply to statistical estimation of the mean of any complex correlation function. 
As long as the asymptotic value of $M_\theta$ is known, Eq.~\eqref{Nscaling} and analogs for other complex correlation 
functions can be used to estimate the required statistical ensemble size necessary to reliably estimate the mean 
correlation function up to a desired time $t$.

\begin{figure}[!ht]
  \centering
  \includegraphics[width=\columnwidth]{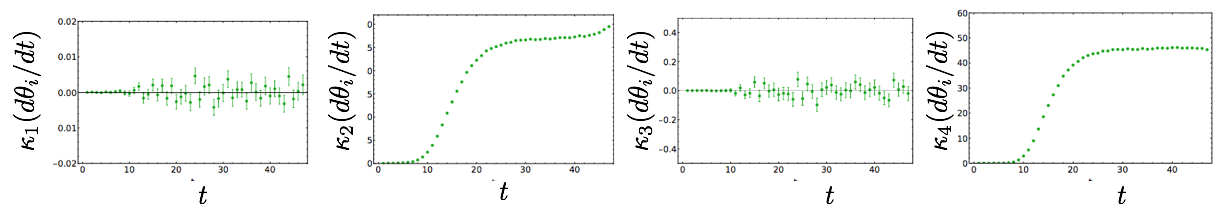}
  \caption{The first four cumulants of $\frac{d\theta_i}{dt}$.}
  \label{dThdtCumulants}
\end{figure}

\begin{figure}[!ht]
  \centering
  \includegraphics[width=\columnwidth]{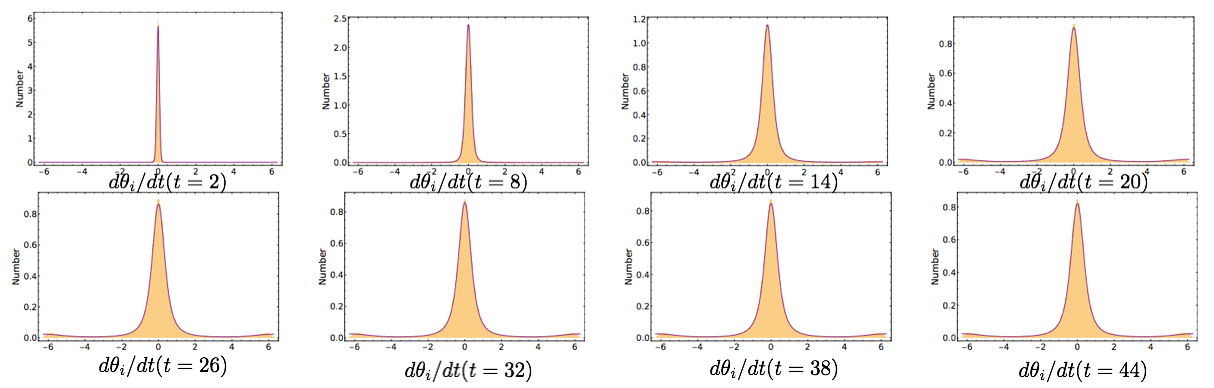}
  \caption{
  Histograms of $\frac{d\theta_i}{dt}$ with fits to  a wrapped stable mixture distribution shown as the purple curves. 
  See the main text for details.
  }
  \label{dThdtHistograms}
\end{figure}

\begin{figure}[!ht]
  \centering
  \includegraphics[width=\columnwidth]{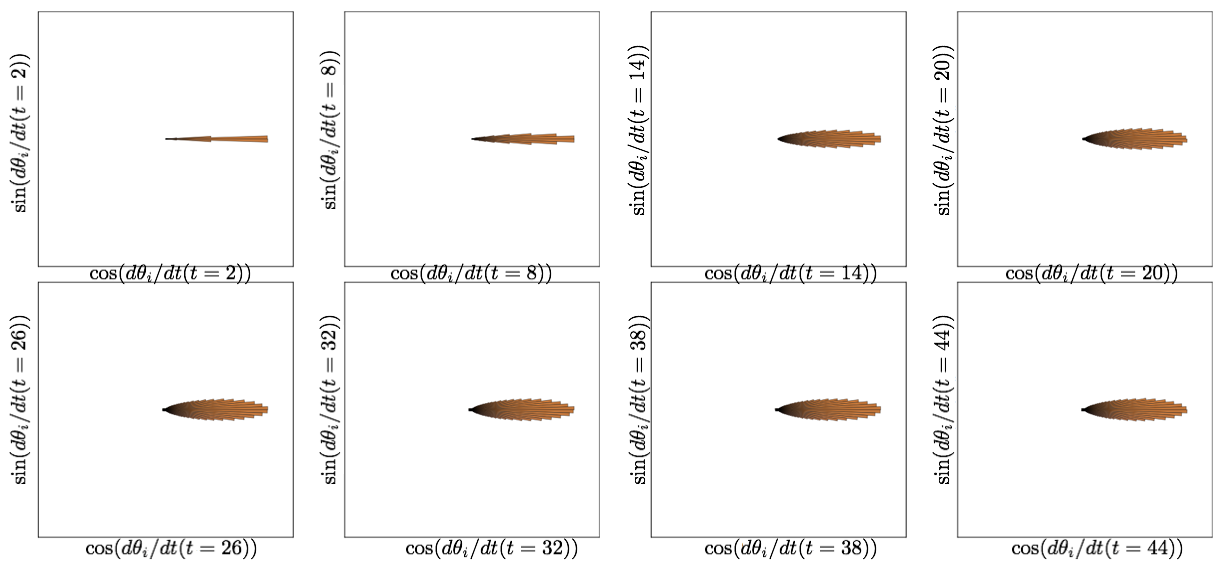}
  \caption{
  Angular histograms of $\frac{d\theta_i}{dt}$. 
  Since $\frac{d\theta_i}{dt}$ is defined on $-2\pi < \frac{d\theta_i}{dt} \leq 2\pi$, 
  normalizations are such that $\frac{1}{2}\frac{d\theta_i}{dt}$ is mapped to the unit circle in analogy to 
  Fig.~\ref{ThAngularHistograms}.}
  \label{dThdtAngularHistograms}
\end{figure}

The pathological features of the large-time distribution of $\theta_i$ are not shared by $\frac{d\theta_i}{dt}$. 
As with the log-magnitude, 
it is useful to 
define general finite differences,
\begin{equation}
  \begin{split}
    \Delta\theta_i(t,\Delta t) = \theta_i(t) - \theta_i(t-\Delta t)
    \ \ \ \  ,
  \end{split}
  \label{DeltaThdef}
\end{equation}
and a discrete (lattice) time derivative,
\begin{equation}
  \begin{split}
    \frac{d\theta_i}{dt} = \Delta\theta_i(t, 1)
    \ \ \ \  .
  \end{split}
  \label{dThdtdef}
\end{equation}
The sample cumulants of $\frac{d\theta_i}{dt}$ are shown in Fig.~\ref{dThdtCumulants}, 
histograms of $\frac{d\theta_i}{dt}$ are shown in Fig.~\ref{dThdtHistograms}, and angular histograms are 
shown in Fig.~\ref{dThdtAngularHistograms}. 
Much like $\frac{dR_i}{dt}$, $\frac{d\theta_i}{dt}$ appears to have a time independent distribution at large times. While $\frac{d\theta_i}{dt}$ is a circular random variable, it's distribution is still well-localized at large times and can be clearly visually distinguished from a uniform distribution. This suggests that statistical inference of $\frac{d\theta_i}{dt}$ should be reliable in the noise region.

\begin{figure}[!ht]
  \centering
  \includegraphics[width=\columnwidth]{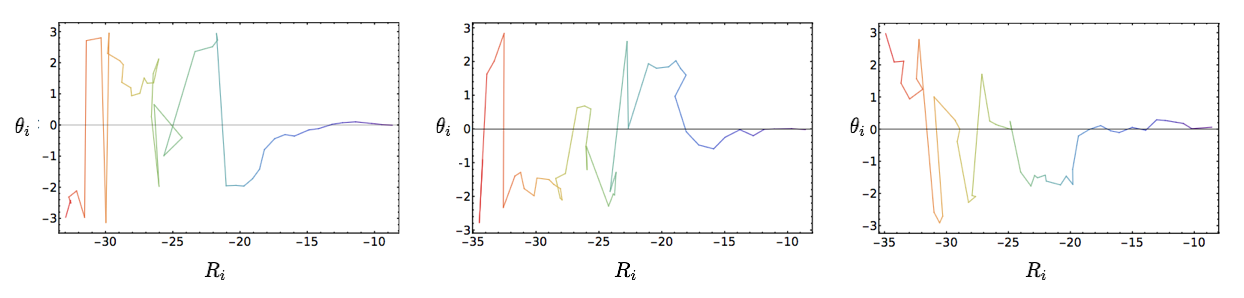}
  \caption{
  Time series showing $R_i(t)$ on the horizontal axis and $\theta_i(t)$ on the vertical axis for three 
  individual nucleon correlation functions,
  where the color of the line shows the time evolution from violet at $t=0$ to red at $t=48$. 
  The evolution of $R_i(t)$ shows a clear drift towards increasingly negative $R_i(t)$. 
  Some large jumps where $\theta_i(t)$ changes by nearly $\pm 2\pi$  
  correspond to crossing the branch cut in $\theta_i(t)$. 
  There are also sizable jumps where $\theta_i(t)$ changes by nearly $\pm\pi$ which likely do not 
 correspond to crossing a branch cut. 
  }
  \label{Tracks}
\end{figure}

Like $\frac{dR_i}{dt}$, $\frac{d\theta_i}{dt}$ shows evidence of heavy tails. 
The time evolution of $R_i(t)$ and $\theta_i(t)$ for three (randomly selected)
correlation functions are shown in Fig.~\ref{Tracks}, 
and exhibit large jumps in both $R_i(t)$ and $\theta_i(t)$ more characteristic of L{\'e}vy flights than Brownian motion,
leading us to consider stable distributions once again.
Wrapped stable distributions can be constructed analogously to wrapped normal distributions as
\begin{equation}
  \begin{split}
    \mathcal{P}_{WS}(\theta_i; \alpha, \beta, \mu, \gamma) &= 
    \sum_{k=-\infty}^\infty \mathcal{P}_S(\theta_i + 2\pi k; \alpha, \beta, \mu, \gamma) \\
    &= \frac{1}{2\pi}\sum_{n=-\infty}^\infty \exp\left( i\mu n - |\gamma n|^\alpha\left[ 1 - i\beta\frac{n}{|n|}\tan(\pi \alpha/2) \right] \right)
    \ \ \   ,
  \end{split}
  \label{PWSdef}
\end{equation}
where, as in Eq.~\eqref{PhiSdef}, 
$\tan(\pi \alpha/2)$ should be replaced by $-\frac{2}{\pi}\ln|n|$ for $\alpha = 1$. 
This wrapped stable distribution is still not appropriate to describe $\frac{d\theta_i}{dt}$ for two reasons. 
First, $\frac{d\theta_i}{dt}$ describes a difference of angles and so is defined on a periodic domain 
$-2\pi < \frac{d\theta_i}{dt} \leq 2\pi$. 
This is trivially accounted for by replacing $2\pi$ by $4\pi$ in Eq.~\eqref{PWSdef}. 
Second, $\theta_i(t)$ is determined from a complex logarithm of $C_i(t)$ with a branch cut placed at $\pm \pi$. 
Whenever $\theta_i(t)$ makes a small jump across this branch cut, $\frac{d\theta_i}{dt}$ will be measured to be around $2\pi$ 
even though the distance traveled by $\theta_i(t)$ along its full Riemann surface is much smaller. 
This behavior results in the small secondary peaks near $\frac{d\theta_i}{dt} = \pm 2\pi$ visible in Fig.~\ref{dThdtHistograms}. 
This can be accommodated by fitting $\frac{d\theta_i}{dt}$ to a mixture of wrapped stable distributions 
peaked at zero and $2\pi$. Since $\theta_i(t)\rightarrow-\theta_i(t)$ symmetry demands that both of these 
distributions are symmetric, a probability distribution able to accommodate all observed features of 
$\frac{d\theta_i}{dt}$ is given by the wrapped stable mixture distribution
\begin{equation}
  \begin{split}
    \tilde{\mathcal{P}}_{WS}(\theta_i; \alpha_1, \alpha_2, \gamma_1, \gamma_2, f) &= 
    \frac{1}{4\pi}\left[ 1 + 2\sum_{n=1}^\infty (1-f)e^{-|\gamma_1 n|^{\alpha_1}}\cos(n\theta_i) 
    + f e^{-|\gamma_2 n|^{\alpha_2}}\cos(n(\theta_i - 2\pi)) \right]
    \ \ \  ,
  \end{split}
  \label{PWSMixdef}
\end{equation}
where $f$ represents the fraction of $\frac{d\theta_i}{dt}$ data in the secondary peaks at $\frac{d\theta_i}{dt} = \pm 2\pi$ 
representing branch cut crossings. Fits of $\frac{d\theta_i}{dt}$ to this wrapped stable mixture model performed with 
maximum likelihood estimation are shown in Fig.~\ref{dThdtHistograms} and are in good qualitative agreement with the LQCD results.

If the widths of the main and secondary peaks in $\frac{d\theta_i}{dt}$  were sufficiently narrow, it would be possible to 
unambiguously associate each $\frac{d\theta_i}{dt}$ measurement with one peak or the other and ``unwrap'' the 
trajectory of $\theta_i(t)$ across its full Riemann surface by adding $\pm 2\pi$ to measured values of $\frac{d\theta_i}{dt}$ 
whenever the branch cut in $\theta_i(t)$ is crossed. 
This should become increasingly feasible as the continuum limit is approached. 
However, the presence of heavy tails in the $\frac{d\theta_i}{dt}$ primary peak prevent unambiguous identification of branch cut 
crossings in the LQCD correlation functions considered here. 
Due to the  power-law decay of the primary peak, there is no clear separation visible between the main and secondary peaks, 
and in particular, points near $\frac{d\theta_i}{dt} = \pm \pi$ cannot be unambiguously identified with one peak or another.

\begin{figure}[!ht]
  \centering
  \includegraphics[width=\columnwidth]{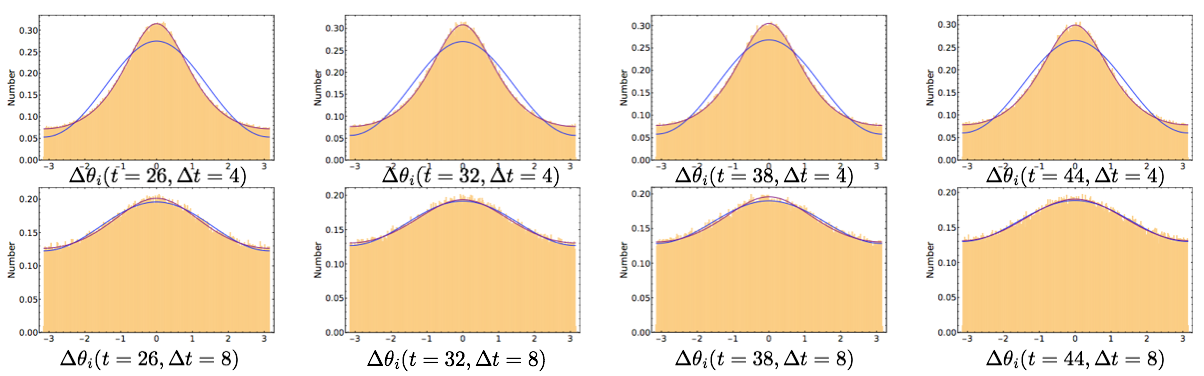}
  \caption{
  Histograms of  $\widetilde{\Delta \theta_i}$ along with fits to wrapped normal distributions in blue and wrapped stable distributions in purple.
  }
  \label{DeltaThHistograms}
\end{figure}

\begin{figure}[!ht]
  \centering
  \includegraphics[width=\columnwidth]{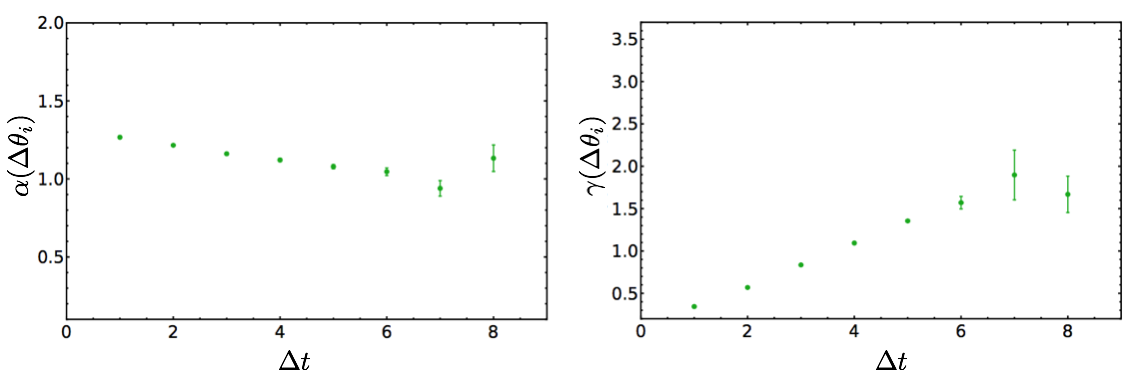}
  \caption{Maximum likelihood estimates for the wrapped stable index of stability $\alpha\left( \widetilde{\Delta \theta_i} \right)$, left, and width $\gamma\left( \widetilde{\Delta \theta_i} \right)$, right extracted from the large-time plateau region as functions of $\Delta t$.
}
  \label{DeltaThStable}
\end{figure}

For descriptive analysis of $\frac{d\theta_i}{dt}$, it is useful to shift the secondary peak to the origin by defining
\begin{equation}
  \begin{split}
    \widetilde{\Delta \theta_i} = \text{Mod}\left( \Delta \theta_i + \pi, 2\pi \right) - \pi
    \ \ \ \ .
  \end{split}
  \label{DeltaThTdef}
\end{equation}
$\widetilde{\Delta \theta_i}$ is well-described by the wrapped stable distribution of Eq.~\eqref{PWSdef}. 
Histograms of the large-time behavior of $\widetilde{\Delta \theta_i}$ are shown in Fig.~\ref{DeltaThHistograms} for $\Delta t = 4,\; 8$ 
and fits of the index of stability of $\widetilde{\Delta \theta_i}$ are shown in Fig.~\ref{DeltaThStable}. 
The large-time distribution of $\widetilde{\Delta \theta_i}(t,\Delta t)$ appears time independent for all $\Delta t$. 
Heavy tails are visible at all times, even as $\Delta t$ becomes large. The large $\Delta t$ behavior visible here is consistent with a wrapped Cauchy distribution.
The estimated index of stability of $\widetilde{\Delta\theta_i}$ differs significantly from that of $\Delta R$, and for $\Delta t = 1$, the large-time behavior is found to have
\begin{equation}
  \begin{split}
    \alpha\left( \widetilde{\Delta\theta_i}(t\rightarrow\infty, \Delta t \sim 0.12\text{ fm}) \right) \rightarrow 1.267(4)(1).
    \ \ \ \  .
  \end{split}
  \label{alpha1}
\end{equation}
This result is consistent with maximum likelihood estimates of $\alpha_1\left( \frac{d\theta_i}{dt} \right)$ in the 
wrapped stable mixture model of Eq.~\eqref{PWSMixdef}. 
$\alpha_2\left( \frac{d\theta_i}{dt} \right)$, associated with the peak shifted from $\theta_i=\pm\pi$
in the wrapped stable mixture model, 
cannot be reliably estimated from the  available LQCD correlation functions. 
The continuum limit index of stability of $\frac{d\theta_i}{dt}$ cannot be determined without additional LQCD studies 
at finer lattice spacings.

As seen in Fig.~\ref{DeltaThStable}, 
the large-time width of $\widetilde{\Delta \theta_i}(t, \Delta t)$ increases with increasing $\Delta t$. 
This behavior is shared by $\Delta \theta_i(t, \Delta t)$. 
In accordance with the observations above that the wrapped normal variance of $\theta_i(t)$ increases linearly with $t$, 
the constant large-time wrapped normal variance of $\Delta\theta_i(t, \Delta t)$ increases linearly with $\Delta t$. 
This is consistent with a pciture of $\Delta\theta_i(t,\Delta t)$ as the sum of $\Delta t$ single time step differences, $\frac{d\theta_i}{dt}$, that make roughly equal contributions to $\Delta \theta_i(t, \Delta t)$.
In accordance with the scaling $\sigma^2(\theta_i(t)) \sim (M_N - \frac{3}{2}m_\pi)t$ discussed previously, 
this linear scaling gives $\sigma^2(\Delta \theta_i(t, \Delta t)) \sim 2(M_N - \frac{3}{2}m_\pi)\Delta t$.

We summarize our observations on the  phase of $C(t)$:
\begin{itemize}
  \item 
  The  phase of the nucleon correlation function is described by an approximately wrapped normal distribution 
  whose width increases with time. At small times the distribution is narrow and resembles a normal distribution. 
  At large times the distribution becomes broad compared to the $2\pi$ range of definition of $\theta_i(t)$ and resembles a uniform distribution. 
  \item 
  The phase effective mass $M_\theta(t)$ appears to plateau to a value close to $M_N - 3/2 m_\pi$. 
  Since $|e^{i\theta_i(t)}|^2 = 1$ is time-independent by construction, this non-zero asymptotic value of 
  $M_\theta$ implies $\theta_i(t)$ has a severe StN problem.
  \item 
  $M_\theta(t)$ can be determined from the time derivative of the wrapped normal variance of $\theta_i(t)$ 
  in analogy to the cumulant expansion. The effective mass extracted from growth of the wrapped 
  normal variance is identical to $M_\theta(t)$ up to statistical fluctuations. 
  This leads to scaling of the wrapped normal variance of $\theta_i(t)$ consistent with 
  $\sigma^2(\theta_i(t)) \sim 2(M_N - \frac{3}{2}m_\pi)t$.
  \item 
  Standard estimators for the wrapped normal variance have a systematic bias and for a sufficiently broad distribution the 
  minimum expected statistical uncertainty is set by finite sample size $1/N$ effects. 
  Once the wrapped normal variance becomes larger than $\ln N$, finite sample size fluctuations 
  become larger than the signal required to extract $M_\theta(t)$. 
  Since the width of $\theta_i(t)$ increases with time, a region where finite sample size errors prevent reliable extractions of $M_\theta(t)$ will inevitably occur at sufficiently large times. This is the noise region empirically identified above. Standard effective mass estimates are systematically biased in the noise region. Exponentially large increases in statistics are necessary to delay the onset of the noise region. 
  \item 
  Finite differences, $\Delta \theta_i(t,\Delta t)$, are described by time-independent distributions at large times. 
  $\Delta \theta_i$ is heavy-tailed for all $\Delta t$ considered here, and  $\frac{d\theta_i}{dt}$ 
  is well-described by a wrapped stable mixture distribution. 
  Further studies will be needed to understand the continuum limit of the  index of stability of $\frac{d\theta_i}{dt}$. 
\end{itemize}

\section{An Improved Estimator}\label{sec:estimator}

The proceeding observations suggest that difficulties in statistical analysis of nucleon correlation functions 
arise from difficulties in statistical inference of $\theta_i(t)$. The same exponentially hard StN and noise region 
problems obstruct large-time estimation of the wrapped normal variance of $\theta_i(t)$ and of $M(t)$. 
Conversely, the width of $\Delta \theta_i(t,\Delta t)$ distributions does not increase with time, and there is 
no StN problem impeding statistical inference of $\Delta \theta_i(t,\Delta t)$. 
This suggests that it would be preferable to construct an effective mass estimator relying on statistical 
inference of $\Delta\theta_i(t,\Delta t)$.

\begin{figure}[!ht]
  \centering
  \includegraphics[width=\columnwidth]{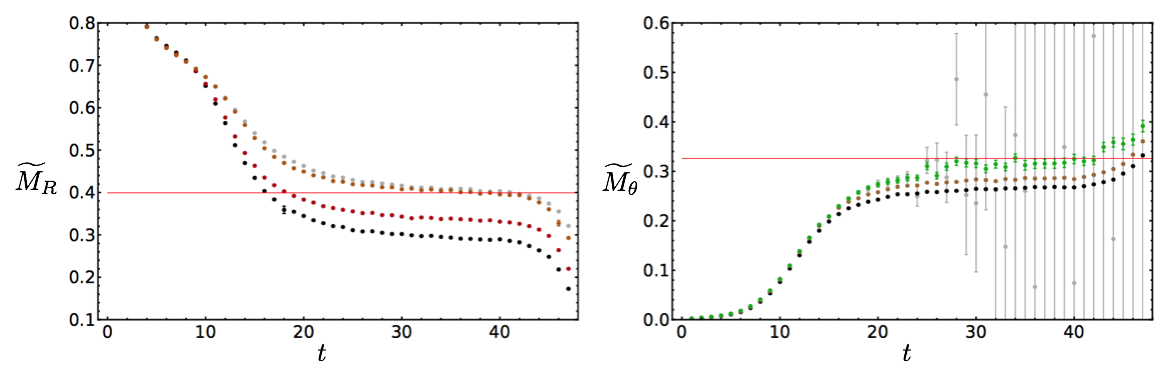}
  \caption{
    Results for the correlation-function-ratio-based estimators $\tilde{M}_R(t, \Delta t)$ and $\tilde{M}_\theta(t, \Delta t)$ with $\Delta t = 1,\;2,\;8$. 
  The left panel shows results for $m_R(t, \Delta t)$ with $\Delta t = 1$ in black, $\Delta t = 2$ in red, and $\Delta t = 8$ in orange. 
  The standard estimator $m_R(t)$ is shown in gray, and a red line is shown for reference at $\frac{3}{2}m_\pi$. 
  The right panel shows results for $m_\theta(t, \Delta t)$ with $\Delta t = 1$ in black, $\Delta t = 2$ in brown, and $\Delta t = 8$ in green. 
  The standard estimator $m_\theta(t)$ is shown in gray and a red line is shown for reference at $M_N - \frac{3}{2}m_\pi$.
  }
  \label{RThWEM}
\end{figure}

First consider the magnitude for simplicity. 
 The mean correlation function magnitude can be expressed in terms of $\Delta R_i$ as
 \begin{equation}
   \begin{split}
     \avg{ e^{R_i(t)} } &= \avg{ \exp\left(R_i(0) + \sum_{t^\prime = 1}^{t} \left.\frac{dR_i}{dt}\right|_{t^\prime} \right)}\\
     &= \avg{ \exp\left(R_i(0) +  \sum_{t^\prime = 1}^{t - \Delta t} \left.\frac{dR_i}{dt}\right|_{t^\prime} \right) \exp\left( \sum_{t^\prime = t- \Delta t + 1}^{t} \left.\frac{dR_i}{dt}\right|_{t^\prime} \right) } \\
     &= \avg{ e^{R_i(0) + \Delta R_i(t - \Delta t, t - \Delta t)}  e^{\Delta R_i(t, \Delta t)}  }
     \ \ \ \   .
   \end{split}
   \label{meanDeltaR}
 \end{equation}
 The last expression above shows that $e^{R_i(t)}$ can be expressed as a product of two factors involving the
 evolution of $R_i(t)$ in the regions $[0, t-\Delta t]$ and $[t-\Delta t, t]$ respectively. 
 Because QCD has a finite correlation length, these two factors should be approximately decorrelated away from the boundary.
 Correlations should only arise from contributions involving points near the boundary at $t- \Delta t$.
 At large times, $t$ can be assumed to be much larger than $\Delta t$ and than any QCD correlation length,
 so boundary effects can be assumed to be negligible for the first region.
 Boundary effects cannot be neglected for the smaller region of length $\Delta t$.
 Treating these boundary effects as a systematic uncertainty allows the correlation function
 to be factorized between the regions $[0, t-\Delta t]$ and $[t-\Delta t, t]$ as
 \begin{equation}
   \begin{split}
     \avg{ e^{R_i(t)} } &= \avg{ e^{R_i(0) + \Delta R_i(t - \Delta t, t - \Delta t)}}  \avg{e^{ \Delta R_i(t, \Delta t) } } 
     \left[ 1   + O\left( e^{-\delta E \Delta t} \right) \right]
     \ \ \ \  .
   \end{split}
   \label{meanDeltaRsplit1}
 \end{equation}
  where $\delta E$ is the smallest energy scale responsible for non-trivial correlations between the factors on the rhs associated with $[0,t-\Delta t]$ and $[t-\Delta t, t]$, and terms suppressed by $e^{-\delta E(t-\Delta t)}$ are neglected.
  If both factors on the rhs of Eq.~\eqref{meanDeltaRsplit1} only receive contributions from the ground state and have single-exponential time evolution,
  then the product of the independently averaged factors on the rhs has the
  same single-exponential behavior as the lhs.
  If excited states make appreciable contributions to either factor on the rhs, then the product of sums of exponentials representing multi-state evolution over $[0,t-\Delta t]$ and $[t-\Delta t, t]$ respectively will not exactly equal the sum of exponentials representing multi-state evolution over $[0,t]$.
  This suggests that $\delta E$ should be set by the gap between the ground state and first excited state with appropriate quantum numbers.\footnote{  It is not proven that the magnitude of a correlation function can be expressed as a sum of exponentials; however,
  the square of the magnitude contributes to the variance correlation function and must have a spectral representation as a sum of exponentials.
  Results of Sec.~\ref{sec:decomposition} demonstrate numerically that the magnitude decays exponentially at large times with a ground-state energy equal to half the ground-state energy of the variance correlation function.
  Eq.~\ref{mRDelta}, which further supposes exponential magnitude excited state contamination, is investigated numerically below, see Fig.~\ref{RThWEM}.}

$e^{R_i(t+1)}$ can similarly be split into an approximately decorrelated product.
Performing this split with regions $[0, t-\Delta t]$ and $[t-\Delta t, t+1]$ gives
\begin{equation}
  \begin{split}
    \avg{ e^{R_i(t+1)} } &= \avg{ e^{R_i(0) + \Delta R_i(t - \Delta t, t - \Delta t)}}  \avg{e^{ \Delta R_i(t + 1, \Delta t + 1) } } 
     \left[ 1   + O\left( e^{-\delta E \Delta t} \right) \right]
     \ \ \ \ .
   \end{split}
  \label{meanDeltaRsplit2}
\end{equation}
The common term in both expressions cancels when constructing the magnitude effective mass, allowing us to define
\begin{equation}
  \begin{split}
    \tilde{M}_R(t, \Delta t) &= \ln\left[ \frac{\avg{e^{\Delta R_i(t, \Delta t)}}}{\avg{e^{\Delta R_i(t + 1, \Delta t + 1 }}} \right]
     = M_R(t) + O\left( e^{- \delta E \Delta t} \right)
     \ \ \ \   .
   \end{split}
  \label{mRDelta}
\end{equation}
Identical steps can be applied to the  phase, leading to
\begin{equation}
   \begin{split}
     \tilde{M}_\theta(t,\Delta t) &= \ln\left[ \frac{\avg{e^{i\Delta \theta_i(t, \Delta t)}}}{\avg{e^{i\Delta \theta_i(t + 1, \Delta t + 1) }}} \right]
     = M_\theta(t) + O\left( e^{- \delta E \Delta t} \right)
    \ \ \ \  .
   \end{split}
   \label{mThDelta}
\end{equation}
The same steps can also be applied to the full correlation function $C_i(t) = e^{R_i(t) + i \theta_i(t)}$.
Noting that
\begin{equation}
   \begin{split}
     e^{\Delta R_i(t,\Delta t) + i \Delta \theta_i(t, \Delta t)} = \frac{C_i(t)}{C_i(t - \Delta t)}
     \ \ \ \   ,
   \end{split}
   \label{decomp}
\end{equation}
the analogous relation for the full effective mass takes the simple form
\begin{equation}
  \begin{split}
     \tilde{M}(t,\Delta t) = \ln\left[ \frac{\avg{C_i(t)/C_i(t - \Delta t)}}{\avg{C_i(t+1)/C_i(t - \Delta t)}} \right]
     = M(t) + O\left( e^{-\delta E \Delta t} \right)
     \ \ \ \  .  
   \end{split}
 \label{mDelta}
\end{equation}
The correlation function ratio effective mass estimator $\tilde{M}(t,\Delta t)$ has different statistical properties than the traditional effective mass $M(t)$ when $\Delta t$ is treated as an independent  $t$.
261 Note that although $\Delta t$ appears in the numerator and denominator of correlation function ratios superficially similarly to $t_J$ in Eq. (6), these two parameters ind    uce quite different statistical behavior.
262 $C_i(t+1)$ in Eq. (47) could be replaced by $C_i(t+t_J)$ (with an appropriate $1/t_J$ overall normalization added).
263 Taking $t_J>1$ increases the time separation between $C_i(t-\Delta t)$ and $C_i(t+t_J)$ in the correlator ratio in the denominator of Eq. (47), resulting in larger statist    ical uncertainties in effective mass results, and will not be pursued further here.

  The approximate factorization leading to Eq.~\eqref{mDelta} can be understood from a quantum field theory viewpoint without reference to the magnitude and phase individually.
  Inserting a complete set of states in a correlation function at $t-\Delta t$ allows the correlation function to be expressed as a sum of exponentials $e^{-E_n \Delta t}$ times prefactors representing the amplitude for the system being in the $n$-th state at time $t-\Delta t$.
  These prefactors for each $e^{-E_n \Delta t}$ term are proportional to $e^{-E_n(t-\Delta t)}$, enhancing the amplitude for finding the system in its ground state at large $t-\Delta t$.
  In this way, the contribution to the correlation function from the region $[0,t-\Delta t]$ can be thought of as an effective source for the correlation function in the region $[t-\Delta t, t]$ whose ground-state overlap is dynamically improved compared to the overlap of the original source at time zero.
  The prefactors for each $e^{-E_n \Delta t}$ will depend on the structure of this effective source, but the exponents are fixed by the QCD spectrum.
  The factor of $C_i(t-\Delta t)^{-1}$ in Eq.~\eqref{mDelta} can be considered to be a modification of the effective source in the region $[0,t-\Delta t]$.
  The presence of $C_i(t-\Delta t)^{-1}$ will modify the prefactor of each $e^{-E_n \Delta t}$ term, but it should not affect time evolution of the system in the region $[t-\Delta t, t]$.
  This suggests that an effective mass designed to extract the ground state energy from the sum of $e^{-E_n \Delta t}$ terms, as in Eq.~\eqref{mDelta}, should provide the exact ground state mass at large $\Delta t$ up to corrections arising from excited state contributions to the $e^{-E_n \Delta t}$ sum.
  These corrections should decrease exponentially with increasing $\Delta t$ at a rate set by the energy gap between the ground and first excited state in the system of interest.
  The size of this energy gap will be set by the lowest-lying excitation consistent with the quantum numbers of the system, a derivatively-coupled pion for the case of the nucleon,\footnote{  Multi-hadron correlation functions contain additional low-lying excitations that may introduce correlation lengths that are larger than $m_\pi^{-1}$ associated for instance with near-threshold bound-states. Such multi-hadron systems are outside the scope of this work.}
  leading to the expectation $\tilde{M}(t, \Delta t) = M(t) + O(e^{-m_\pi \Delta t})$.

  It is not straightforward to construct a representation of $C_i(t-\Delta t)^{-1}$ in terms of local quark and gluon operators that would allow a rigorous proof of these statements, and so numerical LQCD calculations are used to investigate the validity of Eq.~\eqref{mDelta}.
  Exponential reduction of systematic error is numerically demonstrated, but at a faster rate than $m_\pi^{-1}$.
  This suggests that the structure of the effective source plays an important role in determining which $e^{-E_n \Delta t}$ terms are appreciable at the large but finite $\Delta t$ accessible to LQCD calculations in the same way that the structure of the source at time zero determines which excited states make appreciable contributions to the standard effective mass at small $t$.

\begin{figure}[!ht]
  \centering
  \includegraphics[width=.6\columnwidth]{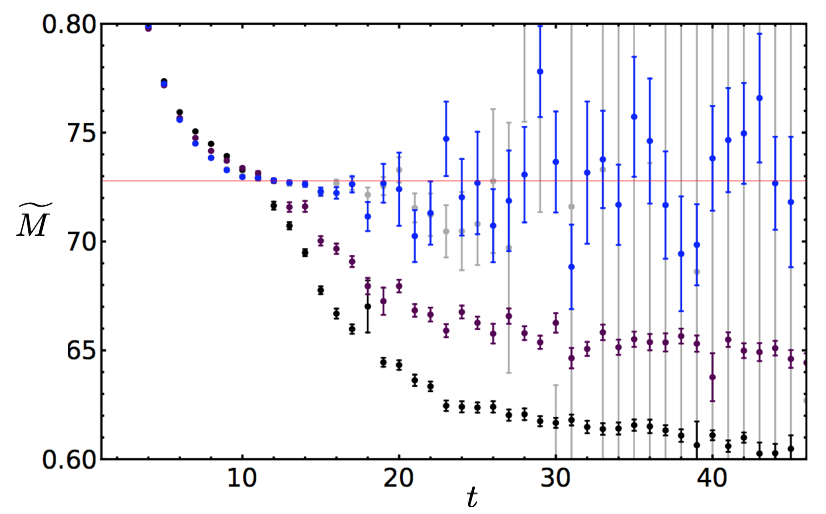}
  \caption{
    Results for the correlation-function-ratio-based estimator $\tilde{M}(t, \Delta t)$. 
  The left panel shows results with $\Delta t =1$ in black, $\Delta t =2$ in purple, and $\Delta t = 8$ in blue, 
  along with the traditional effective mass estimator $M(t)$  shown in gray and a red line at $M_N$ shown for reference. 
  }
  \label{WEM}
\end{figure}
\begin{figure}[!ht]
  \centering
  \includegraphics[width=.6\columnwidth]{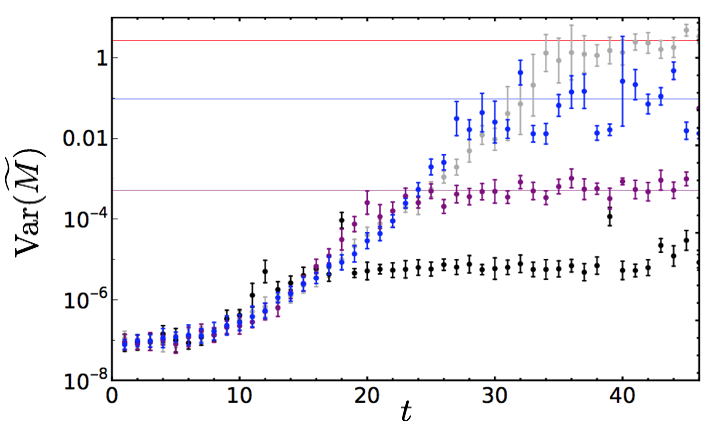}
  \caption{
    Variance in the estimates of $\tilde{M}(t, \Delta t)$ as a function of time $t$ for various choices of $\Delta t$. 
  The black points show $\Delta t = 1$, the purple show $\Delta t = 7$, and the blue show $\Delta t = 15$.
  The gray points show uncertainties in the standard effective mass estimator equivalent to $\Delta t = t$. 
  The purple and blue lines show the expected large-time variance of $\tilde{M}(t, \Delta t)$ with $\Delta t =7,\;15$ 
  predicted by Eq.~\eqref{impscaling} with the overall normalization fixed by the $\Delta t =1$ case. 
  The red line shows the bound of Eq.~\eqref{WEMErrors} with overall normalization again fixed by the $\Delta t =1$ case. 
  Breakdown of statistical inference of broad circular distributions predicts that the large-time variance of $\tilde{M}(t, \Delta t)$ 
  will not systematically rise above the red line for any $\Delta t$.
  }
  \label{WEMErrors}
\end{figure}

The LQCD results for $\tilde{M}_R(t, \Delta t)$ and $\tilde{M}_\theta(t, \Delta t)$ with $\Delta t = 1,\;2\;,8$ are shown in Fig.~\ref{RThWEM}, 
and results for  $\tilde{M}(t,\Delta t)$ are shown in Fig.~\ref{WEM}. 
The statistical uncertainties associated with $\tilde{M}(t, \Delta t)$ are the same as  those of $M(t)$ within the golden window, 
but at large times they become constant in time rather than exponentially increasing. 
This is in accord with our observations about the form of the statistical distributions associated with
$\Delta R_i(t, \Delta t)$ and $\Delta \theta_i(t, \Delta t)$, 
which,  up to small magnitude-phase correlations, indicate that
\begin{equation}
  \begin{split}
    \text{Var}(\tilde{M}(t,\Delta t))  \sim \frac{\text{Var}\left( e^{R_i(t, \Delta t) + i \theta_i(t, \Delta t)} \right)}{\avg{e^{R_i(t, \Delta t) + i \theta_i(t, \Delta t)}}^2} \sim e^{2(M_N - \frac{3}{2}m_\pi)\Delta t}
    \ \ \ \  .
  \end{split}
  \label{impscaling}
\end{equation}
The statistical uncertainties associated with $\tilde{M}(t, \Delta t)$ are constant in $t$, 
although they do increase exponentially with increases in $\Delta t$. 
Since $\Delta \theta_i(t, \Delta t)$ has constant width at large times, 
the inevitable onset of the noise region where statistical inference fails for $\theta_i(t)$ can be avoided. 
The constraint required for reliable statistical inference of $\tilde{M}(t, \Delta t)$ at large times is that the wrapped 
normal variance of $\Delta \theta_i(t, \Delta t)$ can be extracted without large finite sample size errors. 
This constraint can be expressed as a bound on the statistical sample size required for a particular choice of $\Delta t$,
\begin{equation}
  \begin{split}
    N > e^{\sigma^2(\Delta \theta_i(t, \Delta t))} \sim e^{2(M_N - \frac{3}{2}m_\pi)\Delta t}
    \ \ \ \ .
  \end{split}
  \label{WEMstat}
\end{equation}
The statistical uncertainties of $\tilde{M}(t, \Delta t)$ determined from the LQCD correlation functions  are shown in Fig.~\ref{WEMErrors},
from which it can be seen that they become constant at large times for all fixed $\Delta t$. 
For small and moderately large values of $\Delta t = 1,\;7,\;15$, 
the expected exponential increase in large-time statistical uncertainties is observed,
consistent with Eq.~\eqref{WEMstat}. 
Once Eq.~\eqref{WEMstat} is violated, exponential scaling of statistical uncertainties with $\Delta t$ ceases. 
For $\Delta t \lesssim \frac{\ln(N)}{2(M_N - \frac{3}{2}m_\pi)}$, the relative statistical uncertainty in $\tilde{M}(t, \Delta t)$ 
compared to $\tilde{M}(t, \Delta t =1)$ is approximately equal to $N$ rather than $e^{2(M_N - \frac{3}{2}m_\pi)(\Delta t -1)}$.\footnote{These bounds only indicate scaling with $N$. To be made more precise, proportionality constants can be computed using the scaling indicated in Eq.~\eqref{WEMstat}.}
This is seen in Fig.~\ref{WEMErrors} in the large-time behavior of the standard effective mass.

When Eq.~\eqref{WEMstat} is violated, $\Delta \theta_i(t, \Delta t)$ cannot be reliably estimated at large times 
and increasing $\Delta t$ does not improve the accuracy of $\tilde{M}(t, \Delta t)$. 
The standard effective mass estimator can be thought of as evolving with $t \sim \Delta t$, and will become unreliable because of finite sample size effects at large times scaling as $t \gtrsim \ln N / (2(M_N - \frac{3}{2}m_\pi))$.
Similarly, our improved effective mass becomes unreliable for $\Delta t \gtrsim \ln N / (2(M_N - \frac{3}{2}m_\pi))$.
In this extreme case, the bias associated with neglected correlations in $\tilde{M}(t, \Delta t)$ becomes less important than the bias associated with 
statistical inference of overly broad circular random variables. 
In practice, exponential growth of statistical uncertainties with $\Delta t$ suggests that smaller choices of $\Delta t$, 
where Eq.~\eqref{WEMstat} holds, 
likely lead to smaller overall statistical plus systematic uncertainties.

\begin{figure}[!ht]
  \centering
  \includegraphics[width=\columnwidth]{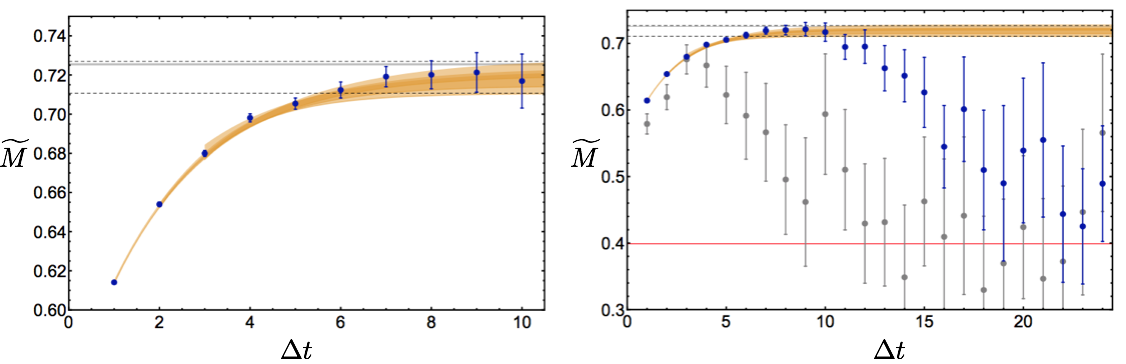}
  \caption{
    In both the left and right panels, results for $\tilde{M}(\Delta t)$ taken from correlated $\chi^2$-minimization fits of $m(t, \Delta t)$ to a constant in the region $25\leq t \leq 40$ with fixed  $\Delta t$ are shown as blue points.
    The tan bands show the results of correlated $\chi^2$-minimization fits of $\tilde{M}(t,\Delta t)$ in various rectangles of $(t,\Delta t)$ to the three-parameter (constant plus exponential) form shown in Eq.~\eqref{mExtrap}.
  The three light-brown bands all use data from $25 \leq t\leq 40$ and then $1\leq \Delta t \leq 10$, $2\leq \Delta t \leq 10$, and $3\leq \Delta t\leq 10$.
  The black dashed lines show the extrapolated prediction for the nucleon mass including statistical errors from the $2\leq \Delta t \leq 10$ fit added in quadrature with a systematic error calculated as half the maximum difference in central values given by the three fits shown.
 The horizontal gray bands show $M_N \pm \delta M_N$ from the precision NPLQCD calculation of Ref.~\cite{Orginos:2015aya}, which used a high-statistics ensemble of correlation functions with optimized sources generated on the same gauge configurations used here.
 The right panel shows a much larger range of $\Delta t$ and also includes results calculated with a smaller ensemble of $N=5,000$ correlation functions as gray points.
 Deviations from the asymptotic prediction due to finite statistics are clearly visible and lead to incorrect results at much smaller $\Delta t$ in the smaller ensemble.
 }
  \label{WEMSysErrors}
\end{figure}

The systematic bias of $\tilde{M}(t, \Delta t)$ can be explored through calculations at various $\Delta t$. 
Fig.~\ref{WEMSysErrors} shows results for with $\Delta t = 1,\dots,9$. 
For $\Delta t \gtrsim 3$, results for $\tilde{M}(t, \Delta t)$ fit during the large-time noise region $25 \leq t \leq 40$ are statistically 
consistent with fits extracted from the golden window $15 \leq t \leq 25$. 
Late-time fits with $\tilde{M}(t, \Delta t)$ have larger statistical uncertainties than golden window fits. 
More precise fits than either could be made by including both the golden window and the noise region in fits of $\tilde{M}(t, \Delta t)$. 
There is only a minor advantage in including the noisier large-time points in fits that include a precise golden window, 
and this exploratory work does not aim for a more precise extraction of the nucleon mass. 
Practical advantages of large-time fits of $\tilde{M}(t, \Delta t)$ compared to golden window fits of $\tilde{M}(t)$ are more likely to 
be found in systems where a reliable golden window cannot be unambiguously identified. 
Large-time fits of $\tilde{M}(t, \Delta t)$ would also be more advantageous for lattices with larger time directions.

Results for a range of $\Delta t$ shown in Fig.~\ref{WEMSysErrors} can also be used to fit the systematic bias in $\tilde{M}(t, \Delta t)$ and formally extrapolate to the unbiased $\Delta t\rightarrow t\rightarrow \infty$ result. 
During the development of a refined version of this improved estimator~\cite{Wagman:2017xfh},
it was realized that the parametric form of the bias can be deduced by 
considering a decomposition of $[0,t]$ into an extended ``source region'' $[0,t-\Delta t]$ involving $C_i(t)$ and $C_i^{-1}(t-\Delta t)$ and an ``evolution region'' $[t-\Delta t, t]$ only involving $C(t)$.
Standard QCD time evolution should apply after the boundary of the source region at $t-\Delta t$, and so at large $\Delta t$ correlation function ratios should scale with $\sim e^{-M_N \Delta t}$ relative to their $t-\Delta t$ boundary values.
Corrections to this ground-state scaling will arise from excited states, which will make contributions to $\left<C_i(t)C_i^{-1}(t-\Delta t)\right>$ scaling as $\sim e^{-(M_N + \delta E)\Delta t}$, where $\delta E$ is the gap between the nucleon ground and first excited state energies.
This allows the dominant contribution to the bias in $\tilde{M}(t,\Delta t)$ to be parametrized as
\begin{equation}
  \begin{split}
    \tilde{M}(t\rightarrow\infty, \Delta t) &= \ln\left[ \frac{e^{-M_N \Delta t}\left( 1 + c\; e^{-\delta E \Delta t} + \dots \right)}{e^{-M_N(\Delta t + 1)}\left( 1 + c\; e^{-\delta E (\Delta t + 1)} + \dots \right)} \right] \\
    & = M_N + c\; \delta E\; e^{-\delta E \Delta t} + \dots,
  \end{split}\label{mExtrap}
\end{equation}
where $c$ is the ratio of excited to ground state overlaps produced by the effective boundary at $t-\Delta t$.
At sufficiently light quark masses and large $\Delta t$, this excited state gap will be set by $m_\pi$.
However, it is noteworthy that Eq.~\eqref{mDelta} involves products of momentum-projected un-averaged correlation functions.
It is familiar from studies of two-baryon correlation functions formed from products of momentum-projected one-baryon blocks that summing over all points in the spatial volume separately for each factor in a product leads to a suppression by $O(m_\pi^{-3}V^{-1})$ in the fraction of points in the product where the nucleons are within one pion Compton wavelength of one another.
It is expected that correlations between $C_i(t)$ and $C_i^{-1}(t-\Delta t)$ described by one-pion excitations will be similarly volume suppressed.
The dominant excited state bias is then expected to arise from excitations that could be produced throughout the lattice volume at the boundary of the source region.
Such excitations are generically far from the nucleon and any other sources of conserved charge, so they should have quantum numbers of the vacuum.
The dominant excited state bias contributing to Eq.~\eqref{mExtrap} is therefore expected to be $e^{-M_\sigma \Delta t}$, where $M_\sigma$ is the mass of the $\sigma$-meson, the lightest excited state with quantum numbers of the vacuum.
Performing a correlated $\chi^2$-minimization three-parameter fit of $\tilde{M}(t, \Delta t)$ to the constant plus exponential form shown in Eq.~\eqref{mExtrap} for noise region data $25 \leq t \leq 40$ gives
\begin{equation}
  \begin{split}
    M_N = 0.7192(49)(42), \hspace{20pt} c = -0.358(26)(17), \hspace{20pt} \delta E = 0.512(65)(73),
  \end{split}\label{mExtrapResults}
\end{equation}
where the first uncertainty is the statistical uncertainty and the second uncertainty is a measure of systematic uncertainty taken from the variation in the central value of the fit as the fitting range in $\Delta t$ is varies. 
The extrapolated result in Eq.~\ref{WEMSysErrors} agrees within uncertainties with the intermediate-time plateau result $M_N = 0.7253(11)(22)$ and with the high-precision GW result $M_N = 0.72546(47)(31)$ of Ref.~\cite{Orginos:2015aya}.
  For the extrapolated large-time result, the total statistical and systematic uncertainties in quadrature is $\delta M_N = 0.0064$, which is larger than the total uncertainty of the plateau region determination $\delta M_N = 0.0025$.
  The large-time plateau considered effectively comprises a two-dimensional region $1 \leq \Delta t \leq 10$ and $25 \leq t \leq 40$ with 150 points.
  The value of the $\chi^2$-minimization fit to this two-dimensional region is most sensitive to points with smaller $\Delta t$ and therefore exponentially smaller uncertainties but is equally sensitive to points with all $t$ that are expected to be approximately decorrelated over intervals $t \gtrsim m_\pi^{-1}$.
  The intermediate-time plateau region $10 \leq t \leq 25$ includes 15 points that are expected to be approximately decorrelated over intervals $t \gtrsim m_\pi^{-1}$.
  The value of the standard effective mass fit is most sensitive to points with smaller $t$ and therefore exponentially smaller uncertainties, though the variance correlation function is not dominated by the three-pion ground state until $t \gtrsim 20$.
  This indicates that results from the intermediate-time plateau have smaller point-by-point uncertainties than points from the large-time noise region.
  The total uncertainty of the noise region result could be reduced by increasing the length of the lattice time direction, while the length of the smaller-time plateau available to standard estimators is restricted by the StN problem.
  The proof-of-principle calculation presented here demonstrates that accurate results can be extracted from the noise region.
  In remains to be seen in future calculations of single- and multi-baryon systems optimized for large-time analysis whether the methods introduced in this work can be used to achieve significantly higher precision with the same resource budget as calculations optimized for smaller-time analysis.

\begin{figure}[!ht]
  \centering
  \includegraphics[width=\columnwidth]{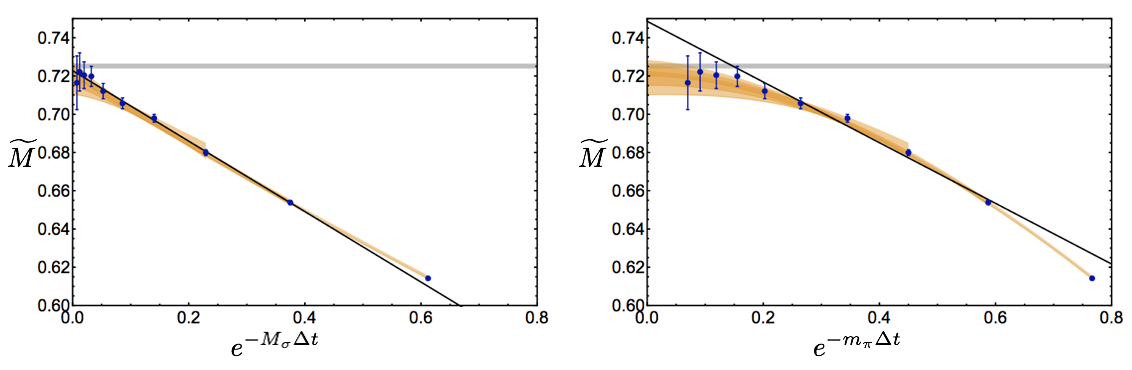}
  \caption{
    The blue points and light-brown bands show the same $\chi^2$-minimization fit results to large-time $\tilde{M}(t,\Delta t)$ plateaus as Fig.~\ref{WEMSysErrors}.
    The horizontal axis has been rescaled to coordinates that would show a linear bias for excited state contributions from $\sigma$-mesons, left, and pions, right.
    Black lines show the central values of $\chi^2$-minimization fits to constrained versions of Eq.~\eqref{mExtrap} where $\delta E$ is fixed to be $M_\sigma$, left, or $m_\pi$, right.
    The horizontal gray bands correspond to $M_N \pm \delta M_N$ from the high-precision NPLQCD calculation of Ref.~\cite{Orginos:2015aya}.
 }
  \label{WEMSysErrors2}
\end{figure}

  The best-fit excitation scale $\delta E = 866(110)(124)$ MeV in Eq.~\ref{WEMSysErrors} can be compared with the $\sigma$-meson mass extracted from mesonic sector calculations to test the heuristic arguments above that lighter excitations will make volume-suppressed contributions.
  Calculations of the $\sigma$-meson face a severe StN problem, 
  particularly at light quark masses where the $\sigma$-meson describes a broad $\pi\pi$ isoscalar resonance rather than a compact QCD bound state,
  but a recent calculation by the Hadron Spectrum collaboration has precisely determined $M_\sigma = 758(4)$ MeV at $m_\pi \sim 391$ MeV where the $\sigma$-meson is weakly bound~\cite{Briceno:2016mjc}.
  Similarly precise results at slightly higher quark masses are not available for interpolation to $m_\pi \sim 450$ MeV, but a crude extrapolation can be made using the Hadron Spectrum result and the (real part of the) physical position of the $\sigma$-meson pole obtained from dispersive analysis of experimental data: $M_\sigma = 457(14)$ MeV~\cite{Caprini:2005zr,GarciaMartin:2011jx}.
  An extrapolation linear in the pion mass gives $M_\sigma \sim 830$ MeV at $m_\pi \sim 450$ MeV, in rough agreement with the best-fit excitation scale determined above.
  This agreement is insensitive to the form of the extrapolation used, as the Hadron Spectrum $\sigma$-meson mass result at $m_\pi \sim 391$ MeV is itself less that one standard deviation smaller than the best-fit nucleon excitation scale.
  Fits where $\delta E = M_\sigma$ is explicitly assumed can be performed more precisely and lead to consistent results with smaller uncertainties for the nucleon mass $M_N = 0.7226(18)$, as shown in Fig.~\ref{WEMSysErrors2}.
  These fits provide another consistency check on $\delta E$ but do not appropriately capture the systematic uncertainties of explicit assumptions about the excited state spectrum.

The improved estimator proposed here exploits physical locality and finite correlation lengths to extract the effective mass
 from the evolution of $C_i(t)$ between times $t - \Delta t$ and $t$ rather than the full evolution between source time $t=0$ and sink time $t$. 
 The correlation function at time $t - \Delta t$ is effectively treated as a new source so that the effective source/sink separation is fixed 
 to be a constant length $\Delta t$ rather than an increasing separation $t$. 
 The effective source at $t - \Delta t$ still incorporates the dynamical evolution of the system between time 0 and $t - \Delta t$, 
 and in particular has exponentially reduced excited state contamination compared to the original source. 
 In principle $t$ can be taken arbitrarily large with $\Delta t$ fixed in order to extract a plateau in $\tilde{M}(t, \Delta t)$ with 
 arbitrarily small excited state contamination and constant statistical uncertainties across the plateau. 
 The length of the lattice time direction becomes the only factor limiting the length of the plateau in this case.

Similar physical ideas underlie the hierarchical integration approach of Ref.~\cite{Luscher:2001up}. 
In that approach, locality is exploited to decompose correlation functions into products of factors that can be computed on 
subsets of a lattice volume with exponentially reduced StN problems. 
Hierarchical integration has been successfully implemented in studies of gluonic observables~\cite{Meyer:2002cd,DellaMorte:2007zz,DellaMorte:2008jd,DellaMorte:2010yp,Vera:2016xpp} and recently explored for baryon correlation functions in the 
quenched approximation~\cite{Ce:2016idq} and beyond~\cite{Ce:2016ajy}. 
For baryon correlators, the method of Ref.~\cite{Ce:2016idq} implements approximate factorization with systematically 
reducible uncertainties, as in the method proposed here. 
The benefits of the two methods are distinct.
  Hierarchical integration also employs standard statistical estimators for observables defined on sub-volumes to determine correlation functions at large $t$ with exponentially slower StN degradation.
  The new estimators introduced here allow data to be extracted from large-$t$ correlation functions with constant StN, but removing all systematic uncertainties requires an extrapolation to large $\Delta t$ with exponential StN degradation of the same severity as the original correlation function.
  Investigations of the compatibility of and relations between these methods are left to future work.
In addition, this method also has similarities to the generalized pencil-of-functions method introduced to LQCD in 
Ref.~\cite{Aubin:2010jc}, where correlation functions involving shifted source and sink times are combined in a variational basis.
 In the generalized pencil-of-functions approach, shifted source and sink times have primarily been investigated to reduce 
 excited-state contamination rather than StN improvement.

In some sense, $\Delta t$ can be considered a ``factorization'' scale in the time direction.
The LQCD calculations are valid for all energy scales below that defined by the inverse lattice spacing, $\pi/a$.
While well-defined, the MC sampling of the path integral and analysis of baryon correlation functions
 fails to converge in the noise region because of the 
quantum fluctuations encountered along the paths from the source to large times, which include many incoherent hadronic volumes.
The new estimator provides exponentially-improved signal extraction at large times through limiting the number of 
contributing hadronic volumes to those within $\Delta t$, 
but  does not provide a complete description of the IR behavior of QCD, introducing a bias in the extracted mass of the nucleon.
An extrapolation in $\Delta t$, using a form motivated by low-energy pion physics, is used to remove this bias.
While  different, this reminds one of matching LQCD calculations to the p-regime of chiral perturbation theory to 
remove finite-spatial-volume effects.
The idea of performing an extrapolation to overcome a sign problem is not new.
It was introduced thirty years ago to deal with the sign problem in MC calculations of modest size nuclei~\cite{Alhassid:1993yd},
and recently used in lattice effective field theory calculations to continuously evolve between the eigenvalues of nuclear many-body systems
described by a Hamiltonian without a sign problem to one that does have a sign problem~\cite{Lahde:2015ona}.

It is not expected that the statistical properties of $\theta_i(t)$ discussed here and, 
in particular the constant large-time width of $\frac{d\theta_i}{dt}$, are unique to single-nucleon correlation functions. 
If analogous statistical properties apply to generic complex correlation functions in 
quantum field theory, then estimators analogous to Eq.~\eqref{mDelta} can be constructed to extract the spectra 
of complex correlation functions and reweighted complex actions without StN problems. 
It remains to be seen if the approaches developed in this work can be fruitfully applied to other systems in 
particle, nuclear, and condensed matter physics that encounter sign and StN problems.

It is remarkable that the Euclidean-time derivate of the logarithm of the correlation function is described by a heavy-tailed 
distribution while the logarithm itself is nearly normally distributed at all times. 
Further studies will be needed to understand the dynamical origin, continuum limit behavior, and universality of 
heavy-tailed Euclidean-time evolution of correlation functions in quantum field theory. 
LQCD calculations at finer lattice spacings are needed to explore the continuum limit of the index of stability 
describing time evolution of the nucleon correlation function. 
Perturbative QCD and model calculations will provide useful insights into the dynamical origin of heavy-tailed time 
evolution of the nucleon correlation function. 
Lattice and continuum studies of other quantum field theories are required to understand the universality of 
heavy-tailed Euclidean-time evolution of correlation functions. Implications for real-time evolution are also left for future investigations.

The properties of the new estimator discussed above may prove practically advantageous in the analysis of LQCD 
calculations of nuclei.
Other types of LQCD calculations may also benefit from the new estimator, for instance in the isoscalar meson 
sector and those at non-zero baryon chemical potential.
Studies of the vacuum channel including glueballs and scalar mesons and analyses of disconnected diagrams provide additional directions for further studies.
Forming ratios of position space, rather than momentum space, correlation functions may be advantageous in future studies.
Preliminary invesitagations in these directions are presented in the next chapters.


\chapter{Phase Reweighting}\label{chap:PR}
 


The signal-to-noise problem leads to exponentially degrading precision in LQCD calculations of multi-baryon systems
that is even more severe than in the single-baryon calculations described above~\cite{Beane:2010em}.
Many of the interesting statistical features~\cite{Parisi:1983ae,Hamber:1983vu,Lepage:1989hd,Beane:2009kya,Beane:2009gs,Beane:2010em,Endres:2011jm,Endres:2011er,Endres:2011mm,Lee:2011sm,DeGrand:2012ik,Grabowska:2012ik,Nicholson:2012xt,Beane:2014oea} of single-nucleon correlator functions are shared by multi-nucleon correlation functions.
In particular the logarithms of generic LQCD correlation functions we have studied exhibit characteristics of L{\'e}vy Flights associated with 
heavy-tailed Stable Distributions.
Generic multi-baryon correlation functions appear to be well-described by an uncorrelated product of a log-normally distributed magnitude and a wrapped normally distributed phase factor.
The average magnitude of a nuclear correlation function is proportional to $\sim e^{-B \frac{3}{2}m_\pi}$, while the average phase factor is proportional to $\sim e^{-B(m_N-3 m_\pi /2)}$.
By the same logic of Sec.~\ref{sec:decomposition}, the StN problem in multi-baryon correlation functions can be identified as arising from reweighting a sign problem.
The baryon number sign problem is spacetime extensive in every form it arises, see Sec.~\ref{sec:corr}, and so can be mitigated by restricting the time interval, $\Delta t$, over which the system contains non-zero particle number prior to measurement.
This restriction, used to construct the improved estimator of Sec.~\ref{sec:estimator}, neglects correlations across distances larger than $\Delta t$
and creates a bias in ground-state energies that decreases exponentially with increasing $\Delta t$.

This chapter introduces a refined estimator for the analysis of LQCD correlation functions that
exploits these ideas and
permits the extraction of 
ground-state energies from the noise region.
Through phase reweighting, this  estimator provides an exponential improvement in the StN ratio, but it also introduces a bias that must be 
systematically removed through extrapolation.  
This technique is similar to that used in Green's Function Monte Carlo (GFMC) methods  applied to  
nuclear many-body systems where the phase 
of the wavefunction is held fixed until the system is close to its ground state, at which 
point the phase is released for final evolution~\cite{Zhang:1995zz,Zhang:1996us,Wiringa:2000gb,Carlson:2014vla}. 
Similar techniques are also used in Lattice Effective Field Theory (LEFT) calculations in which a Wigner-symmetric Hamiltonian, 
emerging from the large-N$_c$ limit of QCD~\cite{Kaplan:1995yg}, 
is used for initial time evolution before asymmetric perturbations  are added that introduce a sign problem~\cite{Lahde:2015ona}.
Phase reweighting shares physical similarities, and possibly formal connections, 
to the approximate factorization of domain-decomposed quark propagators recently suggested and explored by 
C$\grave{e}$, Giusti and Schaefer~\cite{Ce:2016idq,Ce:2016ajy,Ce:2016qto}.

\begin{figure}[!t]
	\includegraphics[width=0.6 \columnwidth]{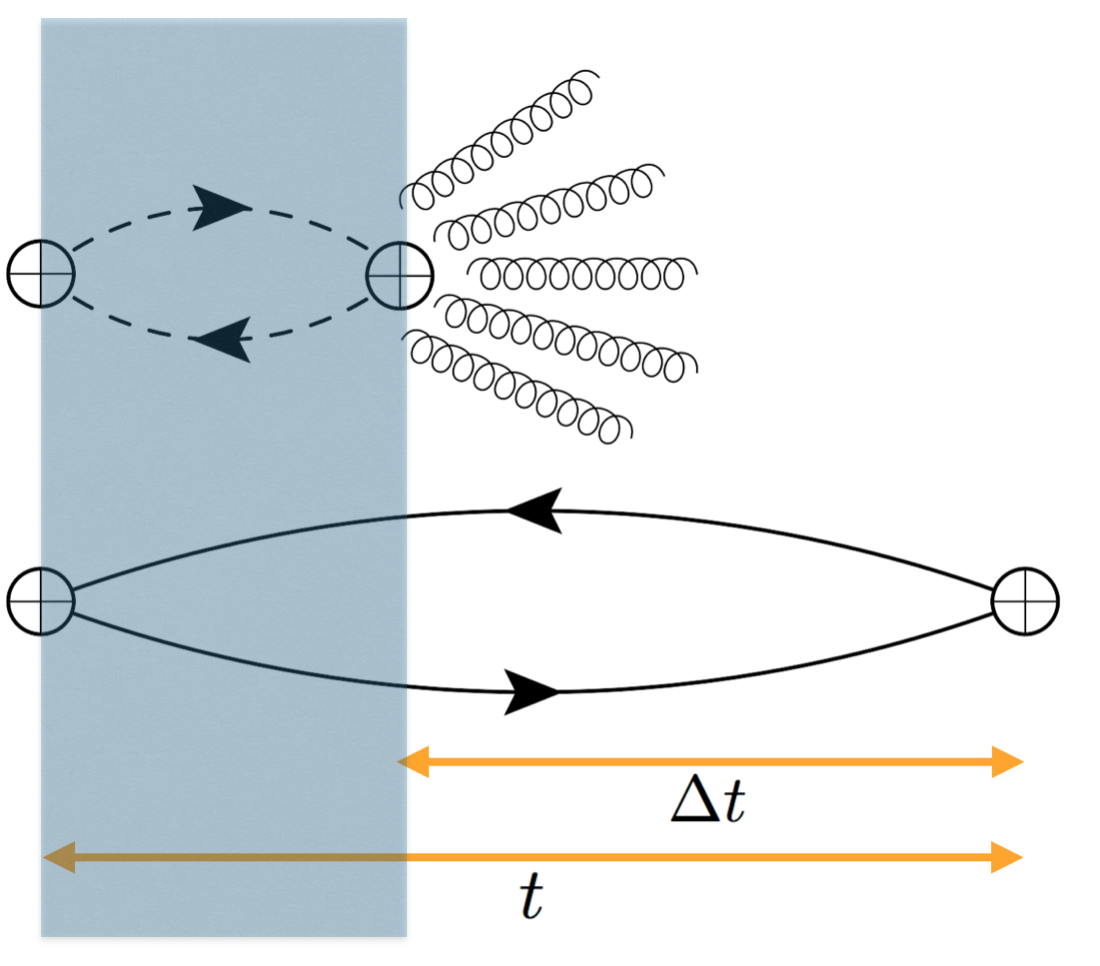}
        \centering
		\caption{
	\label{fig:PRWcorrs} 
  The $\rho^+$-meson phase-reweighted correlation function $G^\theta_\rho(t,\Delta t)$ is a product
  of quark propagators forming $C_i^\rho(t)$, shown as solid lines,
  and a phase factor $e^{-i\theta_i^\rho(t-\Delta t)}$, shown as dashed propagator lines with reversed quark-charge arrows.
        Gluon lines indicate that phase reweighting introduces correlations
        associated with excitations produced at $t-\Delta t$
        and lead to bias when $\Delta t \neq t$.
        For momentum-projected correlation functions, 
        excitations involving correlated interactions between 
        $C_i^\rho(t)$ and $e^{-i\theta_i^\rho(t-\Delta t)}$ are suppressed by the spatial volume.
        $G^\theta_\rho(t,\Delta t)$ effectively includes a non-local source
        whose magnitude is dynamically refined for $t - \Delta t$ steps
        while the phase is held fixed (shaded region)
        before the full system is evolved for the last $\Delta t$ steps of propagation.
	}		
\end{figure}
%


The  phase-reweighted correlation function is defined by
\begin{eqnarray}
G^{\theta}(t,\Delta t) & = & 
\langle e^{-i \theta_i (t-\Delta t)} \ C_i(t)  \rangle
\  \ ,
\label{eq:PRWdef}
\end{eqnarray}
where $\theta_i (t-\Delta t) = \text{arg}[ C_i(t-\Delta t) ]$.
Phase reweighting resembles limiting the approximate L{\'e}vy Flight of the correlation function phase 
to $\Delta t$ steps at large times, suggesting that $G^\theta(t,\Delta t)$ has a StN ratio that decreases exponentially with 
$\Delta t$ but is constant in $t$.
In the limit that $\Delta t \rightarrow t$, the reweighting factor approaches unity and $G^{\theta}(t,t) = G(t)$.
The exact correspondence $G^\theta(t,t)=G(t)$ 
gives phase reweighting
an advantage over our previously suggested estimator~\cite{Wagman:2016bam}  involving 
multiplication by $C_i^{-1}(t-\Delta t)$ rather than $e^{-i\theta_i(t-\Delta t)}$.
Phase reweighting also leads to more precise ground-state energy extractions than estimators involving reweighting with $C_i^{-1}(t-\Delta t)$;
multiplication by the heavy-tailed variable $|C_i(t-\Delta t)^{-1}|$ leads to increased variance.

Dynamical correlations between $C_i(t)$ and $e^{-i\theta_i(t-\Delta t)}$
lead to differences in  ground-state energies extracted from $G^\theta(t,\Delta t)$ and $G(t)$ for $t\neq \Delta t$. 
Locality suggests that these correlations should decrease exponentially with increasing $\Delta t$
at a rate controlled by the longest correlation length in the theory.
At asymptotically large $\Delta t$,
one-pion-exchange correlations are expected to provide the largest contributions to the bias.
These contributions will be
suppressed by factors involving the spatial volume in products of a momentum-projected correlation function with a momentum-projected phase factor.
Excitations involving the $\sigma$ meson, correlated two-pion exchange, and other light
excitations that do not change the quantum numbers of the system
are not volume-suppressed and
may dominate at small $\Delta t$.
Near-threshold bound states may have complicated small $\Delta t$ bias that is sensitive to the size of the spatial volume.

\begin{figure}[!t]
  \centering
	\includegraphics[width=0.7 \columnwidth]{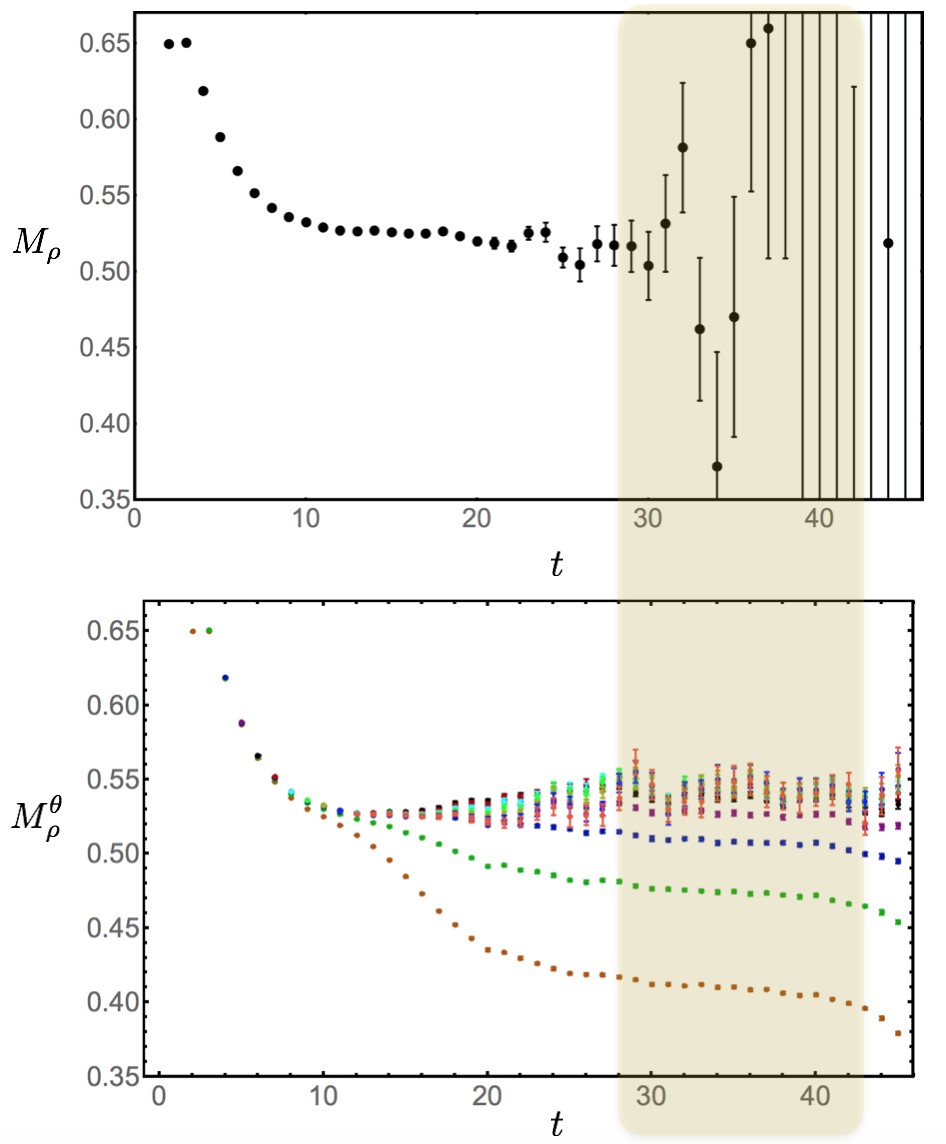}
		\caption{
	\label{fig:emps} 
  The upper panel shows the $\rho^+$ effective mass from the LQCD ensemble of Ref.~\cite{Orginos:2015aya}.
	The lower panel shows $M_\rho^\theta(t,\Delta t)$ with a range of fixed $\Delta t$'s.
        Temporal structure at larger times arises from proximity to the midpoint of the lattice at $t=48$.
        The highlighted interval $t=28\rightarrow 43$ is used for correlated $\chi^2$ minimization fits of $M_\rho^\theta$.
	Masses and times are given in lattice units.
	}		
\end{figure}

The construction of $G^\theta$ is generic for any correlation function,
and is schematically depicted for the $\rho^+$ meson in Fig.~\ref{fig:PRWcorrs}.
In the plateau region of the  $\rho^+$ correlation function, the average of the  magnitude is approximately proportional to $e^{-M_\pi t}$, 
while the average of the phase factor\footnote{
The phases of isovector meson correlation functions are restricted to be discrete values $\theta_\rho = 0,\ \pi$ 
when interpolating operators in a Cartesian spin basis are used.  In forthcoming work, we demonstrate that circular statistics 
applies to real but non-positive isovector meson correlation functions.
} is approximately proportional to  $e^{-(M_\rho-M_\pi) t}$.
$G^\theta(t,\Delta t)$ is a product of these~two averages plus corrections arising from correlations between 
$C_i(t)$ and $e^{-i\theta_i(t-\Delta t)}$, and so at large $t$ and $\Delta t$ it is expected to have the form 
\begin{eqnarray}
G^{\theta}(t,\Delta t) & \sim &
e^{-M_\pi (t-\Delta t) } e^{-M_\rho \Delta t} \left(\alpha+\beta e^{-\delta M_{\rho} \Delta t } +  ... \right),
\;\;\;\;
\label{eq:PRWtdep}
\end{eqnarray}
where $M_\rho +\delta M_{\rho}$ is the energy of the lowest-lying excited state of the $\rho^+$
leading to appreciable correlations between $C_i(t)$ and $e^{-i\theta_i(t-\Delta t)}$,
and $\alpha$ and $\beta$ are overlap factors that cannot be determined with general arguments but can be calculated with LQCD.
The ellipses denote further-suppressed contributions from higher-lying states.
A phase-reweighted effective mass can be defined as
$M^\theta=\log\left( G^{\theta}(t,\Delta t) /G^{\theta}(t+1,\Delta t+1) \right)$, which reduces to the standard effective mass definition when $\Delta t\rightarrow t$.
For the $\rho^+$ meson, the form of the correlation function given in Eq.~(\ref{eq:PRWtdep}) leads to
\begin{eqnarray}
M_\rho^\theta(t, \Delta t) & = & M_\rho \ +\  c\ \delta M_{\rho} e^{-\delta M_{\rho} \Delta t}\ +\ ...
\ \ ,
\label{eq:PRWEMP}
\end{eqnarray}
at large $t$,
where $c=\beta/\alpha$ and the ellipses denote higher order contributions which are exponentially suppressed with
$\Delta t$ and standard excited state contributions that are exponentially suppressed with $t$.

\begin{figure}[!t]
  \centering
	\includegraphics[width=0.7 \columnwidth]{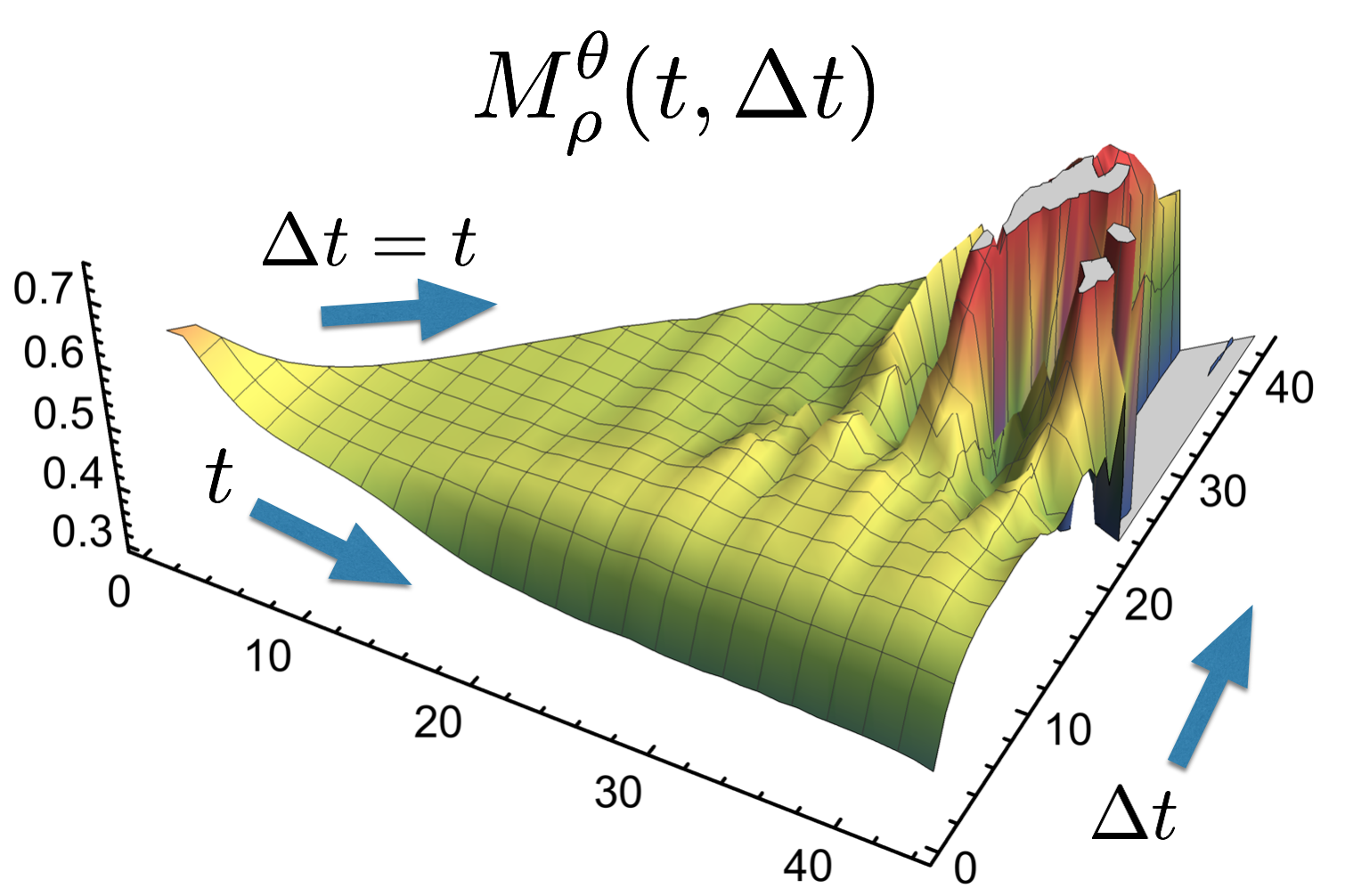}
		\caption{
	\label{fig:2DEMP} 
        The $\rho^+$ meson phase-reweighted effective mass for all $\Delta t \leq t$. 
        The standard effective mass in the upper panel of Fig.~\ref{fig:emps} corresponds to
        $M^\theta_\rho(t,t)$, a projection along the line $t = \Delta t$ indicated.
        The bottom panel of Fig.~\ref{fig:emps} 
        shows $M^\theta_\rho(t,\Delta t)$ on lines of constant $\Delta t$ parallel to the $t$ axis indicated. 
	}		
\end{figure}

\begin{figure}[!t]
  \centering
	\includegraphics[width=0.7 \columnwidth]{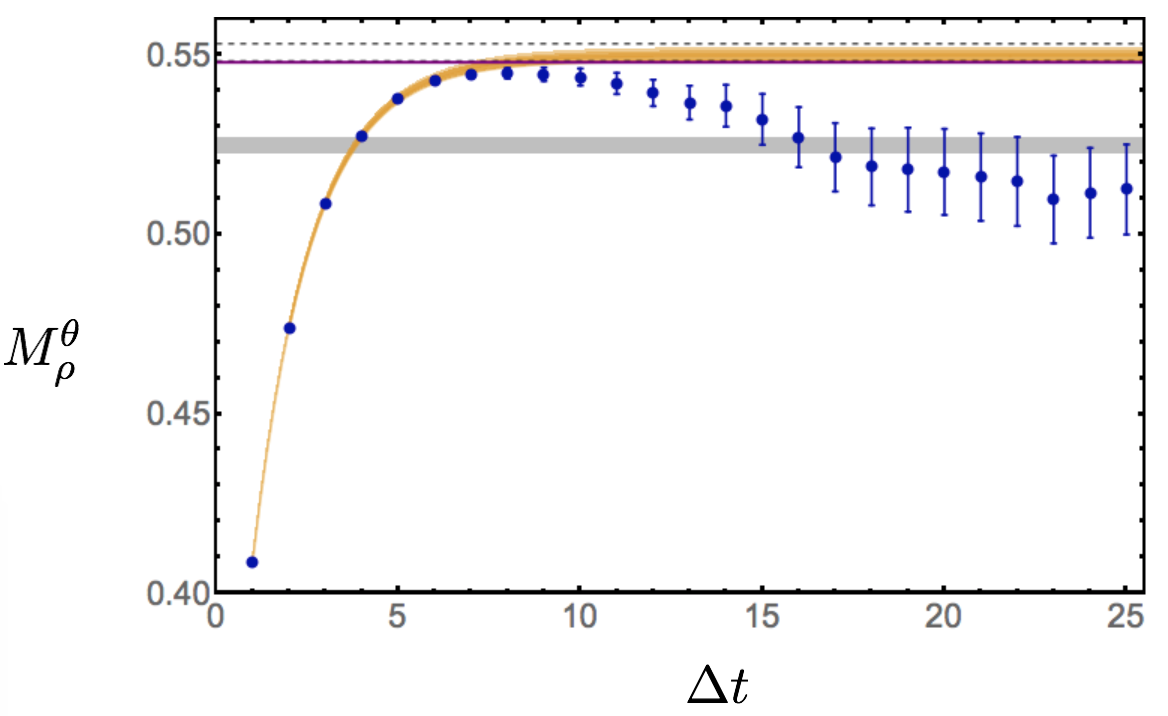}
		\caption{
	\label{fig:Rhoextrap} 
        The $\rho^+$ mass extracted from large-time phase-reweighted correlation functions.
	The light-brown shaded region corresponds to the $68\%$ confidence region associated with 	
        three-parameter (constant plus exponential)  fits to Eq.~\eqref{eq:PRWEMP}.
        The dashed lines show the extrapolated $M_\rho^\theta$ result
        including statistical and systematic uncertainties
        described in the main text.
        The gray horizontal band corresponds to a determination of the $\rho^+$ mass from the plateau region~\cite{Orginos:2015aya}.
        The purple line corresponds to the $\pi\pi$ non-interacting $p$-wave energy.	}		
\end{figure}

LQCD calculations of $M_\rho^\theta$ summarized in Figs.~\ref{fig:emps}-\ref{fig:Rhoextrap}
permit precise numerical study of small $\Delta t$ bias and $\Delta t \rightarrow t$ extrapolation.
These calculations employ $N\sim 130,000$ correlation functions
previously computed by the NPLQCD collaboration~ 
from smeared sources and point sinks on an 
ensemble of 2889 isotropic-clover gauge-field configurations 
at a pion mass of $M_\pi \sim 450~{\rm MeV}$
generated jointly by the College of William and Mary/JLab lattice group and by the NPLQCD collaboration, see Ref.~\cite{Orginos:2015aya} for further details.
The spacetime extent of the lattices is $48^3\times 96$ at a lattice spacing of $a\sim 0.117(1)~{\rm fm}$.
For all of the correlation functions examined in this work,  
momentum projected blocks are derived from quark propagators originating from smeared sources localized about a site in the lattice volume, 
as detailed in previous works by the NPLQCD collaboration, e.g. Ref.~\cite{Beane:2006mx,Orginos:2015aya}.
For instance, the blocks associated with the $\rho^+$ meson are
\begin{eqnarray}
{\cal B}^{(\rho^+)}_\mu ({\bf p},t; x_0) &  = & 
\sum_{\bf x} e^{i{\bf p}\cdot {\bf x}}\ \overline{S}_d({\bf x},t;x_0)\gamma_\mu S_u({\bf x},t;x_0).\;\;
  \label{eq:blockdef}
\end{eqnarray}
Correlations functions are derived by contracting the blocks with local interpolating fields~\cite{Detmold:2012eu},
e.g., 
\begin{eqnarray}
C^{(\rho^+;\mu)}({\bf p},t; x_0) & = & 
{\rm Tr}\left[\ 
{\cal B}^{(\rho^+)}_\mu ({\bf p},t; x_0)  \gamma^\mu
\ \right]
 ,
 \label{eq:blockcon}
\end{eqnarray}
where the trace is over color and spin. It is the phases of contracted momentum-projected blocks that have been used to form phase-reweighted correlation functions.
Expressions similar to those in eqs.~(\ref{eq:blockdef}) and (\ref{eq:blockcon}) are used for the nucleon and two-nucleon 
systems~\cite{Beane:2006mx,Orginos:2015aya}.

At large $t$ and small $\Delta t$, bias in $M_\rho^\theta$ is consistent with Eq.~\ref{eq:PRWEMP}.
At intermediate $\Delta t$, $M_\rho^\theta$ approaches a value consistent  with the $\pi\pi$ non-interacting $p$-wave energy $\sqrt{(2M_\pi)^2 + (2\pi/L)^2}$. 
At large $\Delta t$, $M_\rho^\theta$ approaches a lower-energy plateau consistent with the $\rho^+$ mass extracted from a $t=\Delta t$ plateau $t = 18 \rightarrow 28$.
The suppression of $\rho^+$ bound state contributions  compared to $\pi\pi$ scattering states contributions to $C_i(t)e^{-i\theta_i(t-\Delta t)}$ is found to be less severe in smaller volumes.
The energy gap between the bound and scattering states also increases in smaller volumes.
In accord with these arguments, the non-monotonic $\Delta t$ behavior visible in Fig.~\ref{fig:Rhoextrap} is not seen with $V=32^3$ or $V=24^3$.
$M_\theta^\rho$ is consistent with the $\rho^+$ mass determined in Ref.~\cite{Orginos:2015aya} for $\Delta t \gtrsim 5$ in these smaller volumes.
Variational methods employing phase reweighted correlation functions
with multiple interpolating operators
may be required 
to reliably distinguish closely spaced energy levels with large spatial volumes.

\begin{figure}[!t]
  \centering
	\includegraphics[width=0.7 \columnwidth]{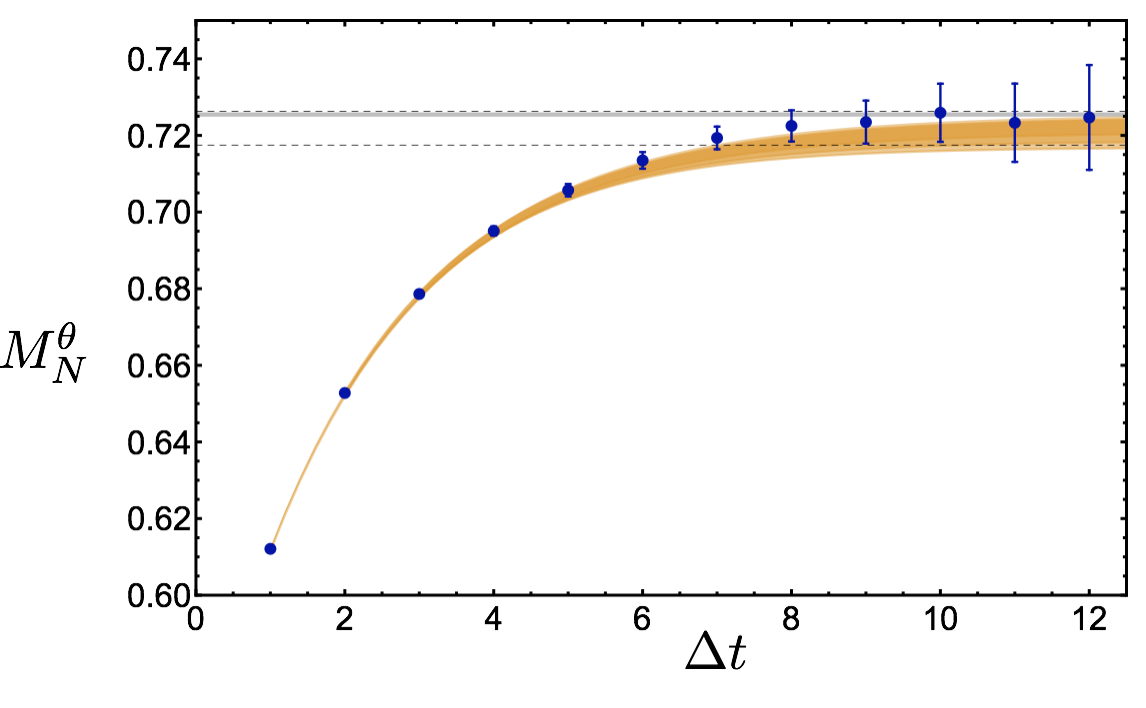}
		\caption{
	\label{fig:Nextrap} 
	The large-time nucleon phase-reweighted effective mass with statistical and systematic extrapolation errors shown
        with light-brown bands and dashed lines as in Fig.~\ref{fig:Rhoextrap}.
        The gray horizontal band corresponds to golden window result of Ref~\cite{Orginos:2015aya} obtained with four times higher statistics.
	}		
\end{figure}

The nucleon mass does not appear to have complications from low-lying excited states
and the large time phase-reweighted nucleon effective mass
derived from $\sim 100,000$ sources with $V=32^3$~\cite{Orginos:2015aya}
approaches its intermediate time plateau value at large $\Delta t$.
Small $\Delta t$ bias is well-described with a constant plus exponential form, and
the nucleon excited state gap can be 
extracted across a range of fitting regions as $\delta M_N = 786(44)(25)$~MeV, 
where the first uncertainty is statistical from a correlated $\chi^2$-minimization fit of $M_N^\theta(t,\Delta t)$ 
to Eq.~\eqref{eq:PRWEMP}
with $\Delta t = 2\rightarrow 10$ and $t = 30\rightarrow 40$  and the second uncertainty is a systematic  
determined from the variation in central value when the fitting region is changed to be $\Delta t = 1\rightarrow 10$ or $\Delta t = 3\rightarrow 10$. 
This result is consistent with a naive extrapolation $M_\sigma \sim 830$ MeV of the $\sigma$-meson mass 
determined at $M_\pi \sim 391$~MeV
~\cite{Briceno:2016mjc}. 
Results for strange-baryon excited-state masses from phase-reweighted effective mass extrapolations  are also 
consistent with the $\sigma$-meson mass in one- and two-baryon systems,
for instance $\delta M_{\Xi} = 822(44)(71)$~MeV and $\delta M_{\Xi\Xi(\si)} = 908(265)(82)$~MeV.

The $\Xi^-\Xi^-(\si)$ has slower StN degradation than a two-nucleon system
and is considered here for a first investigation of phase-reweighted baryon-baryon binding energies.
The $\Xi^-\Xi^-(\si)$ binding energy was
determined 
by
the NPLQCD collaboration
to be
$B_{\Xi\Xi (\si)}= 15.4(1.0)(1.4)~{\rm MeV}$
for the gauge field configurations considered here
using the correlation function production and sink-tuning~\cite{Beane:2009kya,Beane:2009gs,Beane:2010em} described for the deuteron and di-neutron in Ref.~\cite{Orginos:2015aya}.\footnote{$B_{\Xi\Xi(\si)} = -M_{\Xi\Xi(\si)}+2M_\Xi$
approaches the $\Xi\Xi(\si)$ binding energy in the infinite volume limit.
In finite volume $B_{\Xi\Xi(\si)}$ differs from the infinite-volume binding energy
by corrections that are exponentially suppressed by the binding momentum.
}
Results for $\Xi^-\Xi^-(\si)$ using the $\sim 100,000$ correlation function ensemble described above 
for constant 
fits to the phase reweighted binding energy with $t = 28\rightarrow 43$, $\Delta t = 1,2,3\rightarrow 6$ 
give 
$B_{\Xi\Xi(\si)}=15.8(3.5)(2.6)$ MeV.
Consistency between golden window results and phase-reweighted results with large $t$ and all $\Delta t \gtrsim 1$
suggests
a high degree of cancellation at all $\Delta t$ between excited state effects in 
one- and two-baryon phase reweighted effective masses.
$B_{\Xi\Xi(\si)}(t, \Delta t = 0) $, which only involves correlation function magnitudes,
plateaus to $7.1(0.6)(0.8)\text{ MeV}$.
Phase effects modify this magnitude result
by an amount on the order of nuclear energy scales rather than hadronic mass scales,
providing encouraging evidence that extrapolations involving modest $\Delta t$ can accurately determine nuclear binding energies in the noise region.
The precision of phase-reweighted results scales with the number of points in the noise region,
and could be increased on lattices of longer temporal extent then those used in this work ($\sim 11.2 {\rm fm}$).

\begin{figure}[!t]
  \centering
	\includegraphics[width=0.65 \columnwidth]{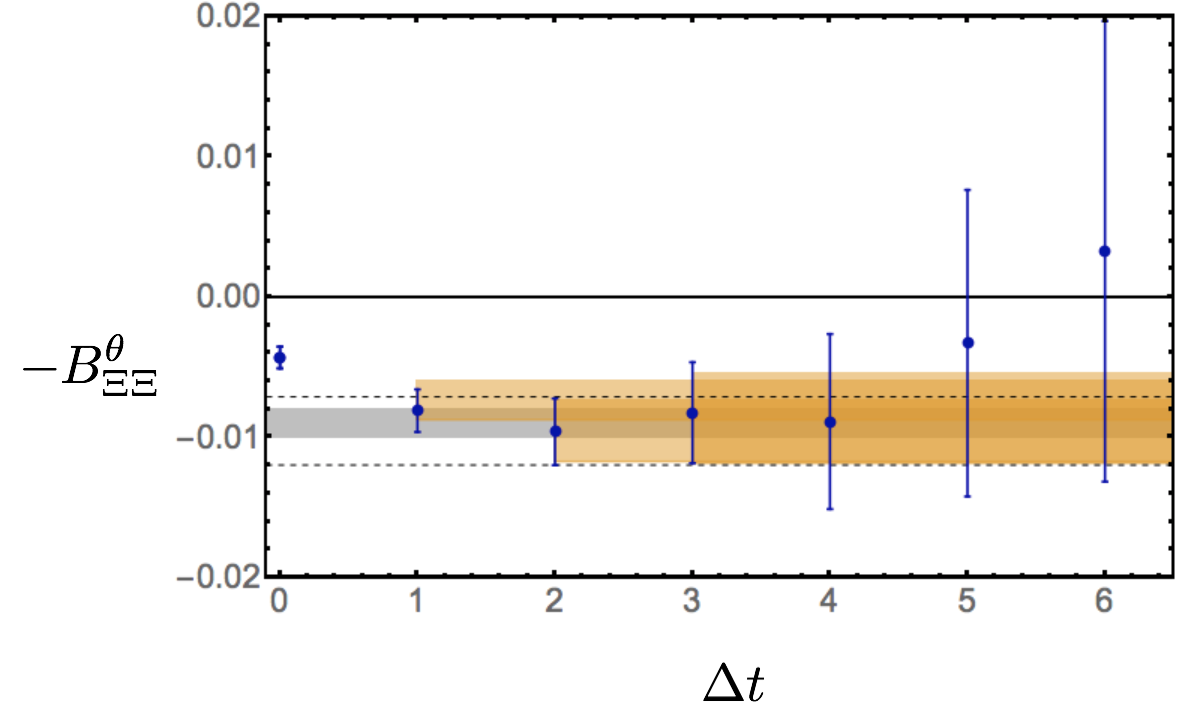}
		\caption{
	\label{fig:XiXiBindextrap} 
	The $\Xi^-\Xi^- (\si)$ phase-reweighted binding energy with statistical and systematic extrapolation errors shown
      with light-brown bands and dashed lines as in Fig.~\ref{fig:Rhoextrap}.
        The gray horizontal band corresponds to the golden window result of Ref.~\cite{Orginos:2015aya}, obtained with four times higher statistics.
	}		
\end{figure}

Phase reweighting 
allows 
energy levels
to be extracted
from LQCD correlation functions
at times larger than the
golden window
accessible to standard techniques involving source and sink optimization~\cite{Beane:2009kya,Beane:2009gs,Beane:2010em,Detmold:2014rfa,Detmold:2014hla}.
It is expected that these methods
will permit the extraction of ground-state energies in systems
without a golden window.
The phase-reweighting method
is equivalent to a dynamical source improvement
in which the phase is held fixed while the magnitude of the hadronic correlation function
is evolved into its ground state,
and then the phase is released to provide a source for subsequent time slices.
The bias introduced by phase reweighting can be removed by extrapolation
but suffers from a StN problem
that can be viewed as arising from evolution of the dynamically improved source.
Generalizations of the phase-reweighting methods presented here
may allow for reaction rates,
operator matrix elements,
and other observables to be extracted from phase-reweighted correlation functions.
Further study is planned of the $\Delta t \rightarrow t$ extrapolation 
and applications of phase reweighting to hadronic and nuclear systems.
Mesonic systems in particular are discussed in the next chapter.

\begin{table}[!ht]
\begin{center}
\begin{minipage}[!ht]{16.5 cm}
\end{minipage}
\setlength{\tabcolsep}{1em}
\resizebox{.7\linewidth}{!}{%
\def\arraystretch{.7}%
\begin{tabular}{|c| l | l | l |}
\hline
$\Delta t$    & \qquad $M^\theta_{\rho^+}$  & \qquad $M_N^\theta$          & \qquad $B_{\Xi\Xi(\si)}$     \\
\hline
\hline
$1$ & \quad 0.40872(21) &\quad 0.61209(50)&\quad  -0.0081(15)           \\
$2$ & \quad 0.47392(30) &\quad 0.65278(66)&\quad  -0.0096(24)           \\
$3$ & \quad 0.50841(40) &\quad 0.67861(88)&\quad   -0.0083(36)          \\
$4$ & \quad 0.52722(52) &\quad 0.6951(12)&\quad    -0.0089(62)         \\
$5$ & \quad 0.53774(67) &\quad 0.7057(16)&\quad   -0.003(11)          \\
$6$ & \quad 0.54284(84) &\quad 0.7135(22)&\quad    0.003(16)          \\
$7$ & \quad 0.5446(11) &\quad 0.7193(30)&\qquad\qquad       -      \\
$8$ & \quad 0.5449(15) &\quad 0.7225(41)&\qquad\qquad       -      \\
$9$ & \quad 0.5446(19) &\quad 0.7235(56)&\qquad\qquad       -      \\
$10$ & \quad 0.5439(23) &\quad 0.7259(76)&\qquad\qquad       -      \\
$11$ & \quad 0.5421(30) &\quad 0.723(10) & \qquad \qquad -      \\
$12$ & \quad 0.5395(37) &\quad 0.725(14) &  \qquad \qquad -      \\
$13$ & \quad 0.5368(47) &\qquad \qquad -  &  \qquad \qquad -      \\
$14$ & \quad 0.5359(58) &\qquad \qquad -  &  \qquad \qquad -      \\
$15$ & \quad 0.5321(71) &\qquad \qquad -  &  \qquad \qquad -      \\
$16$ & \quad 0.5271(83) &\qquad \qquad -  &  \qquad \qquad -      \\
$17$ & \quad 0.5215(95) &\qquad \qquad -  &  \qquad \qquad -      \\
$18$ & \quad 0.519(11) &\qquad \qquad -  &  \qquad \qquad -      \\
$19$ & \quad 0.518(12) &\qquad \qquad -  &  \qquad \qquad -      \\
$20$ & \quad 0.517(12) &\qquad \qquad -  &  \qquad \qquad -      \\
$21$ & \quad 0.516(12) &\qquad \qquad -  &  \qquad \qquad -      \\
$22$ & \quad 0.515(12) &\qquad \qquad -  &  \qquad \qquad -      \\
$23$ & \quad 0.510(12) &\qquad \qquad -  &  \qquad \qquad -      \\
$24$ & \quad 0.512(13) &\qquad \qquad -  &  \qquad \qquad -      \\
$25$ & \quad 0.513(13) &\qquad \qquad -  &  \qquad \qquad -      \\
\hline
PR Ground  & 0.5222(60)(27)   &  0.7220(33)(11) &    -0.0096(22)(11)         \\
PR Excited  & 0.5508(11)(7)   &  \qquad \qquad - &    \qquad \qquad -         \\
\hline 
GW Ground  &  0.5248(14)(15) & 0.72551(35)(26)&    -0.00909(59)(83)         \\
GW $\pi\pi$  &  0.547997(78)(14) & \qquad \qquad - &    \qquad \qquad -         \\
\hline 
\end{tabular}
}
\begin{minipage}[t]{16.5 cm}
\vskip 0.0cm
\tiny
\noindent
  \caption{
Phase-reweighted (PR) effective masses of the $\rho^+$, nucleon and the effective 
energy difference between $\Xi\Xi(\si)$ and two $\Xi$'s
derived from  eq.~(\ref{eq:PRWdef}). 
The extrapolated PR ground values are taken from three-parameter constant plus exponential correlated $\chi^2$-minimization fits for
$M_N^\theta$ and one-parameter constant fits for $B_{\Xi\Xi(\si)}^\theta$
with statistical uncertainties for fits starting at $\Delta t = 2$ and systematic uncertainties
defined from variation of the $\Delta t$ fitting window as described in the main text.
PR data is taken from $t = 28 \rightarrow 43$ for the $\rho^+$ and  $\Xi\Xi(\si)$
and $t = 31 \rightarrow 40$ for the nucleon.
For the $\rho^+$, the region $\Delta t = 2 \rightarrow 10$ is used to constrain the first scattering state for the PR excited state result, 
while the region $\Delta t = 16\rightarrow 25$ is used to constrain the ground state.
Golden window (GW) ground refers to the ground-state energy determinations using the short and intermediate
time plateau regions described in Ref.~\cite{Orginos:2015aya}.
GW $\pi\pi$ refers to the non-interacting $p$-wave energy shift $\sqrt{(2M_\pi)^2 + (2\pi/L)^2}$ using $M_\pi$ and $L$ for the $48^3$ ensemble described in the main text.
}
\label{tab:masses}
\end{minipage}
\end{center}
\end{table}


\chapter{Meson Statistics and Real-Valued Sign Problems}\label{chap:mesons}
 
%

The statistical observations of Chapter~\ref{chap:statistics} begin with a magnitude-phase decomposition of correlation functions.
This is a suitable starting point for baryon and multi-baryon correlation functions, which are generically complex.
It is a less obvious starting point for analyzing correlation functions that are real but non-positive definite.
Such correlation functions face a true ``sign problem'' rather than a ``phase problem.''
Each $C_i$ in a Monte Carlo ensemble of $i=1,\cdots,N$ real but non-positive-definite correlation functions is described by a positive-definite magnitude and a binary valued sign,
\begin{equation}
  \begin{split}
    C_i = |C_i|e^{i\theta_i},\hspace{20pt} e^{i\theta_i} \in \{+1,-1\}.
  \end{split}\label{eq:binary}
\end{equation}
In this chapter, the binary-valued signs of meson correlation functions in QCD are analyzed and found to be consistently described by circular statistics.
Real-valued isovector correlation functions are shown to be statistically well-described by the real part of the complex-log-normal distribution introduced in Chapter~\ref{chap:statistics}.
The correlation-function-ratio estimator introduced in Chapter~\ref{chap:statistics} and phase-reweighted techniques introduced in Chapter~\ref{chap:PR} are both applied to isovector meson correlation functions.
Phase reweighting allows precise ground-state energy results to be extracted from the noise region; analyzing correlation function ratios does not.
As a demonstration of the utility of phase reweighting and exploration of meson phenomenology in the world of heavy quarks where LQCD simulations are currently performed, phase-reweighted ground-state energies are extracted from the noise region for several channels that have proven difficult to access with GW spectroscopy.

The $\rho^+$ correlation function introduced in Eq.~\eqref{eq:blockcon} is part of a more general class of isovector meson correlation functions associated with the pseudoscalar $\pi$, the $1^{--}$ vector $\rho(770)$ the scalar $a_0(980)$, the $1^{++}$ vector $a_1(1260)$, and the $1^{+-}$ vector $b_1(1235)$ resonances~\cite{Olive:2016xmw}. 
At light quark masses, these channels include narrow and broad resonances as well as a large multiplicity of scattering states above $\pi \pi$ and $K\bar{K}$ thresholds. 
The resonance structure of these states has been studied intensively in recent years~\cite{Aoki:2007rd,Dudek:2009qf,Dudek:2010wm,Lang:2011mn,Aoki:2011yj,Edwards:2011jj,Thomas:2011rh,Dudek:2011tt,Dudek:2012ag,Dudek:2012gj,Dudek:2012xn,Pelissier:2012pi,Dudek:2013yja,Wilson:2014cna,Bolton:2015psa,Briceno:2015dca,Wilson:2015dqa,Dudek:2016cru,Briceno:2016kkp,Briceno:2016mjc},
A wide range of single- and multi-particle interpolating operators,
extensive formalism relating scattering parameters to finite volume energy shifts~\cite{Luscher:1990ux,Luscher:1990ck,Rummukainen:1995vs,Lellouch:2000pv,Beane:2003da,Feng:2004ua,He:2005ey,Kim:2005gf,Christ:2005gi,Bernard:2008ax,Lage:2009zv,Bernard:2010fp,Luu:2011ep,Briceno:2012yi,Briceno:2012rv,Hansen:2012bj,Hansen:2012tf,Gockeler:2012yj,Dudek:2012xn,Guo:2012hv,Leskovec:2012gb,Briceno:2013bda,Briceno:2013lba,Briceno:2014oea,Hansen:2014eka,Dudek:2014qha,Wilson:2014cna,Wilson:2015dqa,Bali:2015gji,Briceno:2016xwb,Hansen:2017mnd}
variational methods for identifying excited states~\cite{Michael:1985ne,Luscher:1990ck,Dudek:2007wv,Blossier:2009kd},
and techniques for efficiently computing disconnecting diagrams such as distillation~\cite{Peardon:2009gh},
are needed to extract physical resonance parameters from LQCD.

At larger quark masses, the pion mass is closer to the masses of the other isovector mesons
and channels with multi-pion ground states at light quark masses
may instead have single-hadron ground states associated with stable isovector mesons.
At the heavy quark masses used in this work with $m_\pi \sim 450$ MeV and $m_\pi \sim 800$ MeV,
it is expected that some or all of the low-lying isovector mesons can be described as compact bound states~\cite{Briceno:2016xwb}
and therefore have
finite-volume ground-state energies that are exponentially close to their infinite-volume counterparts.
Extending the use of phase-reweighting to meson spectroscopy at light quark masses
will require new variational methods for phase-reweighted correlation functions 
in conjunction with the existing sophisticated tools of LQCD resonance calculations above.

LQCD results in this chapter employ a larger set of the $m_\pi \sim 450$ MeV correlation function ensembles generated by the College of William and Mary/JLab lattice group and the NPLQCD collaboration previously employed above that includes three spacetime volumes $L^3\times \beta$ of dimensions $24^3\times 64$, $32^3 \times 96$ and $48^3\times 96$, see Ref.~\cite{Orginos:2015aya}. Correlation functions with $m_\pi \sim 800$ MeV and spacetime dimension $32^3 \times 48$ are further employed in this chapter.

\section{A Magnitude-Sign Decomposition}

Isovector meson states can be constructed with simple interpolating operators such as $[\bar{d}\Gamma u](x)$ obeying $[\bar{d}\Gamma u]^\dagger = \bar{d} \Gamma^\dagger u$ where $\Gamma$ is a Dirac spin matrix. 
LQCD calculations require an ensemble of $i=1,\cdots,N$ correlation functions $C_i^\Gamma$ calculated in QCD-vacuum importance sampled gauge field configurations whose average determines the QCD correlation function $G^\Gamma$.
With this choice of interpolating operators, isovector correlation functions are given in terms of quark propagators by
\begin{equation}
  \begin{split}
    G^\Gamma (t) &= \avg{C_i^\Gamma(t)} = \frac{1}{N}\sum_{i=1}^N C_i^\Gamma(t) \\ 
    &= \sum_\v{x} \int \mathcal{D}\bar{q}\mathcal{D}q\; e^{-\sum_y \bar{q}(y)D(U_i,y;y;)q(y)} [\bar{d}\Gamma u](\v{x},t) [\bar{u} \Gamma^\dagger d](0)\\
    &= \sum_\v{x} \tr\left[ S_u(U_i;\v{x},t;0)\Gamma^\dagger S_d(U_i;0;\v{x},t) \right]  \\
    &= \sum_\v{x} \tr\left[ S_u(U_i;\v{x},t;0)\Gamma^\dagger \gamma_5 S_d(U_i;\v{x},t;0)^\dagger \gamma_5 \Gamma \right] ,
  \end{split}\label{Cdef}
\end{equation}
where the trace is over spin and color indices and the last equality uses $\gamma_5$-Hermiticity.
For isoscalar mesons, single propagator traces corresponding to quark-line-disconnected diagrams also contribute to the correlation function. 
Disconnected diagrams have different statistical behavior than the quark-line-connected diagrams considered here, and a detailed discussion of the statistics and phase reweighting of disconnected diagrams is left to future work. 
In the isospin limit where $S_u = S_d \equiv S$,
\begin{equation}
  \begin{split}
    C_i^\Gamma(t)^\dagger = \sum_\v{x} \tr\left[ S(U_i;\v{x},t;0) \gamma_5 \Gamma S(U_i;\v{x},t;0)^\dagger \Gamma^\dagger \gamma_5  \right].
  \end{split}\label{eq:reality}
\end{equation}
Each meson correlation function in a Monte Carlo ensemble is therefore real, provided
\begin{equation}
  \gamma_5 \Gamma \gamma_5 = \pm \Gamma^\dagger,
  \label{g5Hermmat}
\end{equation}
which holds for $\Gamma \in \{\gamma_\mu, \gamma_5, \gamma_\mu \gamma_5\}$ but not for complex linear combinations such as $\gamma_1 + i \gamma_2$. 
Vector-meson correlation functions with interpolating operators corresponding to $J_z = \pm1$ are complex in a generic gauge field configuration while those corresponding to $J_z = 0$ are real, suggesting that reality of isovector meson correlation function is a property of a particular choice of interpolating operator and not a fundamental property of the state. 
One is always free to use complex linear combinations of interpolating operators in a QFT, so real correlation functions can generically be considered to be a special case of complex correlation functions.
Still, the heuristic picture of a complex phase taking a random walk on the unit circle from Chapter.~\ref{chap:statistics} is obscure when applied to signed real numbers and StN expectations associated with this picture should be investigated further.

\begin{figure}[!t] \centering
  \includegraphics[width=.45\columnwidth]{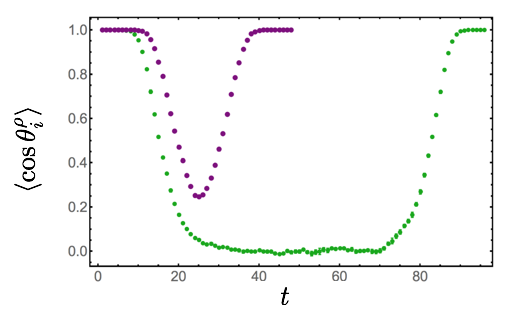} \hspace{10pt}
  \includegraphics[width=.45\columnwidth]{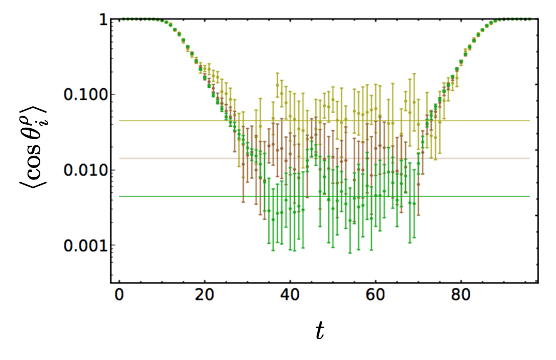}
  \caption{Results for the $\rho^+$ average phase $\avg{cos \theta_i^\rho}$, equal to the fractional difference between the number of positive and negative correlation functions as in Eq.~\eqref{eq:counting}. Results using the $m_\pi\sim 450$ MeV ensemble are shown in green on both plots, with a logarithmic scale used on right. Results for the $m_\pi \sim 800$ MeV ensemble are shown in purple on left. In addition to the green points corresponding to $N= 50,000$ correlation functions, there are also results shown in brown for a subset of $N=5000$ correlation functions and results shown in yellow for $N=500$ correlation functions. Corresponding lines in each color are shown at $1/\sqrt{N}$. The circular statistics bound of Eq.~\eqref{eq:circstatsbound} predicts that estimates of the average phase will be systematically biased for $\avg{\cos \theta_i^\rho} < 1/\sqrt{N}$.
    } 
  \label{fig:RhoThCirc}
\end{figure}

The log-magnitude $R^\Gamma_i$ and phase $\theta^\Gamma_i$ associated with $C_i^\Gamma$ are formally defined by
Effective masses for the magnitude and phase can be defined as in Chapter~\ref{chap:statistics}. 
The phase effective mass $M^\Gamma_\theta = \ln\left(\avg{\cos \theta^\Gamma_i(t)}/\avg{\cos \theta^\Gamma_i(t+1)}\right)$ is determined by the average phase.
If there are $N_+$ positive sign correlation functions and $N_-$ negative sign correlation functions, then the average phase is simply
\begin{equation}
  \begin{split}
    \avg{e^{i\theta_\Gamma(t)}} = \avg{\cos(\theta_\Gamma(t))} = \frac{N_+ - N_-}{N},
  \end{split}\label{eq:counting}
\end{equation}
so the effective mass contribution of a pure sign is simply determined by a counting problem.
The average phase of Eq.~\eqref{eq:counting} for the $\rho^+$ correlation function is shown in Fig.~\ref{fig:RhoThCirc}.
At small times the average phase is nearly one.
At intermediate time the average phase decreases exponentially, and in the $m_\pi\sim 450$ MeV ensemble becomes close to zero at large times.

\begin{figure}[!ht] \centering
  \includegraphics[width=\columnwidth]{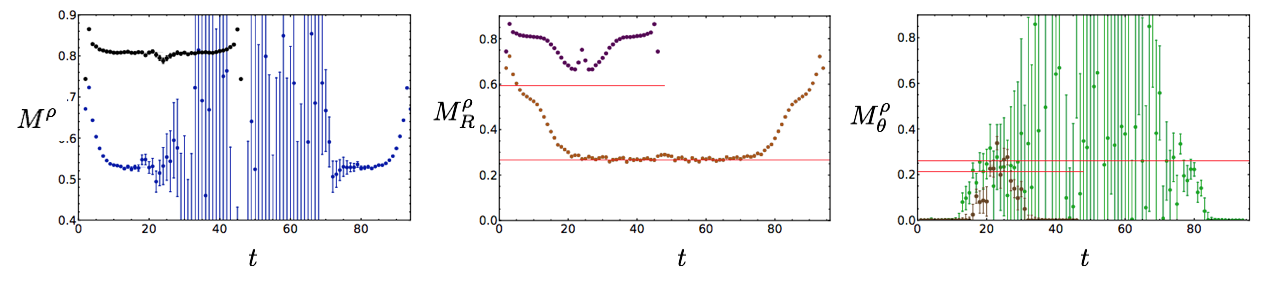}
  \caption{Effective mass and bootstrap uncertainties for the rho magnitude-sign decomposition. The left plot shows $M^\rho(t)$ for the $m_\pi \sim 450$ MeV ensemble in blue and for the $m_\pi \sim 800$ MeV ensemble in black. The middle plot the magnitude effective mass $m_R^{\rho}(t)$ in orange for the $m_\pi \sim 450$ MeV ensemble and in purple for the $m_\pi \sim 800$ MeV ensemble. The right plot shows the phase (sign) effective mass $M_\theta^{\rho}(t)$ in green for the $m_\pi \sim 450$ MeV ensemble and brown for the $m_\pi \sim 800$ MeV ensemble. The middle plot includes red lines at $m_\pi$ for both pion masses, with the shorter line corresponding to the smaller $m_\pi \sim 800$ MeV lattice. The right-hand plot similarly includes red lines at $M_\rho - m_\pi$, with $M_\rho$ the central value from golden widow fits to the full correlation functions shown left.
  }
  \label{fig:RhoEM}
\end{figure}

Parisi-Lepage analysis predicts that all isovector mesons other than the pion face an exponentially hard StN problem,
\begin{equation}
  \begin{split}
    \text{StN}(C_i^\Gamma) \sim e^{-M_\Gamma^{StN}t} \sim \frac{\avg{C_i^\Gamma}}{\sqrt{\avg{|C_i^\Gamma|^2}}} \sim e^{-(M_\Gamma - m_\pi)t}.
  \end{split}\label{eq:mesonstn}
\end{equation}
The nucleon StN problem is identified as a sign problem above by noting that an exponentially-decaying phase inherently faces an exponentially hard StN problem and observing that the average phase of the nucleon correlation function decays at a rate equal to the StN-degradation rate.
Similarly, isovector meson correlation functions can be decomposed into a magnitude and phase (sign) as
\begin{equation}
  \begin{split}
    C^\Gamma_i(t) = e^{R^\Gamma_i(t) + i\theta^\Gamma_i(t)},
  \end{split}\label{eq:mesondef}
\end{equation}
and the Parisi-Lepage StN problem for mesons can be identified as a sign problem if
\begin{equation}
  \begin{split}
    \frac{\avg{e^{i\theta^\Gamma_i(t)}}}{\sqrt{\avg{\left|e^{i\theta^\Gamma_i(t)}\right|^2}}} = \avg{e^{i\theta^\Gamma_i(t)}} \sim e^{-M_\Gamma^{StN} t}.
  \end{split}\label{eq:phaseStN}
\end{equation}
Thermal artifacts associated with ``backwards-propagating'' states have more sizable effects on meson correlation function than baryon correlation functions
because in the meson cause the backwards states are degenerate with the forwards states, see e.g. Ref~\cite{Beane:2009kya},
and it is helpful to define an effective mass taking time reflection symmetry into account as
\begin{equation}
  \begin{split}
    M_\Gamma(t) = \text{arccosh}\left[ \frac{\avg{C_i^\Gamma(t+1)} + \avg{C_i^\Gamma(t-1)}}{2\avg{C_i^\Gamma(t)}} \right].
  \end{split}\label{mGamma}
\end{equation}
Effective masses for magnitude and phase contributions can be defined analogously as
\begin{equation}
  \begin{split}
    M_R^\Gamma(t) &= \text{arccosh}\left[ \frac{\avg{e^{R_i^\Gamma(t+1)}} + \avg{e^{R_i^\Gamma(t-1)}}}{2\avg{e^{R_i^\Gamma(t)}}} \right], \\ 
    M_\theta^\Gamma(t) &=  \text{arccosh}\left[ \frac{\avg{\cos \theta_i^\Gamma(t+1)} + \avg{\cos \theta_i^\Gamma(t-1)}}{2\avg{\cos \theta_i^\Gamma(t)}} \right].
  \end{split}\label{mGammaRTh}
\end{equation}
LQCD results for the $\rho^+$-meson are shown in Fig.~\ref{fig:RhoEM}, and the $\rho^+$ mass can be determined by fitting the correlation function or effective mass in the intermediate-time plateau region $t=13\rightarrow 20$ assuming ground-state dominance.
Results with $m_\pi \sim 450$ MeV show that at large times $M_R^\rho$ approaches $m_\pi$ with no StN problem and $M_\theta^\rho$ approaches $M_\rho - m_\pi$ with a severe StN problem.
Results for $m_\pi \sim 800$ MeV are consistent but do not show a visible plateau and indicate that a larger time direction is needed to observe the approach of the magnitude and phase effective masses to their expected asymptotic values.
Correlations between the magnitude and phase are negligible at small times but statistically significant at large times.
For $t \gtrsim 20$, the ratio
\begin{equation}
  \begin{split}
    \frac{\text{Cov}(e^{R_i^\Gamma},e^{i\theta_i^\Gamma})}{\avg{e^{R_i^\Gamma + i \theta_i^\Gamma}}} = 1 - \frac{\avg{e^{iR_i^\Gamma}}\avg{e^{i\theta^\Gamma_i}}}{G_\Gamma},
  \end{split}\label{eq:magphasecov}
\end{equation}
plateaus to a constant value of $0.319(14)(4)$.
This constant-time behavior is consistent with $M_\Gamma = M_R^\Gamma + M_\theta^\Gamma$, and explicit calculation shows that $M_\Gamma - M_R^\Gamma - M_\theta^\Gamma$ is consistent with zero for $t \gtrsim 20$.
These observations indicate that $M_\Gamma$ can be decomposed into magnitude and phase effective mass contributions, and that the meson StN problem arises from re-weighting the sign problem inherent to calculation of the average meson sign in complete analogy to the case of the nucleon phase.

\begin{figure}[!t] \centering
  \includegraphics[width=\columnwidth]{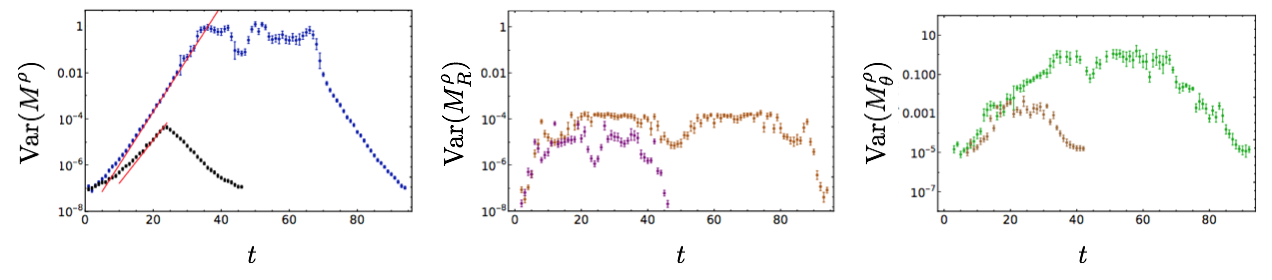}
  \caption{The left panel shows the variance of the effective mass $M_\rho(t)$ shown in Fig.~\ref{fig:RhoEM} determined by bootstrap resampling. Uncertainties on the results shown are determined by further bootstrap resampling of variance results. The middle panel shows the variance of $M_R^\rho(t)$ analogously, and the right panel shows the variance of $M_\theta^\rho(t)$. Results for $m_\pi \sim 450$ MeV and $m_\pi \sim 800$ MeV with $L=32$ are both shown and color-coled as in Fig.~\ref{fig:RhoEM}. The red lines in the left plot show predictions of exponential StN degradation satisfying Parisi-Lepage scaling $M_\rho^{StN} = M_\rho - m_\pi$, where the overall normalization has been fixed by one intermediate time chosen to be $t=22$.
  }
  \label{fig:RhoEMErrors}
\end{figure}

The uncertainties associated with $M_\rho$, $M_R^\rho$, and $M_\theta^\rho$ are shown in Fig.~\ref{fig:RhoEMErrors}.
Agreement between LQCD results and Parisi-Lepage scaling predictions is visible at intermediate times,
but at the largest times the $m_\pi \sim 450$ MeV variance of $M_\rho$ stops growing exponentially and approaches a constant.
This unphysical time-dependence is consistent with observations of the nucleon above
and signals the onset of a noise region where the results of standard estimators systematically deviate from QCD.
The existence of such a noise region is predicted by circular statistics,
and its appearance in $\rho^+$-meson results suggests that the perspective of circular statistics might be useful for real but non-positive correlation functions.

Since $e^{i \theta_i^\Gamma} \in \{\pm 1\}$ is a random variable with unit magnitude, $\theta_i^\Gamma$ can be interpreted as a discrete circular random variable with $\theta_i^\Gamma \in \{0,\ \pi\}$.
Standard theorems of circular statistics apply to discrete circular random variables~\cite{Fisher:1995,jammalamadaka2001topics,Mardia:2009},
including the result that standard parameter inference based on trigonometric moments generically breaks down due to finite-sample-size effects unless
\begin{equation}
  \begin{split}
    \frac{1}{N}\sum_{i=1}^N \cos\theta_i(t) > \frac{1}{\sqrt{N}}.
  \end{split}\label{eq:circstatsbound}
\end{equation}
It is derived explicitly in Chapter~\ref{chap:statistics} that Eq.~\eqref{eq:circstatsbound} must be met in order to reliably calculate the average phase of a random variable drawn from a wrapped normal distribution, which is empirically shown to provide a good fit to baryon correlation functions. 
Eq.~\eqref{eq:circstatsbound} holds for other common distributions in circular statistics, and on general grounds the statistical ensemble size necessary to distinguish any sufficiently broad circular distribution from a uniform distribution grows with the width of the distribution.
Fig.~\ref{fig:RhoThCirc} demonstrates explicitly that the average phase of $\rho^+$ correlation functions in LQCD begin deviating from exponential time-dependence and eventually reach a constant value.
The constant value eventually reached by the phase is seen in Fig.~\ref{fig:RhoThCirc} to decrease as $1/\sqrt{N}$ as the sample size is increased.
This verifies that Eq.~\eqref{eq:circstatsbound} applies to real but non-positive correlation functions with sign problems such as the $C_i^\Gamma$.
Similar results are expected to apply to any quantum Monte Carlo calculation with a sign problem.

\section{Non-Positive Correlation Functions Moments}

The scaling of higher moments of meson correlation functions can be understood analogously to the Lepage-Savage scaling of higher moments of baryon correlation functions.
Even moments of baryon correlation functions $\avg{|C_i^N|^{2n}}$ are associated with states of zero baryon number, as described in Ref.~\cite{Beane:2014oea}.
In the absence of non-zero baryon number charge the lightest states contributing to $\avg{|C_i^N|^{2n}}$ will be multi-pion rather than multi-nucleon states. 
Conversely, odd moments $\avg{|C_i^N|^{2n}C_i^N}$ are associated with states containing one baryon as well as $3n$ pions described by a partial quenched theory of $nN_f$ valence quark flavors.
In general, nucleon correlation function moments have time dependence that in the absence of hadronic interactions is given by
\begin{equation}
  \begin{split}
    \avg{|C_i^N|^{2n} (C_i^N)^B} \sim e^{-3n m_\pi t - B M_N t}.
  \end{split}\label{eq:baryonLS}
\end{equation}
It follows that the real parts of baryon correlation functions have increasingly broad and symmetric distributions at large times.

For $\rho^+$ correlation functions, $G$-parity plays a role analogous to baryon number above and ensures that states with the quantum numbers of an odd number of $\rho^+$ mesons cannot be described as multi-pion states. 
Even moments of $C_i^\rho$ are not distinguished from multi-pion correlation functions by any conserved charge and therefore have large-time scaling controlled by the pion mass,
\begin{equation}
  \begin{split}
     \avg{|C_i^\rho|^{2n}}\sim e^{-2 n m_\pi},
  \end{split}\label{eq:rhoeven}
\end{equation}
By $G$-parity, odd moments of $C_i^\rho$ have large-time scaling influenced by the $\rho^+$-meson mass,
\begin{equation}
  \begin{split}
    \avg{|C_i^\rho|^{2n+1}} \sim e^{-2n m_\pi - M_\rho}.
  \end{split}\label{eq:rhoodd}
\end{equation}
As in the nucleon case, QCD inequalities ensuring the pion is the lightest state in the QCD spectrum  guarantee that $\rho^+$ correlation functions become increasingly broad and symmetric at large times~\cite{Weingarten:1983uj,Witten:1983ut}. 
Other single-meson states distinguishable from single-pion states by some quantum number (parity for the $a_0$ and $a_1$; charge conjugation for the $b_1$) similarly have even moments whose large-time decay is set by the pion mass and odd moments who large-time decay is set by the mass of the meson in question.
Isovector meson correlation functions other than pion correlation functions are therefore increasingly broad and symmetric at large times. 

Lepage-Savage scaling can also be proven generically for the real parts of complex correlation functions.
Consider a correlation function ensemble $C_i$ with $Q$ quark lines and ground-state mass $M$ satisfying $M \geq \frac{Q}{2}m_\pi$, and for the purposes of estimating large-time scaling ignore interactions between hadrons.
Applying trigonometric identities relating $\cos^n(\theta)$ and $\cos(n\theta)$ to the phase gives
\begin{equation}
  \begin{split}
    \avg{(\text{Re} C_i)^n} &= \avg{e^{nR_i}\cos^n(\theta_i)} \\
        &= \begin{cases} \frac{1}{2^{n-1}}\sum_{k=0}^{(n-1)/2} {n \choose k} \avg{e^{n R_i}\cos\left( (n-2k)\theta_i \right)} , 
        & n\text{ odd} \\ 
        \frac{1}{2^n}{n\choose n/2} \avg{e^{n R_i}} + \frac{1}{2^{n-1}}\sum_{k=0}^{(n-2)/2} {n \choose k} \avg{e^{n R_i}\cos\left( (n-2k)\theta_i \right)} , 
        & n\text{ even} \end{cases}.
    \end{split}\label{eq:LSgeneral1}
\end{equation}
Each term in the above sums can be re-grouped as 
\begin{equation}
  \begin{split}
    \avg{e^{2k R_i}e^{(n-2k)R_i}\cos( (n-2k) \theta_i)} = \avg{|C_i|^{2k}C_i^{n-2k}} \sim e^{-k Q m_\pi t - (n-2k) M t},
  \end{split}
\end{equation}
where the right-most relation holds because $|C_i|^{2k}$ contains $kQ$ conserved quarks and $kQ$ separately conserved antiquarks and therefore has time-dependence $\avg{|C_i|^{2k}} \sim e^{-k Q m_\pi t}$.
Since $M \geq \frac{Q}{2}m_\pi$, terms with larger $k$ decay exponentially slower with increasing time than terms with smaller $k$.
The sum is therefore dominated by the term with largest $k$.
For odd $n$, the dominant term at large times has $k= (n-1)/2$ and decays as $e^{-(\frac{n-1}{2}) Q m_\pi t - M t}$,
while for even $n$, the dominant term has $k = (n-2)/2$ and decays as $e^{-(\frac{n}{2} - 1) Q m_\pi t - 2 M t}$.
The additional term besides the sums appearing in Eq.~\eqref{eq:LSgeneral1} dominates for even $n$, giving
\begin{equation}
  \begin{split}
    \avg{C_i^n}  &\sim \begin{cases} e^{-\frac{(n-1)Q}{2}m_\pi t - M t} , & n\text{ odd} \\
      e^{-\frac{Q n}{2} m_\pi t} , & n\text{ even} \end{cases}.
  \end{split}\label{eq:LSgeneral}
\end{equation}
It follows that the real parts of complex correlation functions for generic hadrons in LQCD become increasingly broad and symmetric at late times, with ratios of odd moments to even moments decreasing at a rate fixed (up to hadronic interaction energy shifts) by the number of valence quarks fields used to construct the correlation function.
Analogous results apply to complex or real but non-positive correlation functions in quantum Monte Carlo calculations more generally.

\section{Phase Reweighting Non-Positive Correlation Functions}

\begin{figure}[!t]
  \centering
  \includegraphics[width=\columnwidth]{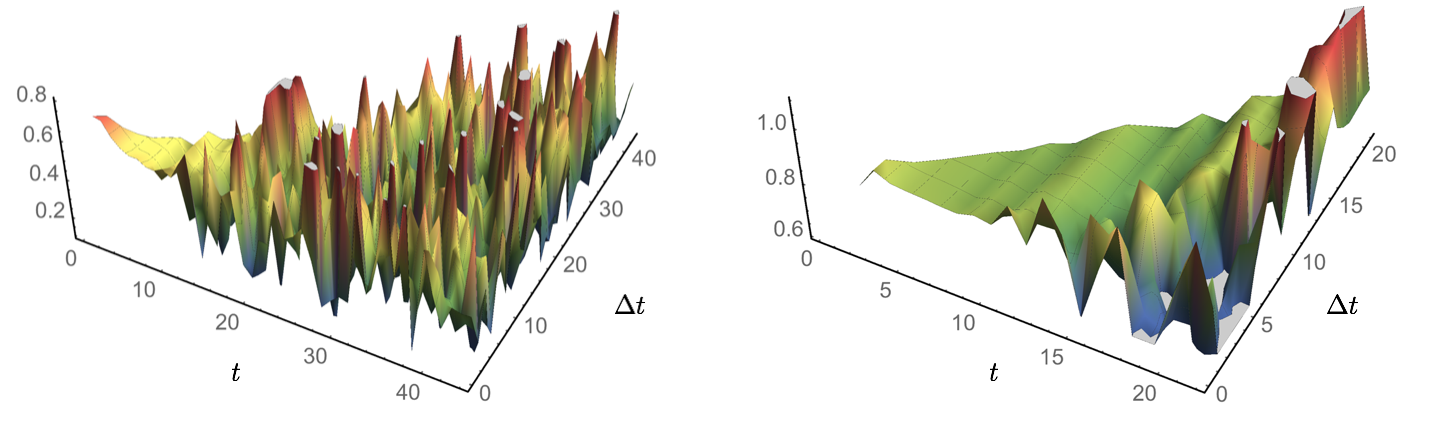} 
  \caption{Results for $\tilde{M}_\rho(t,\Delta t)$ with $m_\pi\sim 450$ MeV, left, and with $m_\pi \sim 800$ MeV, right.}
  \label{Rho2DWEM}
\end{figure}

The previous sections suggest that real but non-positive meson correlation functions have similar statistical behavior to the real parts of baryon correlation functions.
A significant difference between (real) meson and (complex) baryon correlation functions is that ratio-based estimators sampling $C_i(t)/C_i(t-\Delta t)$ analogous to those introduced in Chapter~\ref{chap:statistics} do not remove exponential StN degradation from meson correlations, as shown in
Fig.~\ref{Rho2DWEM}.
The correlation-function-ratio effective mass 
\begin{equation}
  \begin{split}
    \tilde{M}_\Gamma = \ln\left[ \frac{\avg{C_i(t)}}{\avg{C_(t-\Delta t)}} \right] - \ln\left[ \frac{\avg{C_i(t+1)}}{\avg{C_i(t-\Delta t)}}\right]
  \end{split}\label{MGammadef}
\end{equation}
is noisy for $t\gtrsim 15$ when the sign begins contributing appreciably to the effective mass.
A similar problem arises for estimators sampling ratios of the real parts of nucleon correlation functions.
For the real part of the nucleon correlation function, this can be understood because
\begin{equation}
  \begin{split}
    \frac{\text{Re}C_i(t)}{\text{Re}C_i(t-\Delta t)} = e^{R_i(t) - R_i(t-\Delta t)}\left( \frac{e^{i\theta_i(t) - i\theta_i(t-\Delta t)} + e^{-i\theta_i(t) - i \theta_i(t-\Delta t)}}{1 + e^{-2i\theta(t-\Delta t)}} \right).
  \end{split}\label{eq:reratio}
\end{equation}
The first term in the numerator of Eq.~\eqref{eq:reratio} samples a phase difference $e^{i \Delta \theta_i(t, \Delta t)}$ associated with a $\Delta t$ step random walk of the phase.
The second term includes a sum of phases that
constructively rather than destructively interfere
and should have an exponential StN problem in $t$.
It is not obvious that ratios of intrinsically real but non-positive correlation functions must behave similarly,
but Fig.~\ref{Rho2DWEM} shows clear evidence for such an exponential StN problem in $t$.

\begin{figure}[!t]
  \centering
  \includegraphics[width=\columnwidth]{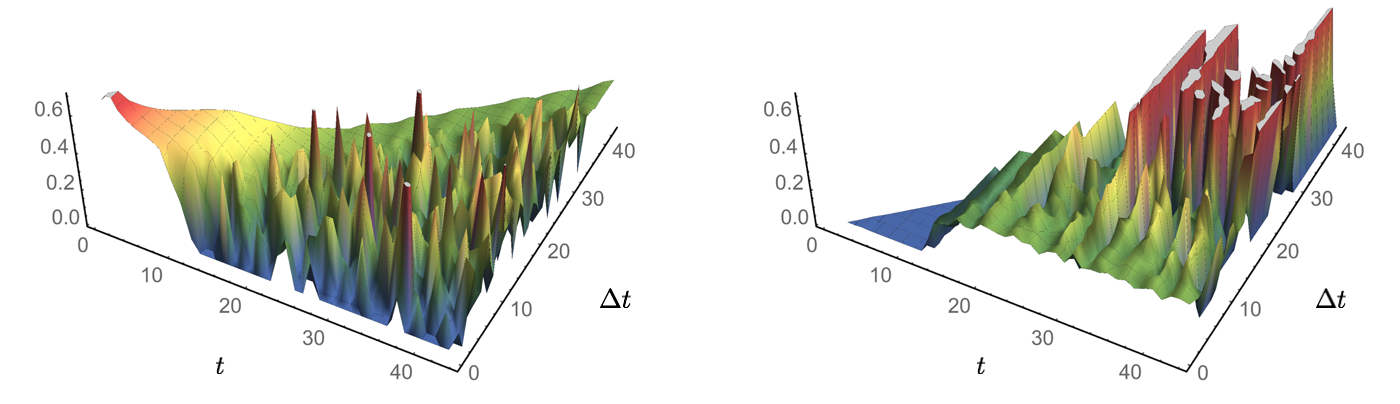} 
  \caption{Results for the $\rho$-meson magnitude and phase ratio-based effective masses $\tilde{M}_R^\rho$, left, and $\tilde{M}_\theta^\rho$, right.}
  \label{Rho2DRThWEM}
\end{figure}

As shown in Fig.~\ref{Rho2DRThWEM}, large noise at large $t$ and small $\Delta t$ in correlation-function-ratio effective masses for $C_i^\rho(t) / C_i^\rho(t-\Delta t)$ is also present in the effective mass for $|C_i^\rho(t)| / |C_i^\rho(t-\Delta t)|$ but not in the effective mass for $e^{i\theta_i^\rho(t) - i\theta_i^\rho(t-\Delta t)}$.
This suggests that explicitly reweighting $C_i^\rho(t)$ by the inverse phase $e^{-i\theta_i(t-\Delta t)}$ (the sign of $C_i(t-\Delta t)$)
faithfully represents a random walk of length $\Delta t$ for the phase,
and motivates the introduction of phase-reweighted correlation functions as defined Chapter~\ref{chap:PR},
\begin{equation}
  \begin{split}
    G^\theta_\Gamma(t) = \avg{ C_i^\Gamma (t) e^{-i\theta_i^\Gamma(t)}} .
  \end{split}\label{eq:mesonPR}
\end{equation}
To partially account for thermal artifacts associated with backwards-propagating states, a symmetrized phase-reweighted effective mass is employed for mesons that is defined as
\begin{equation}
  \begin{split}
    M_\Gamma^\theta(t,\Delta t) = \text{ArcCosh} \frac{G_\Gamma^\theta(t - 1, \Delta t - 1) + G_\Gamma^\theta(t+1, \Delta t + 1)}{G_\Gamma^\theta(t, \Delta t)}.
  \end{split}\label{MGammaACdef}
\end{equation}
As shown in the $\rho^+$-meson results of Chap.~\ref{chap:PR}, phase reweighting tames the StN problem for real but non-positive meson correlation functions as well as complex baryon correlation functions.


\begin{figure}[!t]
  \centering
  \includegraphics[width=\columnwidth]{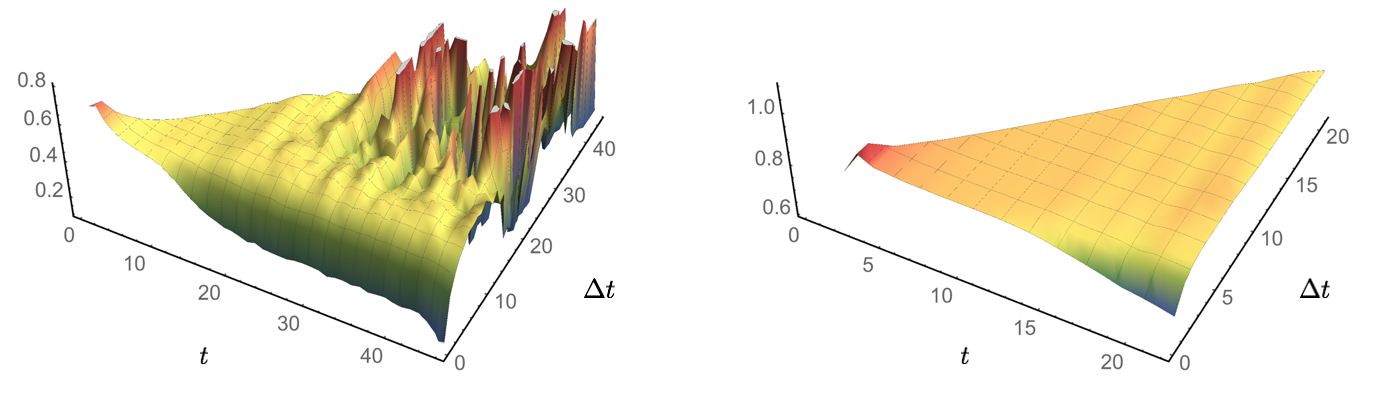} 
  \caption{Results for $M_\rho^\theta(t,\Delta t)$ for the $m_\pi \sim 450$ MeV ensemble left and $m_\pi\sim 800$ MeV ensemble, left. Results should be compared with those for $\tilde{M}(t,\Delta t)$ in Fig.~\ref{Rho2DWEM}.}
  \label{Rho2DEM}
\end{figure}


\begin{figure}[!t]
  \centering
  \includegraphics[width=\columnwidth]{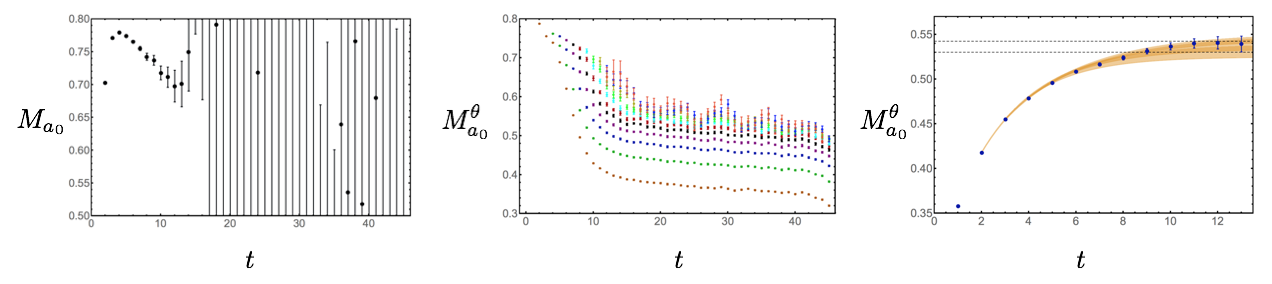} \\
  \includegraphics[width=\columnwidth]{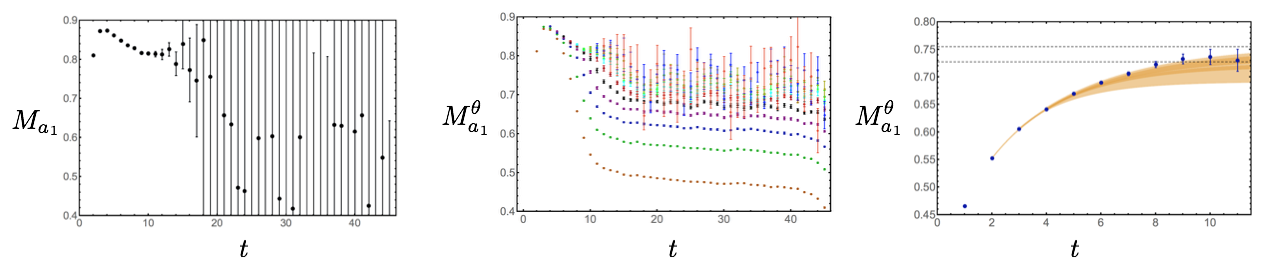} \\
  \includegraphics[width=\columnwidth]{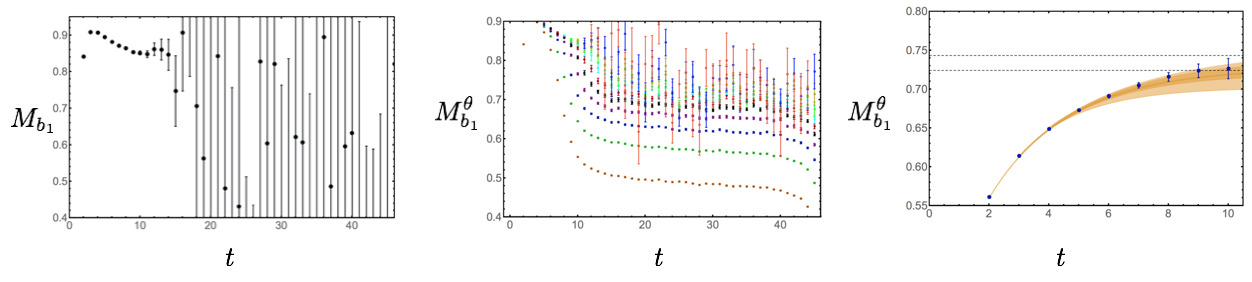}
  \caption{Results for $M_\Gamma(t)$ and $M_\Gamma^\theta(t,\Delta t)$ for the $m_\pi \sim 450$ MeV ensemble for the $a_0$, top, $a_1$, middle, and $b_1$, channels. The left panel shows traditional effective mass $M_\Gamma(t)$ results. The middle panel shows $M_\Gamma^\theta(t,\Delta t)$ as a function of $t$ for a variety of fixed $\Delta t = 1 \rightarrow 9$. The right panel shows large-$t$ plateau results for $M_\Gamma^\theta$ as a function of $\Delta t$ and constant-plus-exponential fits used to extract the $\Delta t \rightarrow t \rightarrow \infty$ mass as in Chapter~\ref{chap:PR}.
    }
  \label{fig:PRmesons}
\end{figure}

\begin{figure}[!t]
  \centering
  \includegraphics[width=\columnwidth]{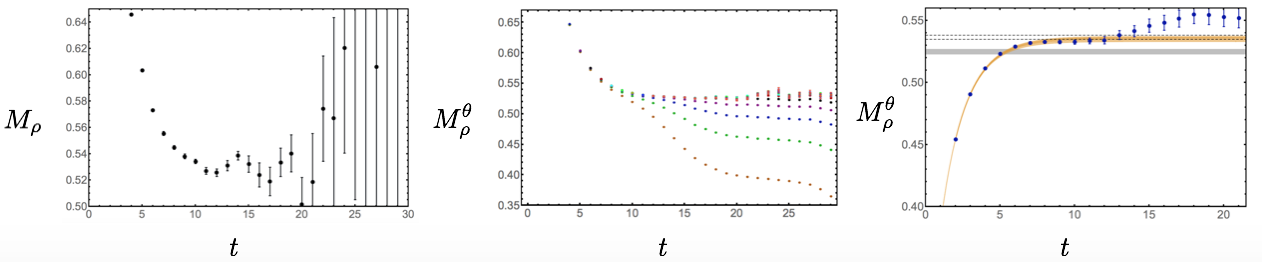} \\
  \includegraphics[width=\columnwidth]{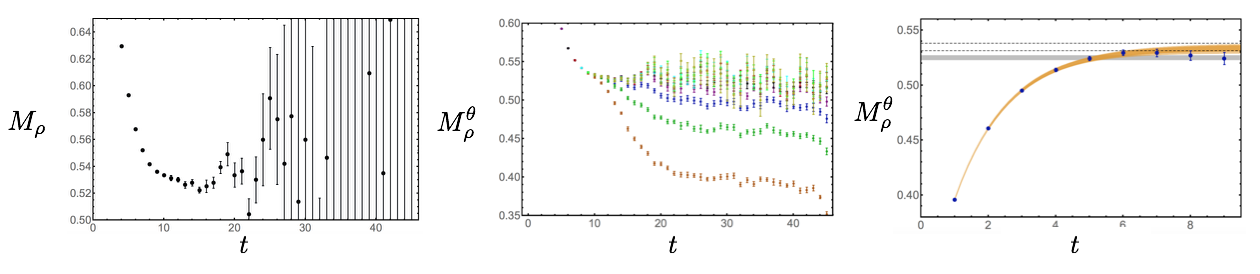} \\
  \includegraphics[width=\columnwidth]{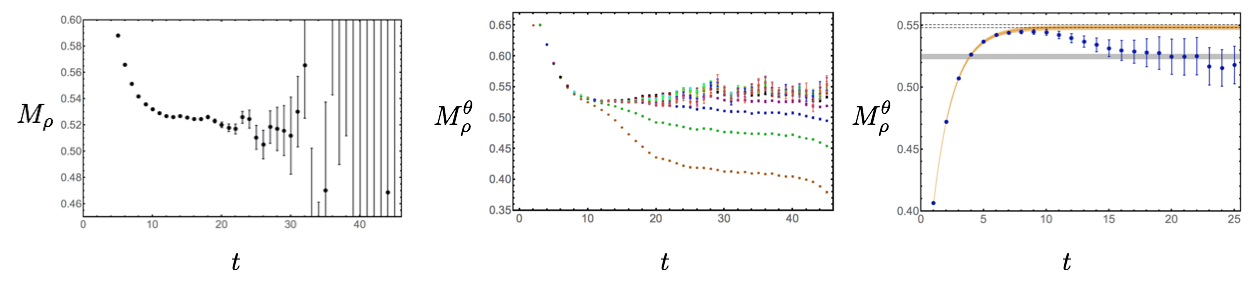}
  \caption{Results for $M_\rho(t)$ and $M_\rho^\theta(t,\Delta t)$ for the $m_\pi \sim 450$ MeV ensemble similar to those of Fig.~\ref{fig:PRmesons} are shown for three different spacetime volumes: $48^3\times 96$, top, $32^3\times 96$, middle, and $24^3\times 64$, bottom. Different sized statistical ensembles are used for each volume, see the main text for more details.}
  \label{fig:PRrho}
\end{figure}

Ground-state saturation is not achieved in standard analyses of isovector meson channels besides the $\pi$ and $\rho^+$ in the LQCD calculations considered here, as seen in Fig.~\ref{fig:PRmesons}.
This suggests that simple $\bar{q}\Gamma q$ interpolating operators have poor onto the signal ground state relative to their overlap onto the noise ground state in these channels.
While this obstructs standard determinations of the ground-state energies of these channels, ground-state saturation of phase-reweighted effective masses occurs after the magnitude and the variance correlation function (dominated by $\avg{|C_i|^2}$) have reached their ground states.
Phase-reweighted ground-state saturation improves with larger overlap onto the variance ground state and should not be obstructed by small relative signal-ground-state overlap to variance-ground-state overlap in $G_\Gamma$.
Figs.~\ref{fig:PRmesons} shows that phase-reweighted ground state saturation can be achieved in all isovector meson channels with large $t\gtrsim 25$ and modest $\Delta t \gtrsim 6 - 10$.

Fig.~\ref{fig:PRrho} shows phase-reweighted results for the $\rho^+$-meson employing the three spacetime volumes $L^3 \times \beta$ of dimension $24^3 \times 64$, $32^3\times 96$, and $48^3\times 96$ with $m_\pi \sim 450$ MeV.
Smaller uncertainties on the $L=48$ ensemble arise from higher statistics $N=600,000$ (including time reversed correlation functions) compared to the $L=32$ ensemble with $N=40,000$.
The $L=24$ ensemble includes a very large sample size of $N=2,880,000$, but
gains less statistical precision by the spatial average involved in momentum projection than the larger volumes.
Larger uncertainties compared to the $L=32$ ensemble also arise from the smaller time direction $\beta = 64$, which restricts the region of usable phase-reweighted correlation function data with $t\gtrsim 25$ to a small number of points.
The fraction of timeslices after variance-ground-state saturation and therefore useful for phase-reweighted calculations increases as the time extent of the lattice is increased.
Since there is no maximum $t$ where phase-reweighted results become overwhelmed by noise, phase-reweighted calculations could be performed more precisely on lattices with larger time extents.

\begin{figure}[!t]
  \centering
  \includegraphics[width=.6\columnwidth]{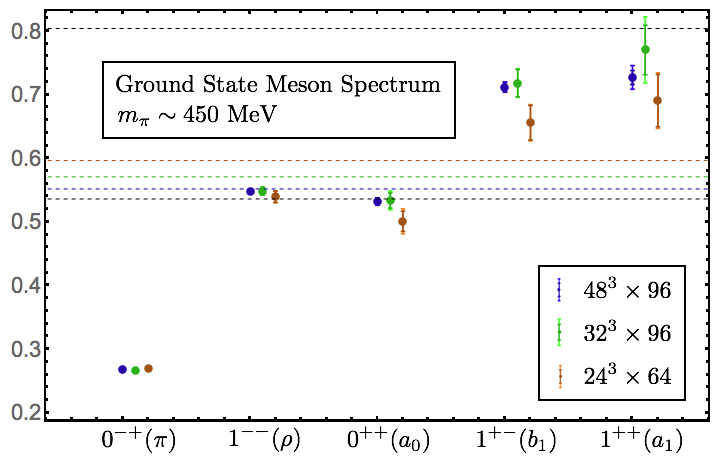}
  \caption{A compilation of $M_\rho^\theta$ results for the $m_\pi \sim 450$ MeV ensemble and all three spatial volumes. Smaller error bars show statistical uncertainty, while large error bars show statistical and systematic uncertainties added in quadrature. Black dashed lines show $2m_\pi$ and $3m_\pi$ for reference, while dashed lines in each color show the non-interacting $p$-wave energy shift $\sqrt{(2m_\pi)^2 + (2\pi/L)^2}$ for the corresponding volume.}
  \label{fig:mesonspec}
\end{figure}

Ground-state isovector meson mass results in all channels from the three volumes are compiled in Fig.~\ref{fig:mesonspec}.
$M_\rho$ is found to be volume-independent within uncertainties over the range of volumes considered here.
In each case $M_\rho$ is below the non-interacting $p$-wave scattering state energy $\sqrt{ (2m_\pi)^2 + (2\pi/L)^2}$ expected for a finite volume state that would associated with an unbound $\pi\pi$ resonance in infinite volume.
The $\rho^+$ is still above the infinite-volume $\pi\pi$ threshold, $M_\rho > 2m_\pi$, and more detailed studies of $\pi\pi$ scattering are required to assess the resonant nature of the $\rho^+$ at $m_\pi \sim 450$ MeV.
Studies by the Hadspec Collaboration in Refs.~\cite{Dudek:2012gj,Dudek:2012xn}
indicate that the $\rho^+$ is slightly unbound at $m_\pi \sim 391$ MeV.
Other meson masses are found to be volume-independent over the range of volumes considered, consistent with the scalings expected for compact bound states.
At these values of the quark masses $M_{a_0} < 2m_\pi$, and therefore the $a_0$ is lighter than the (not precisely known) $K^0\overline{K}^0$ and $\pi\eta$ thresholds.
Furhter studies combining variantional methods, resonance formalism, and phase reweighting are needed
to asses whether the $a_0$-meson is bound at $m_\pi \sim 450$ MeV and better understand light hadron phenomenology with heavier than physical quark masses.

These results demonstrate that phase reweighting can be used to predict ground-state energies of correlation functions without a GW.
The precision of phase reweighted results is limited by the size of the lattice time direction, and in particular no useful results are found for the $m_\pi \sim 800$ MeV ensembles with smaller time directions.
This motivates the generation of LQCD ensembles with very large time directions
where phase reweighting may provide ground-state energy results
for multi-nucleon systems without a GW.
The results of this chapter also suggest that phase reweighting may be applied to real but non-positive correlation functions that appear in GFMC and other many-body methods applicable to particle, nuclear, and condensed matter physics as well as to complex correlation functions describing multi-baryon systems in LQCD.

\chapter{Conclusion and Outlook}\label{chap:conc}

LQCD is emerging as a tool for precise calculations of hadrons and nuclei
that make no uncontrolled assumptions besides the validity of the Standard Model.
Recent calculations of mesons and the nucleon have been performed with physical quark masses,
and in the multi-baryon sector LQCD calculations of electromagnetic and weak fusion reactions have been performed.
In the near future LQCD calculations will be used to predict the structure of nuclei in terms of quarks and gluons at the EIC,
predict electroweak reaction rates relevant for stellar fusion and neutrino-nucleus scattering,
and reliably connect experimental searches for fundamental symmetry violation such as neutrinoless double-beta decay to theoretical bounds on beyond the Standard Model physics.
Matching LQCD calculation to EFTs of nucleons will allow three-body forces and other poorly-known aspects of the nuclear force to be constrained from first principles theory, supporting many-body methods used to study the structure and reactions of larger nuclei.
LQCD calculations can in principle also study large nuclei and the equation of state of dense matter
relevant for understanding the structure of matter in the interior of neutron stars
and the gravitational waves emitted from their collisions.
Practical application of LQCD to systems of large baryon number has been impeded by the sign and StN problems.
This thesis has presented statistical observations of baryon correlation functions relevant for understanding the baryon StN problem as a sign problem afflicting generic complex correlation functions, and has also presented new statistical analysis techniques where
StN degradation is independent of source-sink separation time $t$
and appears instead in a tunable control parameter $\Delta t$.

Chapter~\ref{chap:statistics} presents evidence that nucleon correlation functions are statistically described by an approximately decorrelated product of a log-normal magnitude and wrapped normal phase factor.
The nucleon correlation function magnitude is found to have no StN problem and has the large-time scaling 
$\langle |C(t)| \rangle  \sim e^{-\frac{3}{2}m_\pi t}$. 
The nucleon log-magnitude, $R(t)$, is approximately described by a normal distribution with linearly increasing mean and almost constant variance. 
The complex phase, which gives the direct importance sampling of $C(t)$ a sign problem,  has the large-time scaling of approximately 
$\langle e^{i\theta(t)} \rangle \sim e^{-(M_N - \frac{3}{2}m_\pi)t}$.
This shows that the StN problem arising from solving the sign problem associated with the phase by reweighting is the Parisi-Lepage StN problem.

Building on the observation that $\Delta \theta_i(t,\Delta t)$ has constant width at large times, 
a new estimator for baryon effective masses is studied 
that relies on statistical sampling of correlation function ratios and therefore phase differences.
This estimator has a StN ratio that is constant in $t$, the source-sink separation time, and the StN problem instead leads to an exponentially degrading StN ratio in $\Delta t$, the difference between the numerator and denominator sink times.
  The independence of $t$ and $\Delta t$ in this estimator allows similarly precise results to be extracted from all sufficiently large $t$ rather than from a window of intermediate $t$, as with traditional estimators.
  The new estimator effectively includes $\Delta t$ timesteps of time evolution following $t-\Delta t$ timesteps of dynamical source improvement and it includes a systematic uncertainty that must be eliminated by extrapolating to the limit $ \Delta t \rightarrow t \rightarrow \infty$.
  The systematic uncertainty of the new estimator is expected to decrease as $e^{-\delta E \Delta t}$ for large $\Delta t$, where $\delta E$ is the energy gap between the ground state and the first excited state with appropriate quantum numbers and appreciable overlap with the effective source at $t - \Delta t$.
  Statistical uncertainties increase with increasing $\Delta t$ as $\sim e^{2(M_N - \frac{3}{2}m_\pi)\Delta t}$.
  For $\Delta t \gtrsim \frac{\ln(N)}{2(M_N - \frac{3}{2}m_\pi)}$ additional systematic uncertainties associated with finite-sample-size effects in statistical inference of circular random variables leads to unreliable results in the same way that $t \gtrsim \frac{\ln (N)}{2(M_N - \frac{3}{2}m_\pi)}$ leads to unreliable results in the noise region of standard estimators.

Chapter~\ref{chap:PR} introduces phase reweighting, a refined method of constructing estimators based on sampling phase differences
whose bias is guaranteed to vanish in a well-defined limit.
Reweighting each correlation in a statistical ensemble with a phase factor from the same correlation function at an earlier time
can be intuitively thought of as reducing the number of steps in the random walk of the phase to a fixed interval,
and provides phase-reweighted correlation functions with constant StN ratios in time separation between the source and sink.
Systematic uncertainties in phase reweighting can be understood as arising from excited state contamination, and in particular excitations of the vacuum arising from the boundary introduced by phase reweighting.
Phase reweighting is demonstrated to give accurate results for the $\rho^+$, nucleon, and $\Xi^-\Xi^-$ systems, and in particular the bias in the $\Xi^-\Xi^-$ binding energy is much smaller than the single-hadron bias and consistent with zero within the uncertainties of the calculation.

Chapter~\ref{chap:mesons} applies the statistical tools and phase-reweighting techniques of the previous chapters to isovector meson correlation functions.
These provide an interesting test case because they are real but non-positive.
The moments of generic non-positive correlation functions are proven to have similar scaling properties to the Lepage-Savage scaling of real parts of baryon correlation functions.
The sign of meson correlation functions acts as a discrete circular random variable,
and the bound $\avg{\cos\theta_i^\Gamma} > 1/\sqrt{N}$ for unbiased parameter inference of circular random variables is found to apply to real but non-positive meson correlation functions.
Correlation-function-ratio estimators are found to fail for isovector meson correlation functions in the same way that they fail when applied to only the real parts of baryon correlation functions.
Phase reweighting is further motivated by its comparative success, and is shown to tame the StN problem
facing real but non-positive correlation functions analogously to complex correlation functions.
This extends the scope of possible applications of phase reweighting to many real but non-positive correlation functions with sign problems appearing in a broad array of quantum Monte Carlo calculations in particle, nuclear, and condensed matter physics.

LQCD calculations of multi-baryon systems without a golden window stand to benefit from phase reweighting.
This thesis has discussed re-analysis of existing correlation functions obtained from Monte Carlo ensembles 
optimized for small- and intermediate-time analysis.
The relative precision of phase reweighting compared to the precision of standard techniques is expected to increase with the size of the time direction.
Especially since plateaus in phase-reweighted effective masses are visible only at large $t$ or sometimes not visible at any $t$ in the calculations at hand,
new Monte Carlo ensembles optimized for large-time analysis
will be necessary for phase-reweighted results
to achieve significant gains in precision compared to standard techniques.
This productions and other explorations of complex correlation function statistics and future applications of phase reweighting are underway.

%
\nocite{*}   
\bibliographystyle{plain}
\bibliography{MW_bib}
%
%
\appendix
\raggedbottom\sloppy
 

%
%
%

\end{document}